\documentclass[useAMS,usenatbib,usegraphicx,useepstopdf]{mn2e}

\usepackage[english,english]{babel}
\usepackage{amsmath}
\usepackage{amssymb,amsfonts,textcomp}
\usepackage{supertabular}
\usepackage{hhline}
\usepackage{hyperref}
\usepackage[usenames]{color}
\hypersetup{dvips, colorlinks=true, linkcolor=blue, citecolor=blue, filecolor=blue, urlcolor=blue}
\usepackage[dvips]{graphicx}

\title[AGN feedback for cosmological simulations]
{Self-regulated growth of supermassive black holes by a dual jet/heating AGN feedback mechanism: methods, tests and implications for cosmological simulations}

\author[Y. Dubois et al.]
  {Yohan~Dubois,$^1$\thanks{E-mail: yohan.dubois@physics.ox.ac.uk}
  Julien~Devriendt,$^{1,2}$ Adrianne~Slyz,$^1$ and Romain Teyssier$^{3,4}$ \\
  $^1$Astrophysics, University of Oxford, Denys Wilkinson Building, Keble Road, Oxford, OX13RH, United Kingdom\\
  $^2$Centre de Recherche Astrophysique de Lyon, Universit\'e de Lyon I, CNRS UMR 5574, ENS-Lyon, 9 Avenue Charles Andr\'e,\\ 69561, St-Genis-Laval Cedex, France\\
  $^3$Universit\"at Z\"urich, Institute f\"ur Theoretische Physik, Winterthurerstrasse 190, CH-8057 Z\"urich, Switzerland\\
  $^4$CEA Saclay, DSM/IRFU/SAP, B\^atiment 709, F-91191 Gif-sur-Yvette, Cedex, France}
\date{Accepted 2011 November 18. Received 2011 October 26; in original form 2011 July 30}

\pagerange{\pageref{firstpage}--\pageref{lastpage}} \pubyear{2011}

\def\LaTeX{L\kern-.36em\raise.3ex\hbox{a}\kern-.15em
    T\kern-.1667em\lower.7ex\hbox{E}\kern-.125emX}

\begin{document}

\label{firstpage}

\maketitle

\begin{abstract}
We develop a sub-grid model for the growth of supermassive Black Holes (BHs) and their associated Active Galactic Nuclei (AGN) feedback in hydrodynamical cosmological simulations. This model transposes previous attempts to describe BH accretion and AGN feedback with the Smoothed Particle Hydrodynamics (SPH) technique to the Adaptive Mesh Refinement (AMR) framework. It also furthers their development by implementing a new jet-like outflow treatment of the AGN feedback which we combine with the heating mode traditionally used in the SPH approach. Thus our approach allows one to test the robustness of the conclusions derived from simulating the impact of self-regulated AGN feedback on galaxy formation vis-\`a-vis the numerical method.
Assuming that BHs are created in the early stages of galaxy formation, they grow by mergers and accretion of gas at a Eddington-limited Bondi accretion rate. However this growth is regulated by AGN feedback which we model using two different modes: a quasar-heating mode when accretion rates onto the BHs are comparable to the Eddington rate, and a radio-jet mode at lower accretion rates which not only deposits energy, but also mass and momentum on the grid. In other words, our feedback model deposits energy as a succession of thermal bursts and jet outflows depending on the properties of the gas surrounding the BHs. We assess the plausibility of such a model by comparing our results to observational measurements of the coevolution of BHs and their host galaxy properties, and check their robustness with respect to numerical resolution. We show that AGN feedback must be a crucial physical ingredient for the formation of massive galaxies as it appears to be able to efficiently prevent the accumulation of and/or expel cold gas out of halos/galaxies and significantly suppress star formation.
Our model predicts that the relationship between BHs and their host galaxy mass evolves as a function of redshift, because of the vigorous accretion of cold material in the early Universe that drives Eddington-limited accretion onto BHs. Quasar activity is also enhanced at high redshift. However,  as structures grow in mass and lose their cold material through  star formation and efficient BH feedback ejection, the AGN activity in the low-redshift Universe becomes more and more dominated by the radio mode, which powers jets through the hot circum-galactic medium.
\end{abstract}


\section{Introduction}

Evidence for the ubiquitous presence of supermassive black holes (BH) in the centres of galaxies is overwhelming \citep{kormendy&richstone95}.
BHs spanning a range of masses from a few $10^6\, \rm M_{\odot}$ in the centre of galaxies with small bulges like our Miky-Way~\citep{schodeletal02} up to several $10^9\, \rm M_{\odot}$ for elliptical galaxies in the cores of groups and clusters of galaxies \citep{magorrianetal98} have now been observed.
These supermassive BHs are not only seen in the near universe, but very luminous quasars discovered beyond $z>6$ \citep{fanetal03} suggest that they are already in place during the early stages of galaxy formation. As a consequence, it is now widely accepted that a large variety of galaxies host BHs in their centres, and that these BHs somehow influence the evolution of their host galaxies.

Observations by \cite{magorrianetal98} first pointed out  a relationship  between the central BHs and their host galaxy bulge mass with a quasi-linear scaling \citep{laor01, mclure&dunlop02, marconi&hunt01, haring&rix04}.
A similar, albeit arguably tighter correlation is also found between the BH mass and the stellar velocity dispersion~\citep{ferrarese&merritt00, gebhardtetal00, tremaineetal02, gultekinetal09}, or the S\'ersic index that measures the concentration of the bulge~\citep{graham&driver07}.
These correlations define a BH fundamental plane, similar to the fundamental plane of elliptical galaxies, that links BHs to bulge stellar masses, velocity dispersions and effective radii~\citep{hopkinsetal07fplane}.

These observations led to the suggestion that the growth of BHs is self-regulated by the energy released during their accretion phase. This would be sufficient to unbind the gas in the galaxy and form powerful outflows~\citep{silk&rees98, king03, wyithe&loeb03}. 
There exist abundant observational evidence of  such outflows with direct imaging of X-ray cavities in the vicinity of elliptical galaxies~\citep{boehringeretal93, owenetal00, birzanetal04, mcnamaraetal05, fabianetal06, tayloretal06, dongetal10, dunnetal10} or indirect detection of Broad Line Absorption regions in the spectra of quasars~\citep{chartasetal03, crenshawetal03, poundsetal03}.
These observations are supported by numerical models of the microphysics of BH accretion discs that can drive massive hydro-magnetic outflows and jets~\citep{devilliersetal05, mckinney06, mckinney&blandford09}, and large amounts of heat carried by photons that could potentially ionize the surrounding gas.
Feedback from BHs is commonly called Active Galactic Nuclei (AGN) feedback as the energy is emitted from the centres of galaxies where BH reside. However, this generic 
appellation encompasses various modes of energy release from the central source.

It is commonly believed that two of these modes can describe AGN feedback. These are very similar to the two modes observed for X-ray binaries~\citep{churazovetal05, merloni&heinz08}. The so-called `radio' mode is associated with a strong radio emission filling X-ray depressed cavities with relativistic electrons.  Most of the energy in the radio mode is driven by a mechanical jet feedback mechanism inflating the radio lobe itself.
This mode is equivalent to the `low/hard' state of X-ray binaries that has a hard X-ray power spectrum with a cut-off at a few 100 keV, and gas accretion at low Eddington rates.
The same tendency has been confirmed for supermassive BHs, for which the radio loudness is stronger at lower Eddington accretion ratios~\citep{nagaretal05, chiabergeetal05, churazovetal05}.
This mode is clearly associated with mechanical feedback where most of the energy is powered by the jet mechanism and largely overwhelms the X-ray contribution from the nucleus, as in the well-studied case of M87~\citep{owenetal00}.

A transition to a radio-quiet `quasar' mode occurs at a few $\sim 10^{-2}$ of the Eddington accretion rate for X-ray binaries~\citep{maccarone03}.
Above this threshold, X-ray binaries enter a `high/soft' state emitting a soft and thermal X-ray spectrum with almost no trace of a jet mechanism, which is the equivalent of a quasar spectrum.
The thermal emission is well described by the standard model of optically thick and geometrically thin accretion discs from~\cite{shakura&sunyaev73}, whereas the launching of the jet for the radio mode comes from optically thin and geometrically thick (and radiatively inefficient) accretion discs modeled with the Advection-Dominated Accretion Flow (ADAF) from~\cite{narayan&yi94} or with the Adiabatic Inflow-Outflow Solutions (ADIOS) from~\cite{blandford&begelman99}.

On the other hand, observations of tidally disrupted galaxies often reveal powerful AGN activity~\citep{barnes&hernquist92}.
Mergers between galaxies are invoked to compress the inter-stellar medium (ISM) and provide a fresh flow of material onto the BH in that case.
Theoretical models of galaxy formation that associate the growth of BHs to such events have been successful at reproducing many properties of the population of quasars as well as the BH density seen at low redshift~\citep{cattaneoetal99, kauffmann&haehnelt00, granatoetal01, volonterietal03}. 
Semi-analytic models of galaxy formation require AGN feedback to suppress the cooling catastrophe in massive galaxies, match the bright-end of the galaxy luminosity function, and obtain bulge-dominated galaxies~\citep{crotonetal06, cattaneoetal06, boweretal06, boweretal08, somervilleetal08}.

Important steps have recently been incorporated with hydrodynamical simulation to better quantify the negative AGN feedback effect on star formation and gas properties of galaxies.
These involved simulating idealized disc galaxies and galaxy mergers~\citep{springeletal05, nayakshin&power10, debuhretal10}, or clusters of galaxies~\citep{cattaneo&teyssier07, duboisetal09}. Self-consistent sub-grid models of the AGN feedback heating mode have been introduced in $\Lambda$CDM cosmological simulations using the Smoothed Particle Hydrodynamics (SPH) technique as implemented in the {\sc gadget} code~\citep{sijackietal07, booth&schaye09}. Alternative approaches based on the injection of Cosmic Rays have also been explored~\citep{sijackietal08}.
These cosmological simulations have successfully reproduced relationships between BH masses and galaxy properties~\citep{sijackietal07, dimatteoetal08, booth&schaye09, booth&schaye11}, and suppressed the cooling catastrophe in groups and clusters of galaxies~\citep{puchweinetal08, khalatyanetal08, mccarthyetal10, mccarthyetal11}.
Recently, AGN feedback associated to BH growth has been introduced in Adaptive Mesh Refinement (AMR) cosmological re-simulations of a galaxy cluster with the {\sc ramses} code. These featured either a jet-kinetic mode~\citep{duboisetal10} or a thermal energy input~\citep{teyssieretal11} (see also~\citealp{duboisetal11}), but the cosmic co-evolution of BHs and galaxies has not yet been studied using grid based techniques.

As limitations of the standard SPH technique to capture Kelvin-Helmholtz instabilities have been pointed out \citep{agertzetal07}, and authors like ~\cite{mitchelletal09} have shown that it could have severe consequences on the properties of the intra-cluster gas where AGN-host ellipticals reside, it seems worthwhile to investigate this important issue using a different numerical technique. Therefore, this work uses an Eulerian grid based approach to model AGN feedback from BHs by a dual jet/heating sub-grid model representative of the radio and quasar mechanisms with the self-regulated growth of BHs. We emphasize the importance of these two different modes on the long-term evolution of galaxies, trying to explain why quasars are a common ingredient of the young Universe and why radio jets are more commonly observed in late and massive structures such as clusters of galaxies.

The paper is organised as followed. In section~\ref{subgrid_physics}, we detail the numerical technique used for following the BH growth along with its associated AGN feedback, and the standard models employed for galaxy formation (cooling, star formation, supernovae feedback, etc.). In section~\ref{simulations}, we describe the set of simulations employed to test the AGN feedback model with the {\sc ramses} code. In section~\ref{parameters_resolution}, we present a parameter study of the AGN feedback model and assess the convergence of the results vis-a-vis resolution. Section~\ref{BHevolution} scrutinizes what drives the domination of the quasar and the radio mode at different epochs.  Finally, we comment on the results in section~\ref{conclusion}.

\section{Modelling the physics of galaxy formation}
\label{subgrid_physics}

\subsection{BH growth and AGN feedback}

Sink particles were first introduced by \cite{bateetal95} in a SPH code. Sinks are massive particles that capture gas particles in their surroundings. They mimic the formation of unresolved 
compact objects, e.g. proto-stellar cores in the ISM, black holes in the ISM, central SMBHs in galaxies etc.  Due to the very Lagrangian nature of the sink particle technique, it has been extensively and exclusively used in SPH codes until \cite{krumholzetal04} extended its use to grid codes. The version in {\sc ramses} \citep{teyssier02} is strongly inspired by the \cite{krumholzetal04}  numerical implementation, and has already been presented in \cite{duboisetal10} and \cite{teyssieretal11}, but we reproduce here the details of the numerical implementation to facilitate the discussion of our results.

\subsubsection{Seeding galaxies with BHs}

There are at least two scenarios for the formation of  seed BHs. 
The first one invokes population III stars with zero metallicity. These can produce BH remnants as massive as $10^2$--$10^3\,\rm M_{\odot}$~\citep{madau&rees01, heger&woosley02, schneideretal02} that will eventually rapidly merge in their primordial halo to reach even larger masses.
Another channel of BH formation is the direct collapse of matter from halos with very low angular momentum generating BHs as massive as $10^5\,\rm M_{\odot}$~\citep{loeb&rasio94, bromm&loeb03, begelmanetal06}.
With the kpc-scales typically used in cosmological simulations of galaxy formation, it is pointless to try to follow the formation of these first seeds since this occurs on much smaller scales, but we can take these scenarios as guidelines for a sub-grid generation of seed BHs.

BHs represented by sink particles are created in regions where the Jeans criterion is violated, i.e. in regions where the maximum level of refinement is reached and
where the gas density is large enough to potentially produce a numerical instability, in other words where:
\begin{equation}
{\Delta x\over 4} >  \lambda_{\rm J}=\sqrt{\pi c_s^2\over G \rho}\, .
\label{L_Jeans}
\end{equation}
Here $\Delta x$ is the size of the smallest cell, $\lambda_{\rm J}$ the Jeans length, $c_s$ the sound speed and $\rho$ the gas density. According to
\cite{trueloveetal97}, the numerical stability of a gravitationally bound object is ensured if it is resolved with at least 4 cells. With a mixed composition
of matter (dark matter, gas, stars), Jeans stability is not trivial anymore, but we can reasonably assume that gas is the dominant source of gravitational
potential inside dense collapsed objects, like galaxies, in our case.

For numerical stability, each time that the Jeans criterion is violated we should spawn a sink particle with a mass corresponding to the depleted mass.
However, in cosmological simulations this leads to excessively large sink masses. The reason is that the gas is concentrated in galactic structures that are poorly
resolved with kpc-scale resolution. As a result an entire galactic disk can be defined by only a few Jeans-violating cells leading to excessively  massive sink particles.
To form sufficiently small seed BHs in the centres of the galaxies, we prefer to choose their initial mass, $M_{\rm seed}$, thereby introducing a free parameter. 
We set $M_{\rm seed}=10^5 \, \rm M_{\odot}$ as the default value of our model in agreement with previous cosmological simulations (e.g. \citealp{booth&schaye09}).
Despite choosing the seed mass, BHs are still spawned only in cells belonging to the maximum level of refinement and that verify equation~\ref{L_Jeans}. In 
section~\ref{parameters_resolution}, the importance of the choice of the initial seed mass will be tested.
One consequence of this self-controlled formation of the seed BHs is that they are not allowed to accrete gas when the Jeans criterion is violated. They can only accrete gas by a reasonable physical process such as Bondi accretion. With this prescription for initializing the mass of the seed BHs, it is conceivable
that gas could be numerically violently Jeans unstable, but this issue is partially solved by the consumption of gas in the star forming process that temporarily
restores gravitational stability.

To avoid formation of sink particles in low density regions that are Jeans-unstable, we set a minimum threshold for the density $\rho>\rho_0$ of gas that
can create a new sink, where $\rho_0$ is the same density threshold that we use for star formation. To make sure that sink BHs do not form before 
the very first stars form, we check that the local star density $\rho_*$ calculated with a Cloud-in-Cell (CIC) interpolation verifies
\begin{equation}
f_*={\rho_*\over \rho_*+\rho} > 0.25 \, ,
\label{fraction_star}
\end{equation}
before a new sink particle is spawned, where $\rho$ is the local gas density. Note that these criteria are very similar to those employed by~\cite{bellovaryetal10}, as they also confine the formation of seed BHs to cold, dense, metal poor gaseous regions at the centre of galaxies.

To obtain one BH per massive galaxy only, a halo finder is usually run on-the-fly during the simulation to check if candidate galaxies already host a BH \citep{dimatteoetal05, booth&schaye09}.
We prefer a simpler, more direct, and computationally cheaper approach. To avoid creating multiple BHs inside the same galaxy, we ensure that
each time a cell could potentially produce a sink particle (i.e. it verifies eq.~(\ref{L_Jeans})), it is farther than a minimum radius $r_{\rm min}$
from all other pre-existing sink particles.
This distance has to be larger than the typical size of galactic discs and smaller than the typical average inter-galactic distance.
Test runs suggest that the choice $r_{\rm min}=50$ kpc produces very satisfactory results.

In summary, a sink particle forms out of gas satisfying criteria on:
Jeans instability, gas density threshold, stellar fraction threshold, and minimum distance from other BHs. Once the sink particle is created, it is split into several cloud particles with equal mass. Cloud particles are spread over a $4 \Delta x$ radius sphere
and positioned every $0.5 \Delta x$ in (x,y,z). The exact number of cloud particles in this configuration is therefore $n_{\rm cloud}=2109$ per sink.  
These cloud particles are essentially created to probe the evolution of the region around the BH and provide spatially averaged quantities for the Bondi accretion formula. They move around on the finest time step scale (corresponding to the highest  spatial resolution) and are destroyed and re-created around their sink particles with a given distribution at the beginning of every coarse time step (corresponding to lowest spatial resolution).

On the other hand, sink particles are only updated every coarse time step with quantities that have been evolved through the intermediate calculation with cloud particles. 
They are merged together (mimicking the BH merger) if they stand at a distance closer than $4\Delta x$ from each other. Mass is conserved in this process and momentum vectors of the old sink particles are simply added to compute the momentum of the new sink particle. They are also the source AGN feedback.

Finally we insist on the fact that BH positions and velocities are updated in the classical way used to update standard particles such as DM particles, i.e. using the Particle-Mesh solver of {\sc ramses} with CIC interpolation of particle masses into cells.
No correction on their positions and velocities is done to force them to stay near their host galaxy (as could be done with a halo
finder approach). Thus, weakly bound objects, such as BHs in galaxy satellites of large groups and clusters, may be
stripped from their host galaxy. These BHs behave like star particles that tidal forces compel to populate the stellar halo of massive galaxies.

\subsubsection{Accretion rate}

Since we do not resolve the accretion disks around BHs, whose sizes are sub-parsec even for the most massive ones ($\sim 10^{-3}$ pc according to \citealp{morganetal10} from micro-lensing estimates), we use the most common prescription that these BHs accrete gas at a Bondi-Hoyle-Lyttleton rate \citep{bondi52}
\begin{equation}
\dot M_{\rm BH}={4\pi \alpha G^2 M_{\rm BH}^2 \bar \rho \over (\bar c_s^2+\bar u^2) ^{3/2}}\, ,
\label{dMBH}
\end{equation}
where $\alpha$ is a dimensionless boost factor ($\alpha\ge 1$), $M_{\rm BH}$ is the BH mass, $\bar \rho$ is the average gas density, $\bar c_s$ is the average sound speed,
and $\bar u$ is the average gas velocity relative to the BH velocity. One of the major difficulties encountered with the computation of the relative gas velocity is that in cosmological runs, the
ISM is poorly resolved and leads to galaxy scaleheights comparable to the resolution which is much larger than the scaleheights of galaxies in nature. Moreover, due to limited sampling of the gravitational force in the
galactic disc, BHs can wander in their host galaxy. For this reason a BH close to the centre of a galaxy can feel the infalling material coming from the halo or the ISM
at a relative velocity much higher than the typical velocity inside the bulge.
Therefore because $\bar u$ is not a reliably measured  quantity, we enforce the relative velocity to be no larger than an average gas velocity dispersion in the ISM which is assumed constant and equal to $u_{\rm max}=10\, \rm km.s^{-1}$ for our fiducial model \citep{dibetal06}.
We will test the effect of varying this maximum allowed value on the properties of BHs.

The average density $\bar \rho$ and sound speed $\bar c_s$ are computed around the BH using the cloud particles for this operation, as mentioned in the previous section.
To compute the averages, the cell in which each cloud particle sits is assigned a weight given by a kernel function $w$, similar to the one used in \cite{krumholzetal04}:
\begin{equation}
w\propto\exp \left( -r^2/r_K^2\right )\, ,
\end{equation}
where $r$ is the distance from the cloud particle to the sink particle and $r_K$ is the radius defined as
\begin{equation}
r_K =
\left\{
\begin{array}{lr}
\Delta x/4 & \,\, r_{\rm BH}<\Delta x /4\, ,\\
r_{\rm BH}&\,\, \Delta x /4 \le r_{\rm BH} \le 2 \Delta x\, , \\
2\Delta x & \,\, r_{\rm BH}>2\Delta x\, .
\end{array}
\right.
\end{equation}
The Bondi-Hoyle radius $r_{\rm BH}$ is given by:
\begin{equation}
r_{\rm BH}={GM_{\rm BH} \over c_s^2}\, ,
\end{equation}
where $c_s$ is the exact sound speed in the cell where the sink lies.

The accretion rate onto the sink is finally limited by its Eddington rate
\begin{equation}
\dot M_{\rm Edd}={4\pi G M_{\rm BH}m_{\rm p} \over \epsilon_{\rm r} \sigma_{\rm T} c}\, ,
\label{dMEdd}
\end{equation}
where $\sigma_{\rm T}$ is the Thompson cross-section, $c$ is the speed of light, $m_{\rm p}$ is the proton mass, and $\epsilon_{\rm r}$ is the radiative efficiency, assumed to be equal to $0.1$ 
for the \cite{shakura&sunyaev73} accretion onto a Schwarzschild BH.

The accretion rate is computed at each time step and a fraction $\dot M_{\rm BH} \Delta t / n_{\rm cloud}$ of gas mass is depleted from the cell where the cloud particle lies and is added to
that cloud particle and to the sink particle. For each coarse time step of the simulation, cloud particles are re-scattered with equal-mass $M_{\rm BH}/n_{\rm cloud}$.
As the timestep does not depend on the accretion speed onto BHs and as low-density cells can be close to high density cells, a BH might remove more mass than is acceptable.
To avoid dealing with negative or extremely low gas densities and numerical instabilities arising from this, we do not allow any cloud particle to deplete more than 25\% of the gas content in a cell.

With large-scale cosmological simulations and the limited typical kpc-scale resolution, we cannot resolve the scale and the clumpiness of the ISM that require parsec-like resolution~\citep{powelletal11}. To prevent the collapse of the gas from numerical
instabilities and to take into account the mixing of the different phases in the ISM (cold and warm components), we use the polytropic EoS described in 
section \ref{sec:sf}. Applying this EoS means that it is impossible to track the `true' density and 
sound speed in the ISM, thus the accretion rate onto BHs must be modified. Early works modelling the accretion rate onto BHs with such a polytropic EoS
set the boost factor to a constant value $\alpha=100$ \citep{springeletal05, sijackietal07, dimatteoetal08}. Here we follow the prescription from \cite{booth&schaye09} who show
that $\alpha = (\rho/\rho_0)^2$ for $\rho > \rho_0$ where $\rho_0=0.1\, \rm H.cm^{-3}$ is the threshold for star formation, and $\alpha = 1$ for $\rho \leq \rho_0$ give reasonable results compared to observational predictions.

We insist on the fact that this polytropic EoS (equation~\ref{polytropic_EoS}) has important consequences on the accretion rate onto BHs in high gas density regions: equation~(\ref{dMBH}) turns into
$\dot M_{\rm BH}\propto{M_{\rm BH}^2 \rho^{5/2}}$, and the temperature dependence vanishes. On the other hand, as soon as the cold gas component
has been evaporated by star formation or feedback mechanisms giving $\rho \leq \rho_0$ in massive galaxies, the accretion rate of the BH is, by definition,
the proper ($\alpha = 1$) Bondi accretion rate. This $\alpha$ boost of the accretion rate is an artificial way of modeling the very fast accretion of gas within
cold and gas-rich galaxies at early epochs, where the clumpiness of the ISM due to gas-disc fragmentation is unresolved in large-scale cosmological simulations.

\subsubsection{AGN feedback: Quasar and Radio mode}
\label{agn_fbk:quasar_radio}
It is believed that the feedback from AGN can proceed in two distinct modes. 
The quasar mode is essentially seen in the high redshift Universe and proceeds by emitting large amounts of radiation that can photo-ionize and heat gas.
It is assumed in reionisation models of the IGM that quasars are an important contribution to the UV background~\citep{haardt&madau96}.
The radio mode of AGN feedback, on the other hand, proceeds at lower redshifts in the cores of massive galaxy halos.
The typical signatures of this radio mode are inflated cavities with strong magnetic fields and high levels of cosmic ray energy.

Our aim is to treat self-consistently both modes in the simulation according to very simple prescriptions.
It is believed that the radio mode is preferentially triggered during low accretion-rate episodes, and that the quasar mode occurs when gas accretion takes place at rates comparable to the Eddington limit \citep{churazovetal05, merloni&heinz08}.
We use the ratio of accretion rate to its Eddington limit 
\begin{equation}
\chi={\dot M_{\rm BH}\over \dot M_{\rm Edd}}
\end{equation}
as the criterion to determine which of the two AGN modes is active.
Following \cite{merloni&heinz08}, we take  $\chi_{\rm radio}=10^{-2}$ as the value dividing the radio from the quasar mode.
Above $\chi>\chi_{\rm radio}$, the AGN undergoes quasar-like activity with energy mostly emitted by photons. We model this mode by thermal injection of energy.
Below $\chi \le \chi_{\rm radio}$, BHs smoothly accrete gas and provide a radio-mode feedback which is modeled by our kinetic jet implementation.
We point out that a similar approach has been taken by~\cite{sijackietal07}, but they treat both modes as thermal inputs of energy with different injection radii.

For both modes, we assume that a fraction $\epsilon_{\rm f}$ of the radiated energy, $L_r$, is released to the ambient gas
\begin{equation}
\dot E_{\rm AGN}=\epsilon_{\rm f} L_r=\epsilon_{\rm f} \epsilon_{\rm r} \dot M_{\rm BH}c^2\, ,
\label{E_BH}
\end{equation}
where $\epsilon_{\rm f}$ is a free parameter that depends on the mode that is triggered by the accretion. As the energy is continuously released, time scales for dissipating energy by cooling can be sometimes far smaller than the hydro time step. This problem is often encountered in SN feedback modelling \citep{navarro&white93} and leads to a null dynamical impact on the surrounding gas. So as to impact the ambient medium, some authors release the AGN heating energy only when a sufficient amount of gas has been accreted \citep{sijackietal07, booth&schaye09}.
Our modeling of the quasar mode ($\chi>\chi_{\rm radio}$) as a heating mode is very similar to the approach adopted by~\cite{booth&schaye09} (see also \citealp{teyssieretal11}): 
we store the rest mass energy of gas accreted onto the BH until it would be enough to raise the temperature of the gas around the BH above $T=10^7$K. At that point we release the
energy as thermal energy within a bubble of radius $r_{\rm AGN}$ around the BH with efficiency $\epsilon_{\rm f}=0.15$.  

For the radio mode ($\chi \le \chi_{\rm radio}$) we model the AGN feedback with a jet-like outflow with the same profile as in \cite{ommaetal04} (see also \citealp{duboisetal10}). Mass, momentum and energy are spread over a small cylinder of radius $r_{\rm J}$ and height 2$h_{\rm J}$ multiplied by a kernel window function
\begin{equation}
  \psi \left (r_{\rm cyl} \right)={1 \over 2\pi r_{\rm J}^2} \exp \left( -{r_{\rm cyl}^2\over 2 r_{\rm J}^2 } \right) \, ,
\end{equation}
where $r_{\rm cyl}$ is the cylindrical radius, and where we impose $r_{\rm J}=h_{\rm J}=r_{\rm AGN}$. The size of the jet,  $r_{\rm AGN}$, for the radio mode and the size of the bubble,
$r_{\rm AGN}$, for the quasar mode are parameter choices which we will test in section~\ref{AGN_input_size}. Note that once the radius is chosen, it remains fixed for the duration of the simulation. By contrast, \cite{sijackietal07} use prescriptions where the size of the radio bubbles depends
 on the amount of energy released and the gas density (see also \citealp{barai08}). This different choice stems from the fact that these authors model the formation of bubbles by assuming they are the result of large radio cocoons inflated by jets, whereas we attempt to directly model the jet.
 
The mass deposition for the radio mode follows
\begin{equation}
\dot M_{\rm J} \left (r_{\rm cyl} \right)={\psi \left (r_{\rm cyl} \right) \over \| \psi \|} \eta \dot M_{\rm BH} \, ,
\end{equation}
where $\| \psi \|$ is the integrated value of $\psi$ over the whole cylinder, and $\eta=100$ is an arbitrary value that represents the mass loading factor of the jet on unresolved scales. The value of $\eta$ adopted here corresponds to a sub-relativistic bipolar outflow with velocity ~10 000 km/s, rather than that which is expected from a radio loud relativistic jet launched from the BH horizon. 
Thus our modelling should be interpreted as a tentative description of the wind arising from a larger region surrounding the BH, and is expected to carry most of the momentum (see Omma et al 2004
for a more thorough discussion of this issue). Such a choice also allows one to keep the Courant time step of the simulation under control whist retaining a physically motivated model of the the jet outflow propagation on kpc scales.

Mass is transferred from the central cell (where the BH lies) to all the cells enclosed within the jet. Momentum, $q$, is deposited in outflowing opposite directions from the centre along the jet axis, according to
\begin{equation}
\|Ê\mathbf {\dot q_{\rm J} }\| \left (r_{\rm cyl} \right)
= \dot M_{\rm J} \left (r_{\rm cyl} \right)\| \mathbf{u_{\rm J}\|}
={\psi \left (r_{\rm cyl} \right) \over \| \psi \|} \dot M_{\rm BH} \sqrt{2 \epsilon_{\rm f}\epsilon_{\rm r}\eta} c {\mathbf{j}.{\rm d}\mathbf{r}\over \| {\rm d}\mathbf{r} \| } \, ,
\end{equation}
where $\| \mathbf{u_{\rm J}} \| =(2\epsilon_{\rm f}\epsilon_{\rm r}/\eta)^{1/2} c$ is the velocity of the jet, ($\| \mathbf{u_{\rm J}} \| \simeq 9487 \, \rm km.s^{-1}$ for $\epsilon_{\rm f}=1$ and $\eta=100$), $\mathbf{j}$ is the spin vector of the BH and defines the jet axis, and ${\rm d}\mathbf{r}$ is the distance vector from the centre of the BH. $\mathbf{j}$ is computed by adding the different contributions from the neighboring cells, sampled with the cloud particles, to the total angular momentum
\begin{equation}
\mathbf{J}=\sum_{i=1}^{n_{\rm clouds}} m_{i} {\rm d}\mathbf{r_{i}}\times \mathbf{u_{i}} \, ,
\end{equation}
where $m_{i}$ and $\mathbf{u_{i}}$ are the mass and velocity of the gas in the cell harbouring the cloud particle, i
so that $\mathbf{j}=\mathbf{J}/\| \mathbf{J}\|$. Finally the kinetic energy released within a single cell is
\begin{equation}
\dot E_{\rm J}\left (r_{\rm cyl} \right)
={\mathbf {\dot q_{\rm J} }^2 \left (r_{\rm cyl} \right) \over 2 \dot M_{\rm J}\left (r_{\rm cyl} \right)}
={\psi \left (r_{\rm cyl} \right) \over \| \psi \|} \dot E_{\rm AGN} \, .
\end{equation}
Integrating the energy over all the cells within the jet, we verify that the energy is strictly equal to $\dot E_{\rm AGN}$ given in eq.~\ref{E_BH}. Energy efficiency $\epsilon_{\rm f}$ is a free parameter which we take to have different values depending on the AGN feedback mode, with  $\epsilon_{\rm f, r}=1$ and $\epsilon_{\rm f, q}=0.15$  our fiducial values for the radio and  quasar mode respectively. 

High values of $\epsilon_{\rm f, r} \sim 1$ for the radio mode of AGN feedback are consistent with relativistic MHD simulations of BH accretion discs \citep{devilliersetal05, hawley&krolik06} for
 maximally spinning BHs. Using analytic arguments, \cite{benson&babul09} argue that balancing the spin up of BHs caused by gas accretion by angular momentum extraction through a jet, naturally 
 leads to an equilibrium high spin value of the BHs, i.e. $a/M=0.93$ corresponding to typical efficiencies $\epsilon_{\rm f,r} \sim 0.1$ and which seem to agree quite well with local observations \citep{allenetal06}. Of course such a picture is simplistic because of potentially rapid change of BHs spin during mergers, and the accretion mode of BHs (thin disks or ADAFs) is uncertain. However, it has the merit of yielding a very  straightforward prediction of what should be the spin (and, thus the efficiency) of an isolated BH with an ADAF-like accretion disc, model which is consistent with our wind feedback assumptions.

Unlike in \citet{duboisetal10}, \citet{duboisetal11}, in this implementation of kinetic AGN feedback (the radio mode), we allow for a time delay between the energy release and its mass accretion similar to what \cite{sijackietal07} do for thermal bubbles in their radio mode.
The idea is that the energy is released into a kinetic jet when the BH has grown by more than $\Delta M_{\rm d} \%$ of its mass.
This parameter, $\Delta M_{\rm d}$,  gives a relative, but artificial, control on the timescales of AGN feedback and their duty cycle, by allowing long periods during which the AGN is off and short periods during which the energy is released. 

Our approach for this dual radio/quasar mode of AGN feedback is obviously and voluntarily more simplistic than reality.
As said before, the quasar mode involves the release of soft X-ray photons for which radiative transfer effects are not negligible.
Also even during the quasar mode it is possible to get a faint radio detection, though it is not completely clear whether such a signal would be coming from the remnant of a 
previously active radio mode or whether it is an intrinsic signal of the quasar mode.
Jets are filled with a non-thermal cosmic ray component and strong magnetic fields that could have important consequences on the dynamics of the jet.
Moreover, the transition from the radio mode to the quasar mode very likely takes place at a different accretion ratio $\chi$ than the transition from the quasar to the radio mode, reflecting the changing nature of the accretion disc.

\subsection{Modelling star formation and stellar feedback}
\label{sec:sf}

Gas in our simulation radiates energy by atomic collisions in a H/He primordial composition gas \citep{sutherland&dopita93} down to $T_0=10^4$ K, so that it can collapse into DM
potential wells to form galaxies.
We also account for the enhancement of cooling by metals released in SN explosions from massive stars.
The metals are passively advected with the gas and a Solar composition of heavy elements is assumed.
The minimum temperature $T_0$ reached is not modified by the presence of metals but they allow for a more efficient cooling.
Heating from a UV background is considered following \cite{haardt&madau96} during and after the redshift of reionisation which we take to be $z_{reion}=10.5$.

Star formation occurs in high-density regions with gas density $\rho> \rho_0= 0.1\, \rm H.cm^{-3}$ using a random Poisson process to spawn star cluster particles, according to a Schmidt-Kennicutt law 
\begin{equation}
\dot \rho_*= \epsilon_* {\rho \over t_{\rm ff}}\, ,
\end{equation}
where $\dot \rho_*$ is the star formation rate density, $\epsilon_*$ the star formation efficiency, and $t_{\rm ff}$ the gas fee-fall time.
In these simulations we set the efficiency of star formation to $\epsilon_*=0.02$ in good agreement with observational surface density relationships of galaxies \citep{kennicutt98}, and local giant molecular clouds \citep{krumholz&tan07}.
Each star cluster particle has a mass of $m_*=\rho_0\Delta x^3$, reaching $3.6 \,10^5\, \rm  h^{-1}.M_{\odot}$ for our most resolved simulation with $\Delta x=0.38\, \rm  h^{-1}.kpc$.
For numerical stability, we check that no more than 90\% of the gas in a cell is depleted during the star formation process for numerical stability.

We account for the mass and energy release from type II supernovae (SNe) assuming a Salpeter Initial Mass Function (IMF).
Using this IMF, $10\%$ of the stars more massive than  $10\, \rm M_{\odot}$ end their life as type II SN releasing $10^{51}$ erg per $10 \, \rm M_{\odot}$.
Direct thermal input of energy from SN has been identified as an inefficient way of returning energy back into the ISM because thermal energy is efficiently radiated away by gas cooling
in high density regions  \citep{navarro&white93}.
Approaches to circumvent this include temporarily switching off gas cooling to allow the blast wave to propagate or kinetic energy feedback.
The method from \cite{dubois&teyssier08winds} implements the SN energy input by releasing mass, momentum, and kinetic energy locally into the surrounding gas according to a Sedov blast wave solution.
The explosion takes place 10 Myr after the birth of a star cluster particle and a fraction of the gas in the cell where the star particle resides is carried into the neighboring cells with a mass loading factor $f_w=1$.
We assume that type II SNe release all their mass into the gas (no stellar remnant) with a $y=0.1$ stellar yield, which is the fraction of primordial gas transformed into heavy elements and released back into the ISM.
Our prescription does not take into account the energy and mass release from stellar winds (AGB stars), nor from long-lived type Ia SNe.

In order to take into account the thermal impact of the heating of the ISM by SNe, we modify the temperature at high density $\rho> \rho_0$ with a polytropic EoS
\begin{equation}
T=T_0 \left ( {\rho \over \rho_0} \right )^{p-1}\, , 
\label{polytropic_EoS}
\end{equation}
where $p$ is the polytropic index of the gas.
The adopted value of $p=4/3$ is comparable to the value obtained in \cite{springel&hernquist03} through analytic considerations of the multiphase structure of the ISM with stellar heating.
This value of $p=4/3$ does not rigorously ensure that gas will not fragment because of  numerical instabilities \citep{trueloveetal97}, as with this polytropic index the Jeans length is proportional to the gas density
\begin{equation}
\lambda_{\rm J} = 10.7 \left( { \rhoÊ\over 0.1\, \rm H.cm^{-3} }\right )^{-1/3} \, \rm kpc.
\end{equation}
This last formula shows that at very high gas densities the Jeans length can be smaller than our minimum resolution and would cause spurious fragmentation of the gas.
Fortunately, the gas cannot infinitely condense because of force resolution sampling and the star formation process that removes gas in the ISM.
We did not choose the steeper, but safer, polytropic EoS, $p=2$, because it leads to very thick galactic discs in simulations with kpc resolution.

\section{Numerical aspects}
\label{simulations}

\begin{table*}
\caption{Simulations performed with different sub-grid galactic models, different parameters for the AGN feedback mode, and different resolutions. (a) Name of the simulation. (b) Number of DM particles. (c) Mass resolution of a DM particle. (d) Size of the simulation box. (e) Minimum resolution reached at $z=0$. (f) Presence of feedback from SNe. (g) Presence of AGN feedback: ``BH'' stands for the formation and growth of BHs without AGN feedback, ``Jet'' stands for the radio mode only, ``Heat'' stands for the quasar mode only, and ``JET/HEAT'' stands for the quasar and radio mode both triggered in the same simulation (see text (section~\ref{agn_fbk:quasar_radio}) for details). (h) AGN feedback efficiency. (i) AGN energy delay. (j) Maximum relative velocity of the gas to the BH. (k) Mass loading factor of the jet. (l) Initial BH mass. (m) Size of the region for AGN energy input.}
\label{tabnames}
\begin{tabular}{@{}|l|c|c|c|c|c|c|c|c|c|c|c|c|}
  \hline
  (a) & (b) & (c) & (d) & (e) & (f) & (g) & (h) & (i) & (j) & (k) & (l) & (m) \\
  Name & $N_{\rm DM}$ & $M_{\rm DM}$ & $L_{\rm box}$ & $\Delta x$ & SN & AGN & $\epsilon_f$ & $\Delta M_{\rm d}$ & $u_{\rm max}$& $\eta$ & $M_{\rm seed}$ & $r_{\rm AGN}$ \\
   & & ($\rm M_{\odot}/h$) & ($\rm Mpc/h$) &  ($\rm kpc/h$) &  &  &  & $\%$ & (km/s) & &($\rm M_{\rm \odot}$) &  \\
  \hline
  \hline
  256L12noAGN & $256^3$ & $6.9\,10^6$ & 12.5 & 0.38 & Yes & No & -- & -- & -- & -- & -- & -- \\
  256L12JH & $256^3$ & $6.9\,10^6$ & 12.5 & 0.38 & Yes & Jet/Heat & 1/0.15 & 0/-- & 10 & 100/-- & $10^5$ & $\Delta x$ \\
  \hline
  \hline
  64L25JH & $64^3$ & $3.5\,10^9$ & 25 & 3.04 & Yes & Jet/Heat & 1/0.15 & 0/-- & 10 & 100/-- & $10^5$ & $\Delta x$ \\
  \hline
  128L25noAGN & $128^3$ & $4.4\,10^8$ & 25 & 1.52 & Yes & No & -- & -- & -- & -- & -- & -- \\
  128L25BH & $128^3$ & $4.4\,10^8$ & 25 & 1.52 & Yes & BH & -- & -- & 10 & -- & $10^5$ & -- \\
  128L25J & $128^3$ & $4.4\,10^8$ & 25 & 1.52 & Yes & Jet & 1 & 0 & 10 & 100 & $10^5$ & $\Delta x$ \\
  128L25Je0.15 & $128^3$ & $4.4\,10^8$ & 25 & 1.52 & Yes & Jet & 0.15 & 0 & 10 & 100 & $10^5$ & $\Delta x$ \\
  128L25Je0.01 & $128^3$ & $4.4\,10^8$ & 25 & 1.52 & Yes & Jet & 0.01 & 0 & 10 & 100 & $10^5$ & $\Delta x$ \\
  128L25Jm1 & $128^3$ & $4.4\,10^8$ & 25 & 1.52 & Yes & Jet & 1 & 1 & 10 & 100 & $10^5$ & $\Delta x$ \\
  128L25Jm10 & $128^3$ & $4.4\,10^8$ & 25 & 1.52 & Yes & Jet & 1 & 10 & 10 & 100 & $10^5$ & $\Delta x$ \\
  128L25Jv100 & $128^3$ & $4.4\,10^8$ & 25 & 1.52 & Yes & Jet & 1 & 0 & 100 & 100 & $10^5$ & $\Delta x$\\
  128L25Jv1000 & $128^3$ & $4.4\,10^8$ & 25 & 1.52 & Yes & Jet & 1 & 0 & 1000 & 100 & $10^5$ & $\Delta x$\\
  128L25J$\eta$10 & $128^3$ & $4.4\,10^8$ & 25 & 1.52 & Yes & Jet & 1 & 0 & 10 & 10 & $10^5$ & $\Delta x$ \\
  128L25J$\eta$1000 & $128^3$ & $4.4\,10^8$ & 25 & 1.52 & Yes & Jet & 1 & 0 & 10 & 1000 & $10^5$ & $\Delta x$ \\
  128L25Js0.1 & $128^3$ & $4.4\,10^8$ & 25 & 1.52 & Yes & Jet & 1 & 0 & 10 & 100 & $10^4$ & $\Delta x$\\
  128L25Js10 & $128^3$ & $4.4\,10^8$ & 25 & 1.52 & Yes & Jet & 1 & 0 & 10 & 100 & $10^6$ & $\Delta x$\\
  128L25J2dx & $128^3$ & $4.4\,10^8$ & 25 & 1.52 & Yes & Jet & 1 & 0 & 10 & 100 & $10^5$ & $2\Delta x$ \\
  128L25J4dx & $128^3$ & $4.4\,10^8$ & 25 & 1.52 & Yes & Jet & 1 & 0 & 10 & 100 & $10^5$ & $4\Delta x$ \\
  128L25H & $128^3$ & $4.4\,10^8$ & 25 & 1.52 & Yes & Heat & 0.15 & -- & 10 & -- & $10^5$ & $\Delta x$ \\
  128L25H2dx & $128^3$ & $4.4\,10^8$ & 25 & 1.52 & Yes & Heat & 0.15 & -- & 10 & -- & $10^5$ & $2\Delta x$ \\
  128L25H4dx & $128^3$ & $4.4\,10^8$ & 25 & 1.52 & Yes & Heat & 0.15 & -- & 10 & -- & $10^5$ & $4\Delta x$ \\
  128L25JH & $128^3$ & $4.4\,10^8$ & 25 & 1.52 & Yes & Jet/Heat & 1/0.15 & 0/-- & 10 & 100/-- & $10^5$ & $\Delta x$ \\
  \hline
  256L25noSNAGN & $256^3$ & $5.5\,10^7$ & 25 & 0.76 & No & No & -- & -- & -- & -- & -- &-- \\
  256L25noAGN & $256^3$ & $5.5\,10^7$ & 25 & 0.76 & Yes & No & -- & -- & -- & -- & -- &-- \\
  256L25JH & $256^3$ & $5.5\,10^7$ & 25 & 0.76 & Yes & Jet/Heat & 1/0.15 & 0/-- & 10 & 100/-- & $10^5$ & $\Delta x$ \\
  \hline
  \hline
  128L50noAGN & $128^3$ & $3.5\,10^9$ & 50 & 3.04 & Yes & No & -- & -- & -- & -- & --  &--\\
  128L50JH & $128^3$ & $3.5\,10^9$ & 50 & 3.04 & Yes & Jet/Heat & 1/0.15 & 0/-- & 10 & 100/-- & $10^5$ & $\Delta x$ \\
  \hline
  256L50noAGN & $256^3$ & $4.4\,10^8$ & 50 & 1.52 & Yes & No & -- & -- & -- & -- & -- &-- \\
  256L50JH & $256^3$ & $4.4\,10^8$ & 50 & 1.52 & Yes & Jet/Heat & 1/0.15 & 0/-- & 10 & 100/-- & $10^5$ & $\Delta x$ \\
  \hline
\end{tabular}
\end{table*}

\begin{figure*}
  \centering{\resizebox*{!}{8.5cm}{\includegraphics{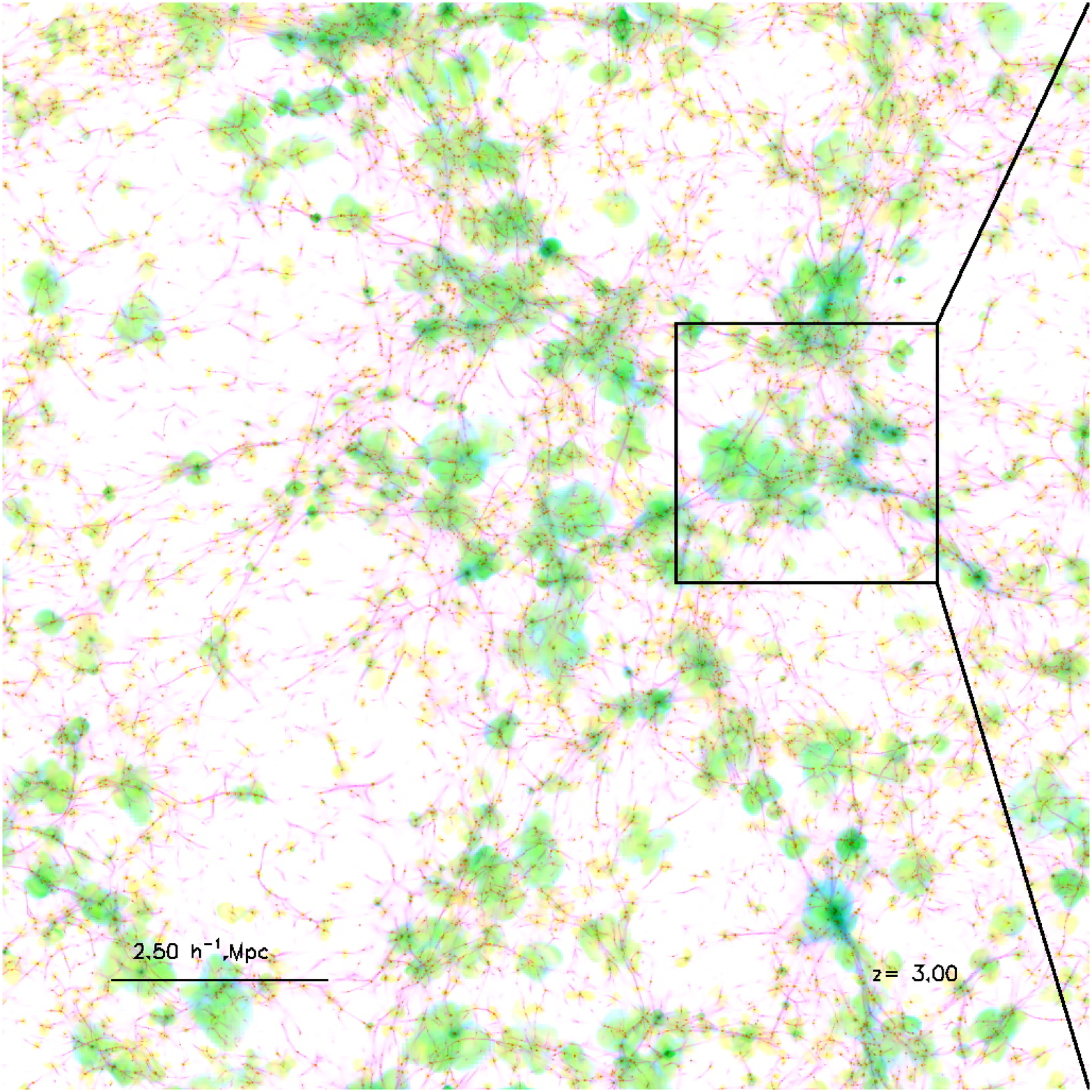}}\hspace{-0.09cm}}
  \centering{\resizebox*{!}{8.5cm}{\includegraphics{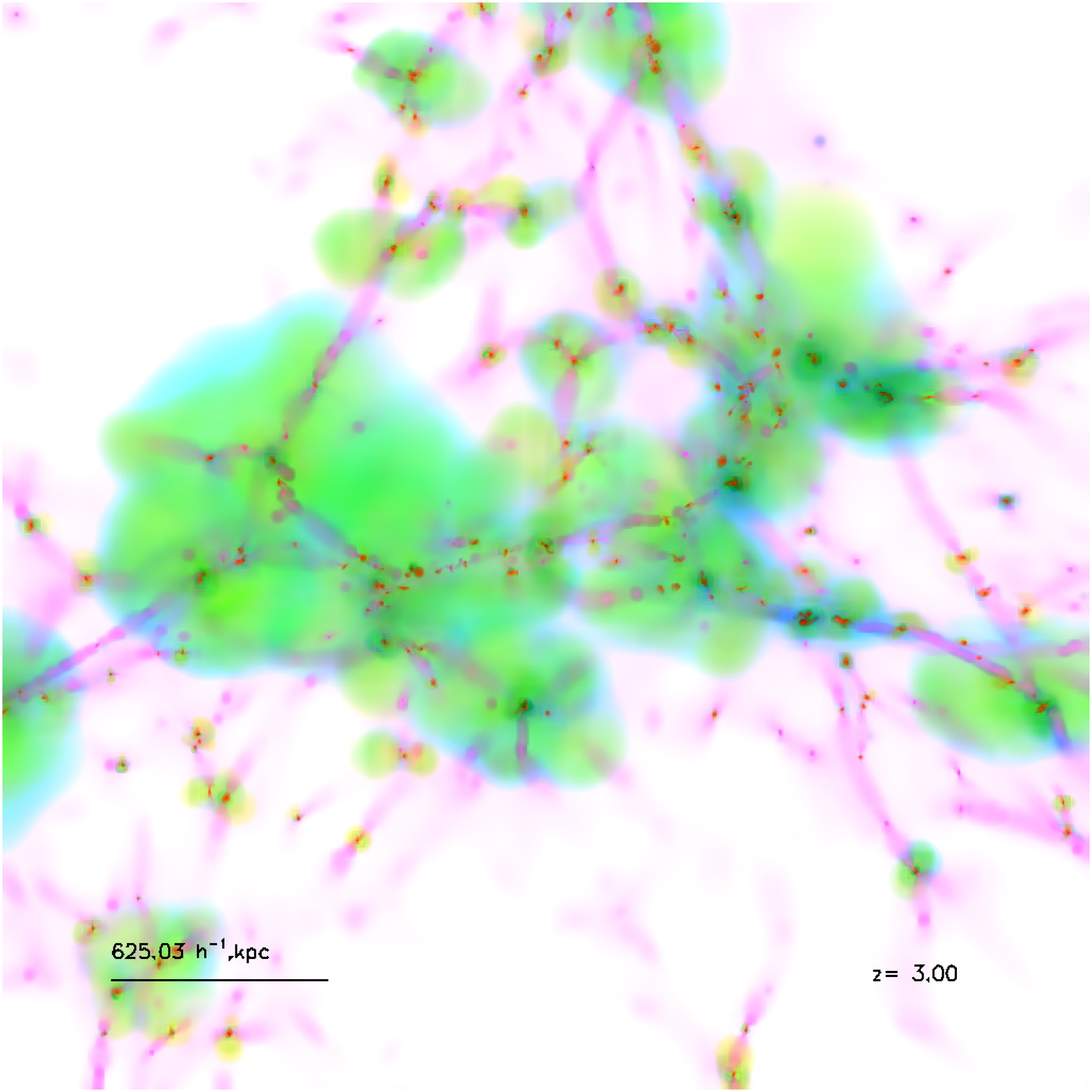}}\hspace{-0.09cm}}
  \centering{\resizebox*{!}{8.5cm}{\includegraphics{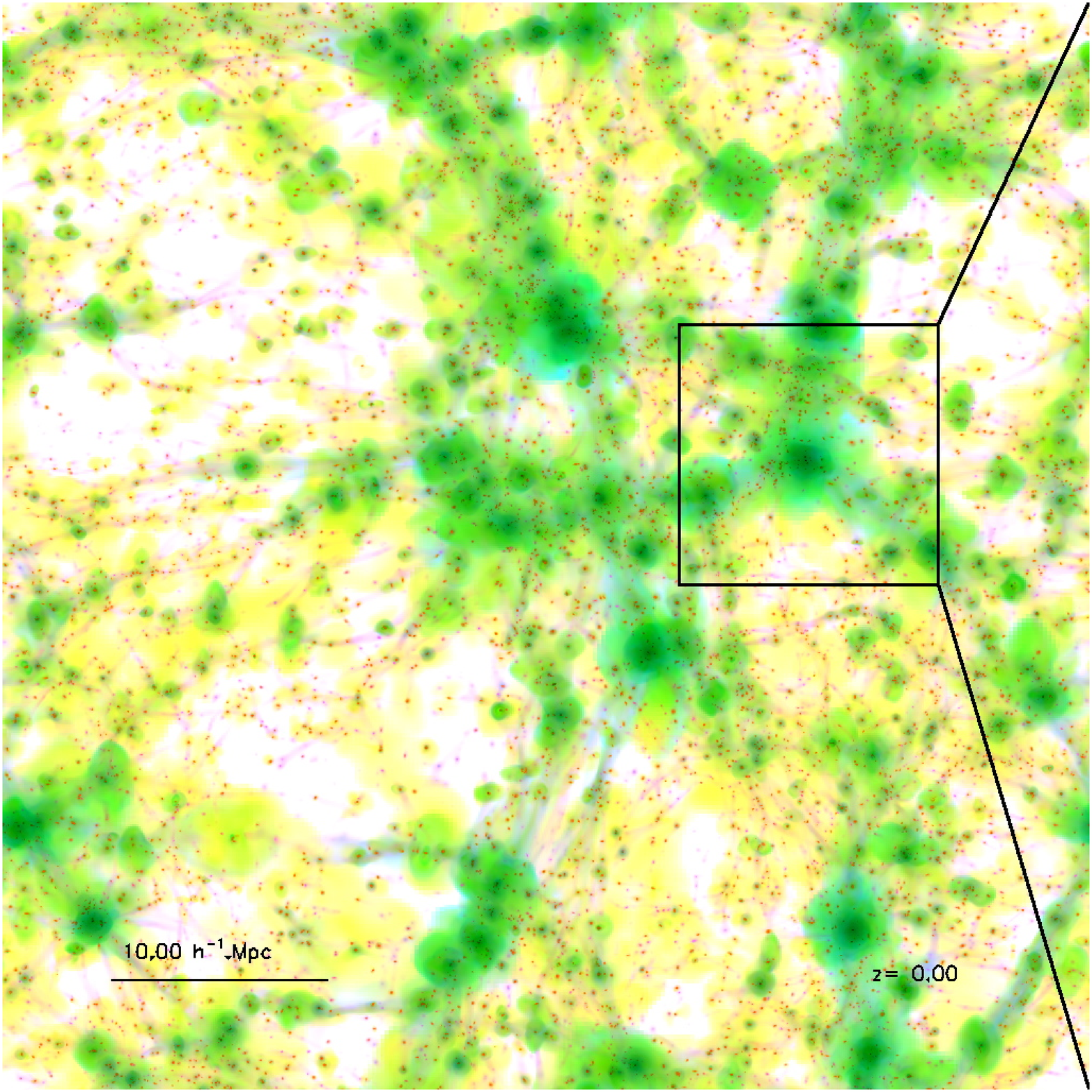}}\hspace{-0.09cm}}
  \centering{\resizebox*{!}{8.5cm}{\includegraphics{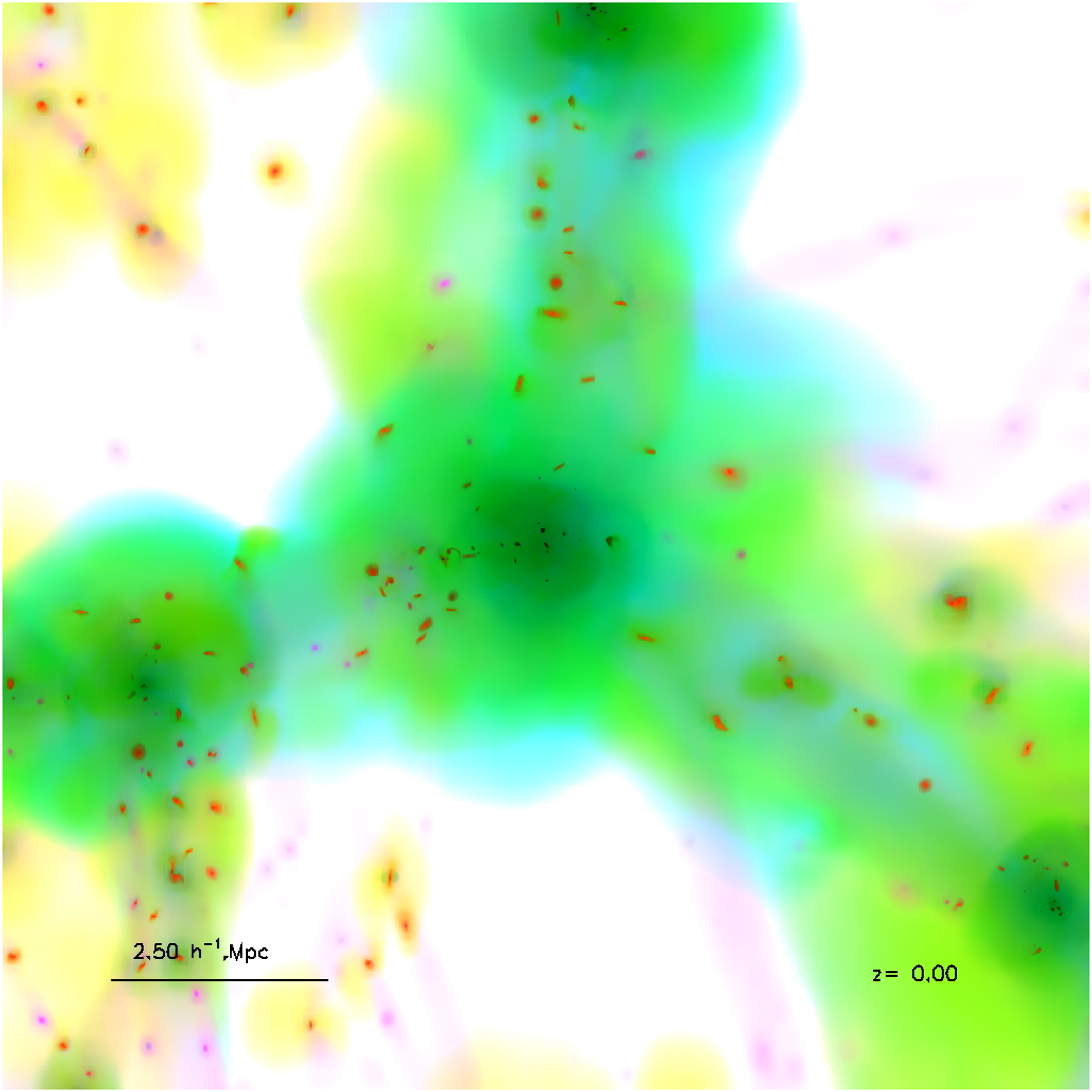}}\hspace{-0.09cm}}
  \caption{Three color image of the 256L50JH simulation (see table~\ref{tabnames}) at $z=3$ (upper panels) and $z=0$ (bottom panels) with a zoom on the largest halo (top and bottom right panels). Gas density is colour coded in magenta, gas temperature in cyan and gas metallicity in yellow. }
    \label{image_nice}
\end{figure*}


We assume a flat $\Lambda$CDM cosmology with total matter (baryons$+$DM) density $\Omega_{m}=0.26$, baryon density $\Omega_b=0.045$, dark energy density $\Omega_{\Lambda}=0.74$, fluctuation amplitude at $8 \, h^{-1}.\rm Mpc$ $\sigma_8=0.80$ and Hubble constant $H_0=70\, \rm km.s^{-1}.Mpc^{-1}$ consistent with WMAP 5-year data \citep{dunkleyetal09}.
We use several simulations with different box sizes $L_{\rm box}$, number of initial DM particles $N_{\rm DM}$, and minimum cell sizes $\Delta x$ in order to test the convergence of our AGN feedback model with resolution.
For a given $L_{\rm box}$ size, we generate our most resolved initial conditions (ICs) and degrade them to obtain lower resolution ICs, so that for the same box size, the ICs will produce the same structures but with different numbers of DM particles.  

These simulations are run with the AMR code {\sc ramses} \citep{teyssier02}. The evolution of the gas is followed with a second-order unsplit Godunov scheme for the Euler equations. The Riemann solver for the flux computation at the cell interface  uses a first-order MinMod Total Variation Diminishing scheme to reconstruct the interpolated variables from their cell-centered values. Collisonless particles (dark matter, stellar and sink particles) are evolved using a particle-mesh (PM) solver with a Cloud-In-Cell interpolation.

Simulations refine the initial mesh by as many as 7 levels of refinement, reaching  a physical cell size of $\Delta x=0.38\, \rm h^{-1}.kpc$ for our most resolved ICs and smallest box size $L_{\rm box}=12.5 \, \rm h^{-1}.Mpc$ (simulations 256L12noAGN, 256L12JH).
Note that the $\ell_{\rm max}=14$ level of refinement is only reached at $a_{\rm exp}=(1+z)^{-1}=0.8$ for our most resolved simulations. Because we enforce a nearly constant physical resolution (rather than a constant
comoving resolution), the highest refinement level triggered for a given redshift increases as the expansion factor grows with time, i.e. $\ell_{\rm max}-2$ at $a_{\rm exp}=0.2$, $\ell_{\rm max}-1$ at $a_{\rm exp}=0.4$, etc.
A cell is refined following a quasi-Lagrangian criterion: if more than 8 dark matter particles lie in a cell, or if the baryon mass exceeds 8 times the initial dark matter mass resolution.
Lower thresholds for triggering refinement can be adopted to resolve the smallest halos with $10-1000$ DM particles, using sufficient force resolution \citep{osheaetal05}.
However, these can lead to excessive amplification of noise discreetness effects \citep{romeoetal08}.

Our whole set of simulations using different box sizes, resolutions and sub-grid physics is summarized in table~\ref{tabnames}.

Fig.~\ref{image_nice} shows the gas density, temperature and metallicity through a three-color composite image at two different redshifts for our most resolved simulation of a $50h^{-1}$ Mpc box (256L50JH). with cooling, star formation, supernova feedback and our fiducial model of AGN feedback, i.e. the dual quasar and radio modes.
One can see the formation of hot and metal-rich bubbles at the intersection of filaments where galaxies collapse, already at high redshift $z=3$, suggesting a pre-heating of halos before a cluster forms. The bubbles extend further into the IGM as the simulation proceeds.

\section{Constraining  the AGN feedback model}
\label{parameters_resolution}

In order to test the model for BH growth and its associated AGN feedback, we measure different statistically meaningful quantities and compare them to their observational equivalents.
Quantities such as the cosmological density of BHs and the \cite{magorrianetal98} relationships that link BH masses to their host galaxy properties provide good constraints on the coevolution of BHs and galaxies.
We also test the effect of varying the AGN feedback parameters on the cosmic star formation rate (SFR per unit comoving volume).

The procedure adopted here is in many ways very similar to the one used in \cite{booth&schaye09} to constrain their AGN sub-grid model based on thermal energy inputs.
As we share some common parameters with them, like the accretion rate boost factor $\alpha$, and the thermal input of energy for the quasar mode of AGN feedback, most of the parameter tests
for the quasar mode will not be repeated, except for the AGN bubble size $r_{\rm AGN}$ because SPH and AMR codes treat their gas elements with a different numerical approach.
Thus we set up our thermal (quasar) mode for AGN feedback to the values of \cite{booth&schaye09}'s best-fitting model: we assume an $\epsilon_{\rm f}=0.15$ efficiency, which is also the value employed in \cite{teyssieretal11}.
In this paper rather than focusing on the quasar mode, we concentrate on testing our radio mode based on bipolar kinetic outflows, which is a modified version of the bipolar kinetic outflows in \cite{duboisetal11} and has never been modeled before in cosmological simulations of galaxy formation.

To test the AGN sub-grid model, we fix the `standard' galactic sub-grid models for star formation and SN feedback (see section~\ref{sec:sf}, e.g.
star formation threshold $\rho_{0} = 0.1$\rm H.cm$^{-3}$, star formation efficiency $\epsilon_{*} = 0.02$, Salpeter IMF, mass loading factor $f_w$=1, stellar yield y=0.1, polytropic index p=$4/3$)
and we vary the parameters of the AGN feedback modeling ($\epsilon_f$, $\delta M_{\rm d}$, u$_{\rm max}$, $\eta$, $M_{\rm seed}$, r$_{\rm AGN}$, mode of AGN feedback).
These tests are performed on a $L_{\rm box}=25\, \rm h^{-1}.Mpc$ simulation box with $128^3$ DM particles (i.e. simulations with names starting with the prefix '128L25' in Table~\ref{tabnames}) that are  sufficient to resolve halos with masses as small as $\sim10^{11}\, \rm h^{-1}.M_{\odot}$ ($\sim$100 particles) and as large as several $\sim10^{13}\, \rm h^{-1}.M_{\odot}$.
This choice of box size and resolution is a good compromise between affordable computational resources and resolution requirements.
Finally convergence tests are performed with larger box sizes (128L50JH, 256L50JH) and more (256L12JH, 256L25JH) and less resolved simulations (64L25JH) for our radio/quasar mode with parameters from our best fitting model (simulation 128L25JH, see table~\ref{tabnames}).

In order to compare the simulated BH masses to their host galaxy bulge stellar mass as has been done for observations, we must decompose the stellar surface density profiles into an inner bulge and an outer disc component.
Our bulge-disc decomposition is detailed in Appendix~\ref{AppendixA}. 

Fig.~\ref{bhdensity_comp} compares the BH comoving density as a function of redshift while varying the parameters of the AGN feedback model.
The value of the BH density in our local Universe from \cite{shankaretal04} is overplotted with its $3\sigma$ uncertainty.
Any AGN feedback model that pretends to model the cosmological growth of BHs should remain close to this data point at $z=0$.
However, the latter does not ensure that their growth is correct, as the observational point is only for redshift $z=0$ and the data at larger redshifts is more sparse.

Fig.~\ref{magorrian_comp} shows the relationships between BH mass and their host galaxy stellar bulge mass as well as with the host's stellar velocity dispersion. These are good constraints for testing the coevolution of BHs and their galaxy mass content.
We represent the average value of the distribution of stellar bulge masses for a given bin of BH mass.
Observational fits to BH mass ($M_{\rm BH}$) versus stellar mass ($M_{\rm s}$) \citep{haring&rix04}, and on BH mass ($M_{\rm BH}$) versus stellar velocity dispersion ($\sigma_{\rm s}$) \citep{tremaineetal02} are overploted with $3\sigma$ uncertainties.
Fig.~\ref{sfr_comp} shows the cosmic SFR to test the impact of varying AGN feedback parameters on the history of mass assembly of galaxies. 

For these comparisons, we take simulation 128L25J as our reference model for the radio AGN feedback mode (see table~\ref{tabnames}), and each parameter of the AGN feedback model is varied one-by-one.
Even though this procedure does not explore the full parameter space, and does not ensure that another set of parameters is possible, it allows us to test the validity of our fiducial model as well as the dependency of the results on the variation of its parameters.

\subsection{Parameter study}

\begin{figure*}
  \centering{\resizebox*{!}{5.cm}{\includegraphics{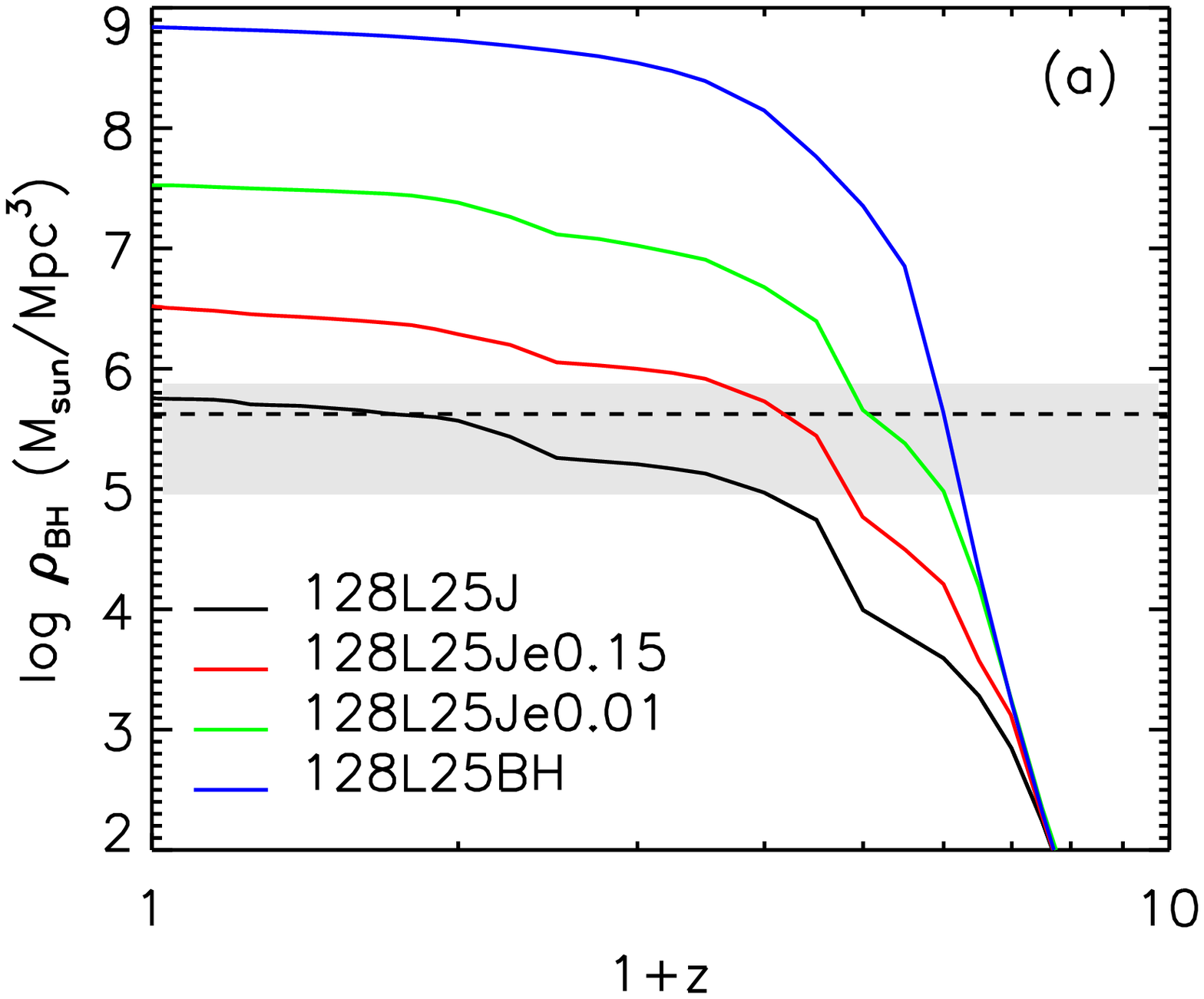}}\hspace{-0cm}}
  \centering{\resizebox*{!}{5.cm}{\includegraphics{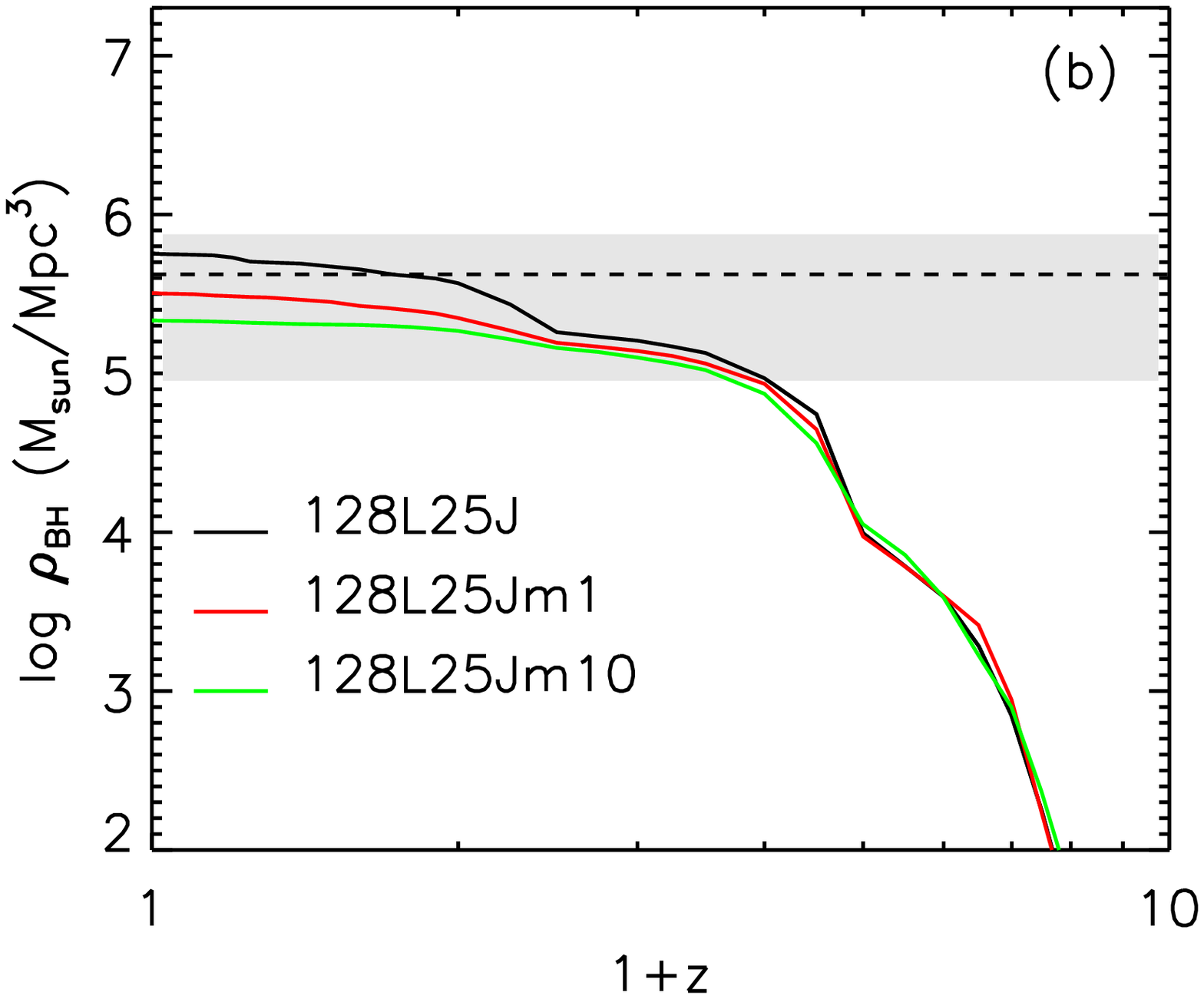}}\hspace{-0cm}}
  \centering{\resizebox*{!}{5.cm}{\includegraphics{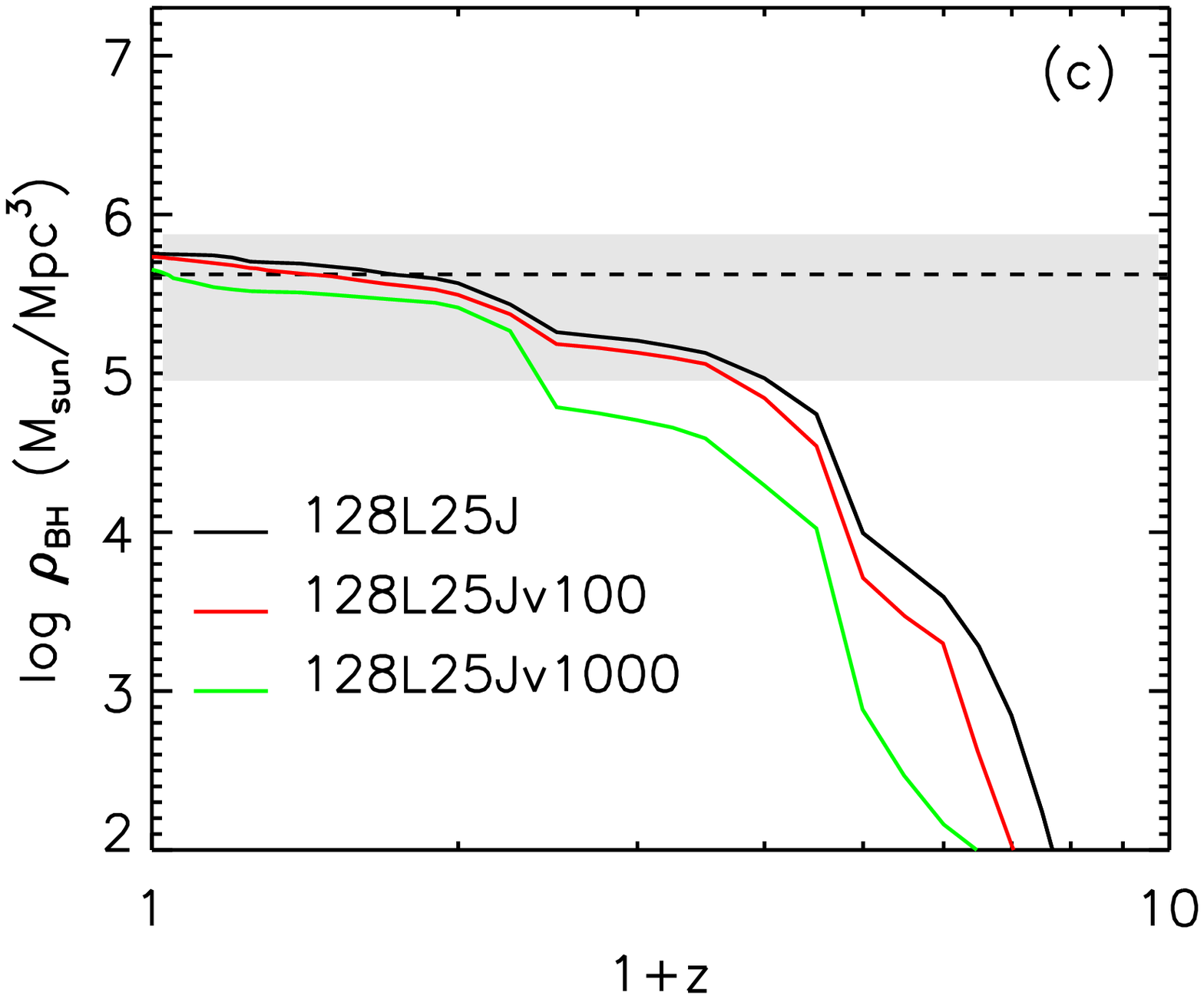}}\hspace{-0cm}}
  \centering{\resizebox*{!}{5.cm}{\includegraphics{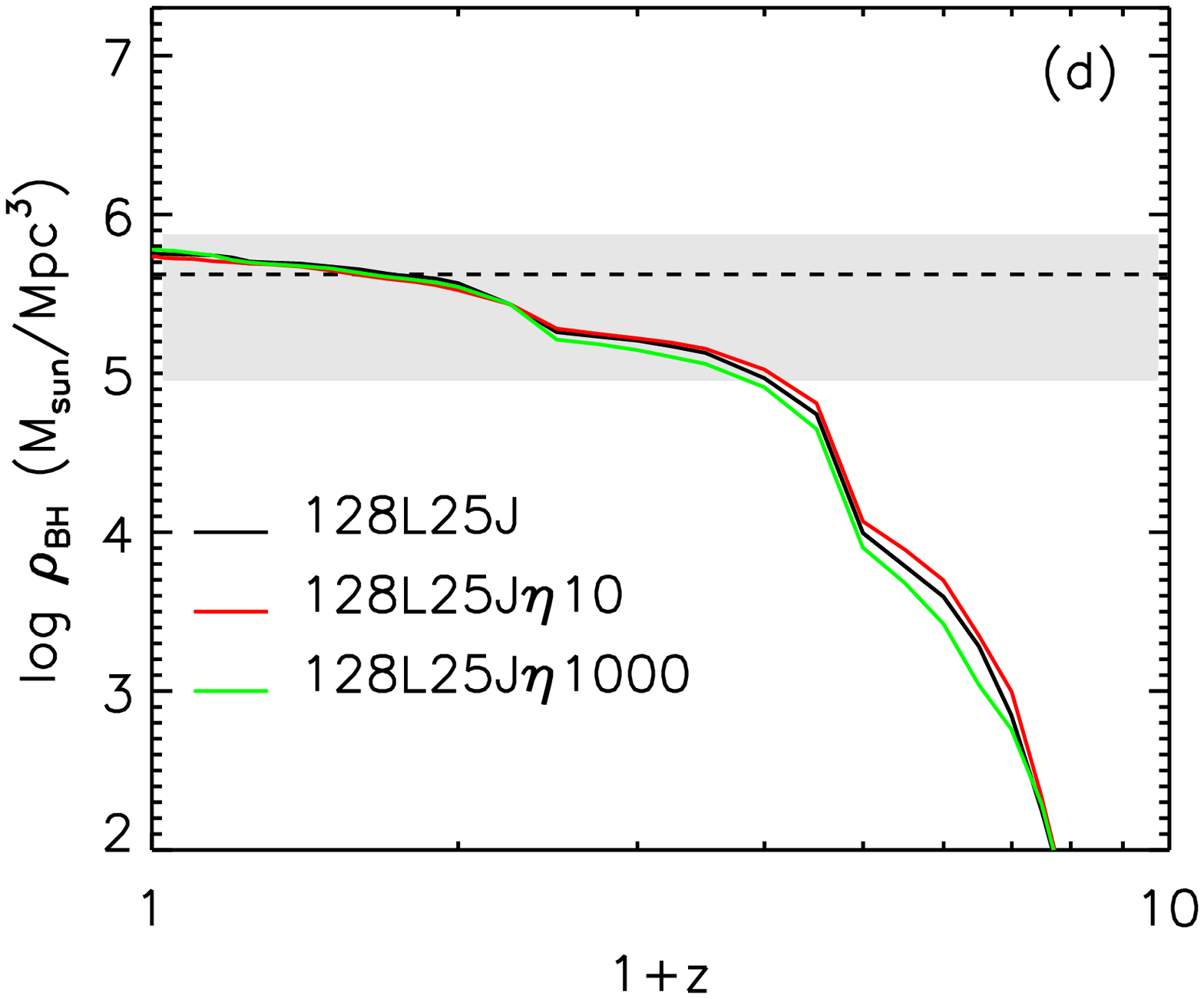}}\hspace{-0cm}}
  \centering{\resizebox*{!}{5.cm}{\includegraphics{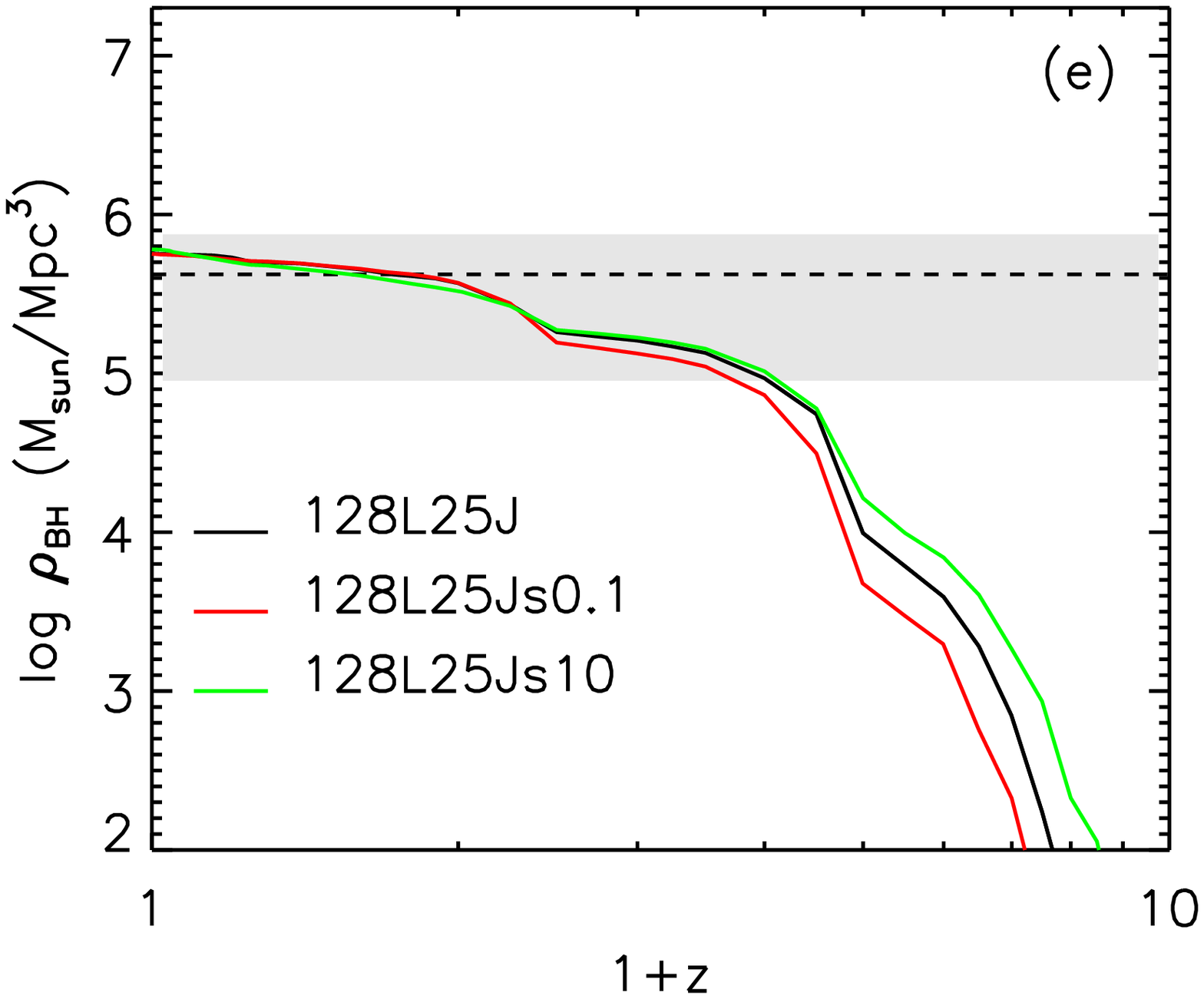}}\hspace{-0cm}}
  \centering{\resizebox*{!}{5.cm}{\includegraphics{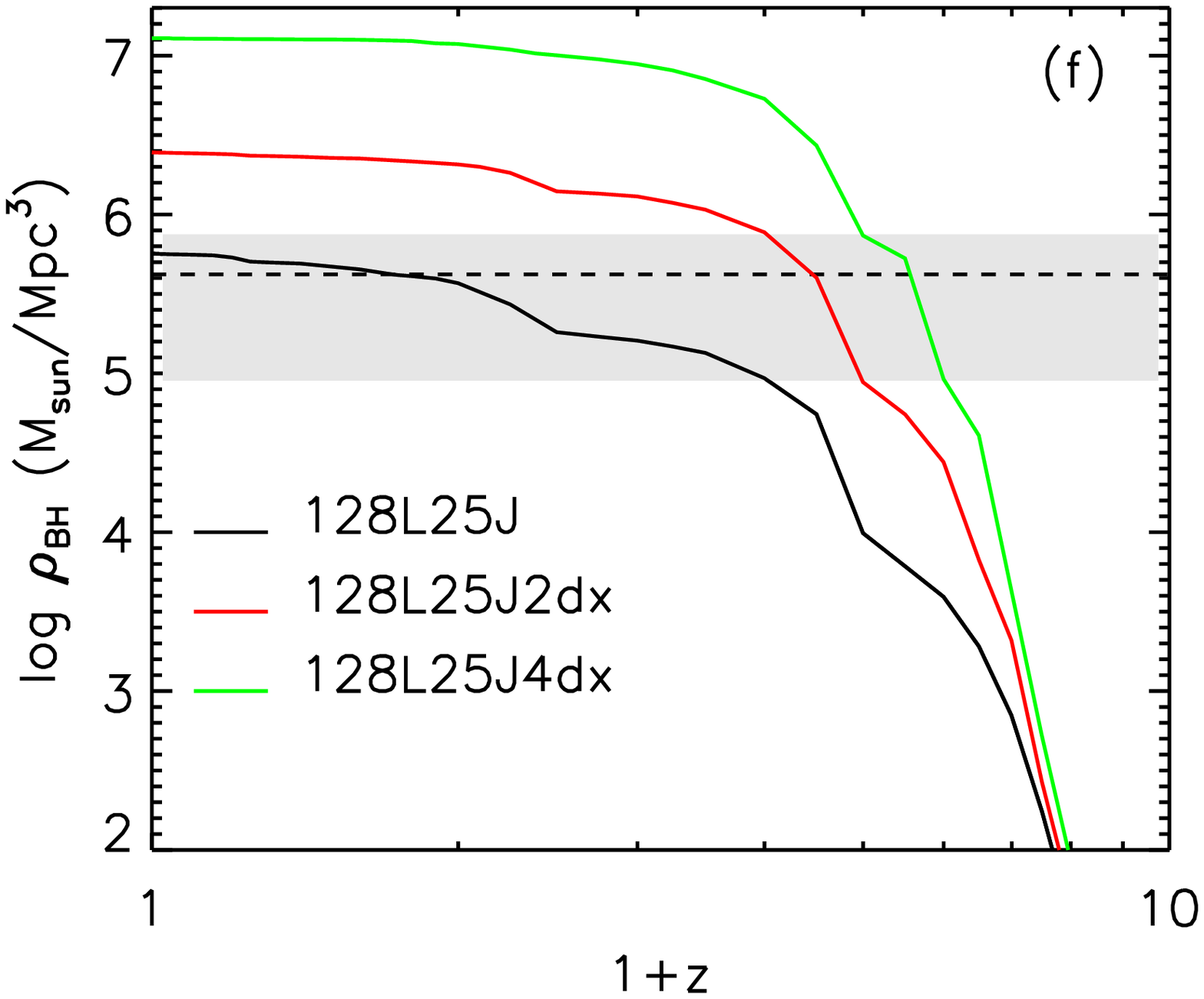}}\hspace{-0cm}}
  \centering{\resizebox*{!}{5.cm}{\includegraphics{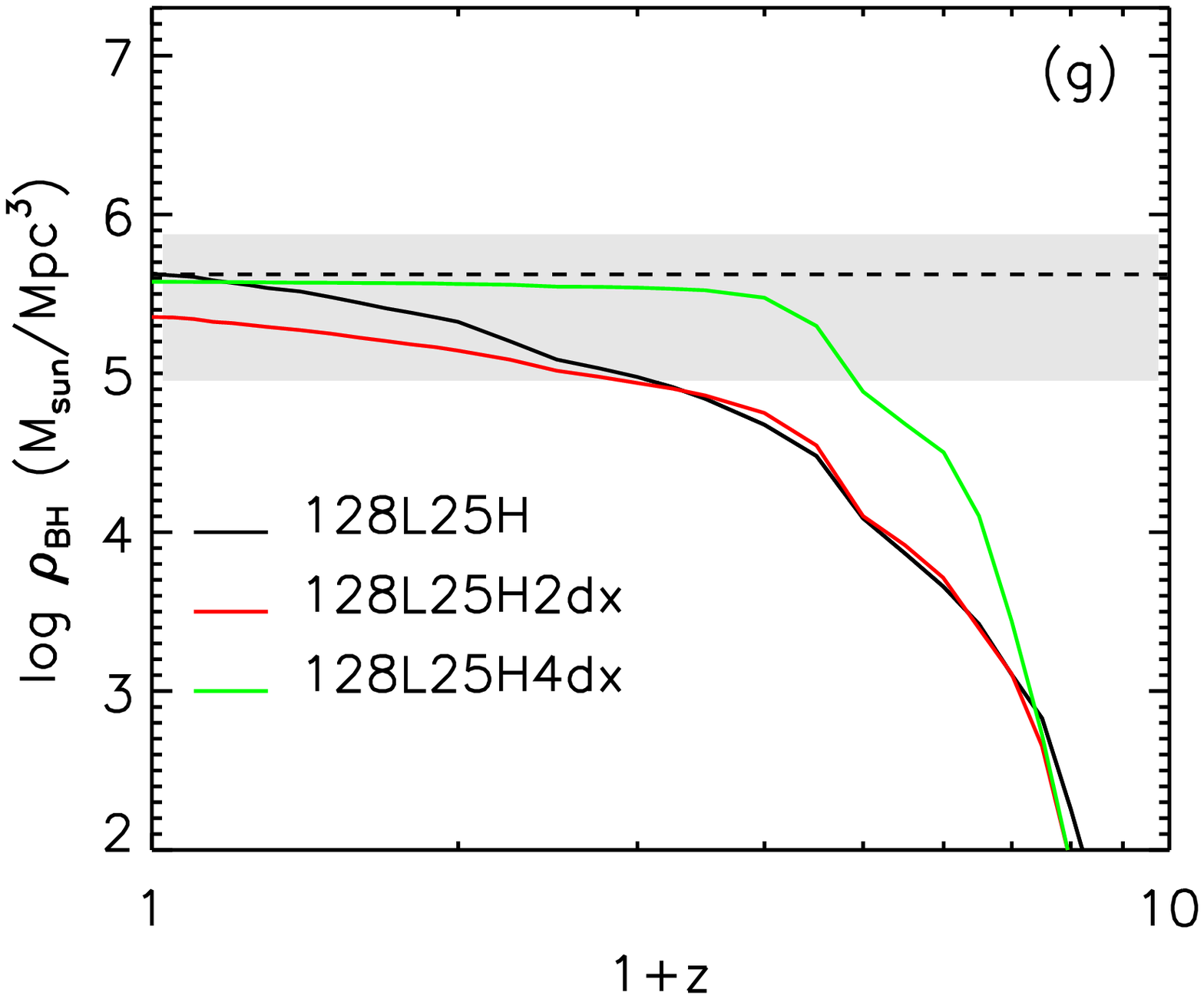}}\hspace{-0cm}}
  \centering{\resizebox*{!}{5.cm}{\includegraphics{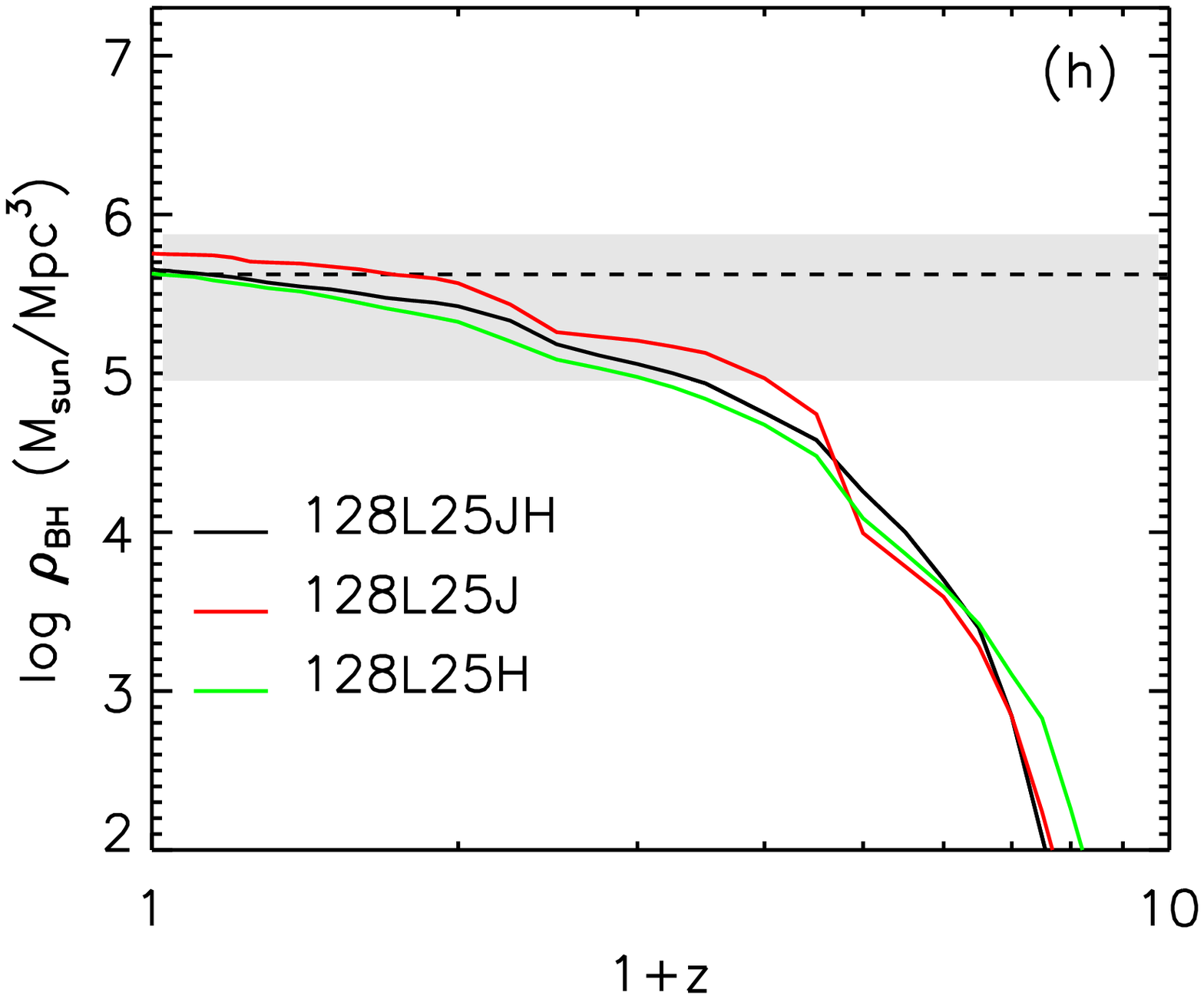}}\hspace{-0cm}}
  \caption{Comoving black hole mass density as a function of redshift. (a) Varying the efficiency $\epsilon_{\rm f}$ for the jet mode. (b) Varying the AGN energy delay $\Delta M_{\rm d}$ for the jet mode. (c) Varying the maximum relative velocity $u_{\rm max}$ for the jet mode. (d) Varying the mass loading factor for the jet mode. (e) Varying the BH initial mass $M_{\rm seed}$ for the jet mode. (f) Varying the the AGN input size $r_{\rm AGN}$ for the jet mode. (g) Varying the the AGN input size $r_{\rm AGN}$ for the heating mode. (h) Varying the mode of the AGN feedback. The dashed line is the average black hole mass density in our local Universe with its 3$\sigma$ uncertainty (grey shaded area) from \citet{shankaretal04}. }
    \label{bhdensity_comp}
\end{figure*}

\begin{figure*}
  \centering{\resizebox*{!}{5.cm}{\includegraphics{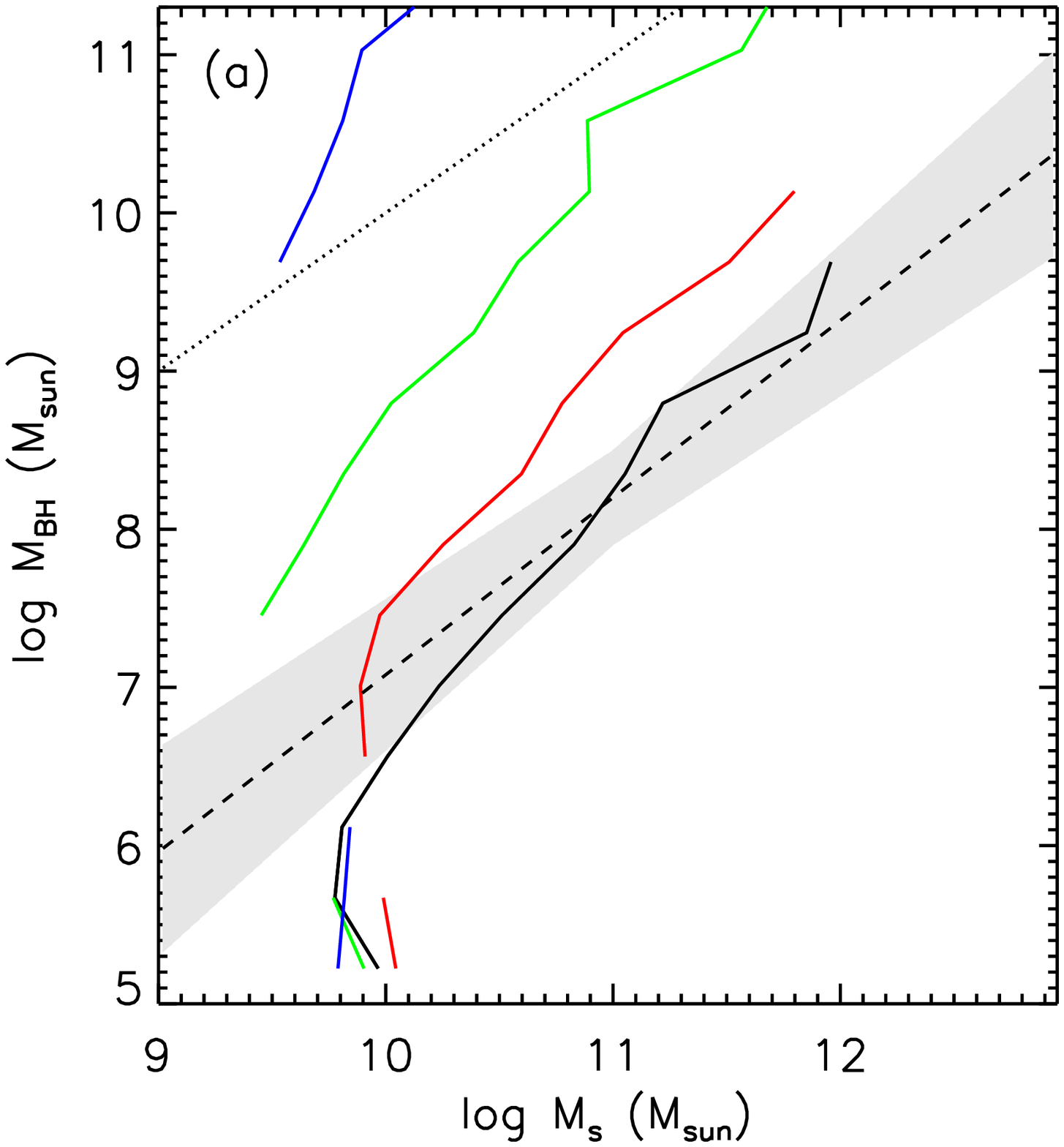}}\hspace{-1.54cm}}
  \centering{\resizebox*{!}{5.cm}{\includegraphics{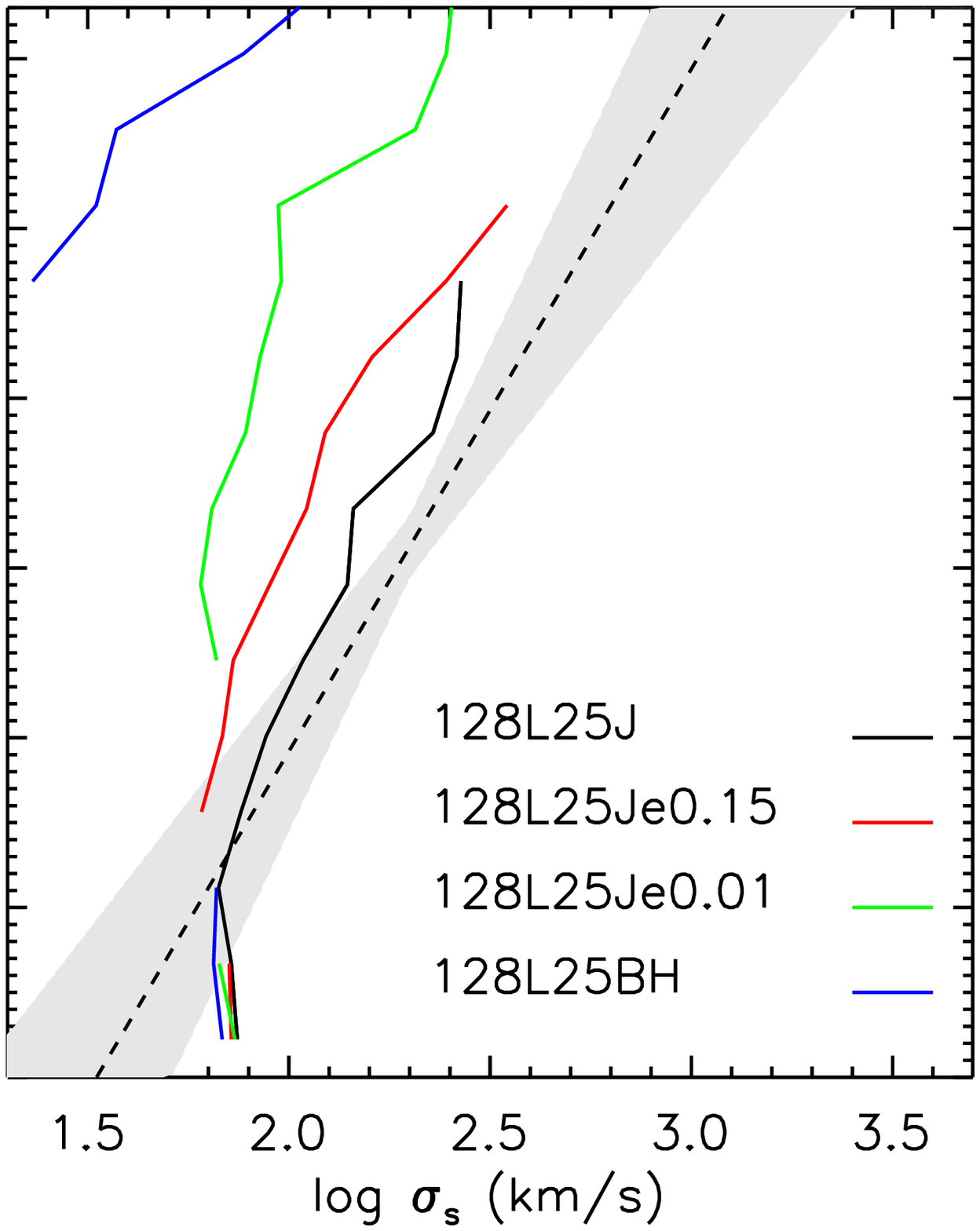}}}
  \centering{\resizebox*{!}{5.cm}{\includegraphics{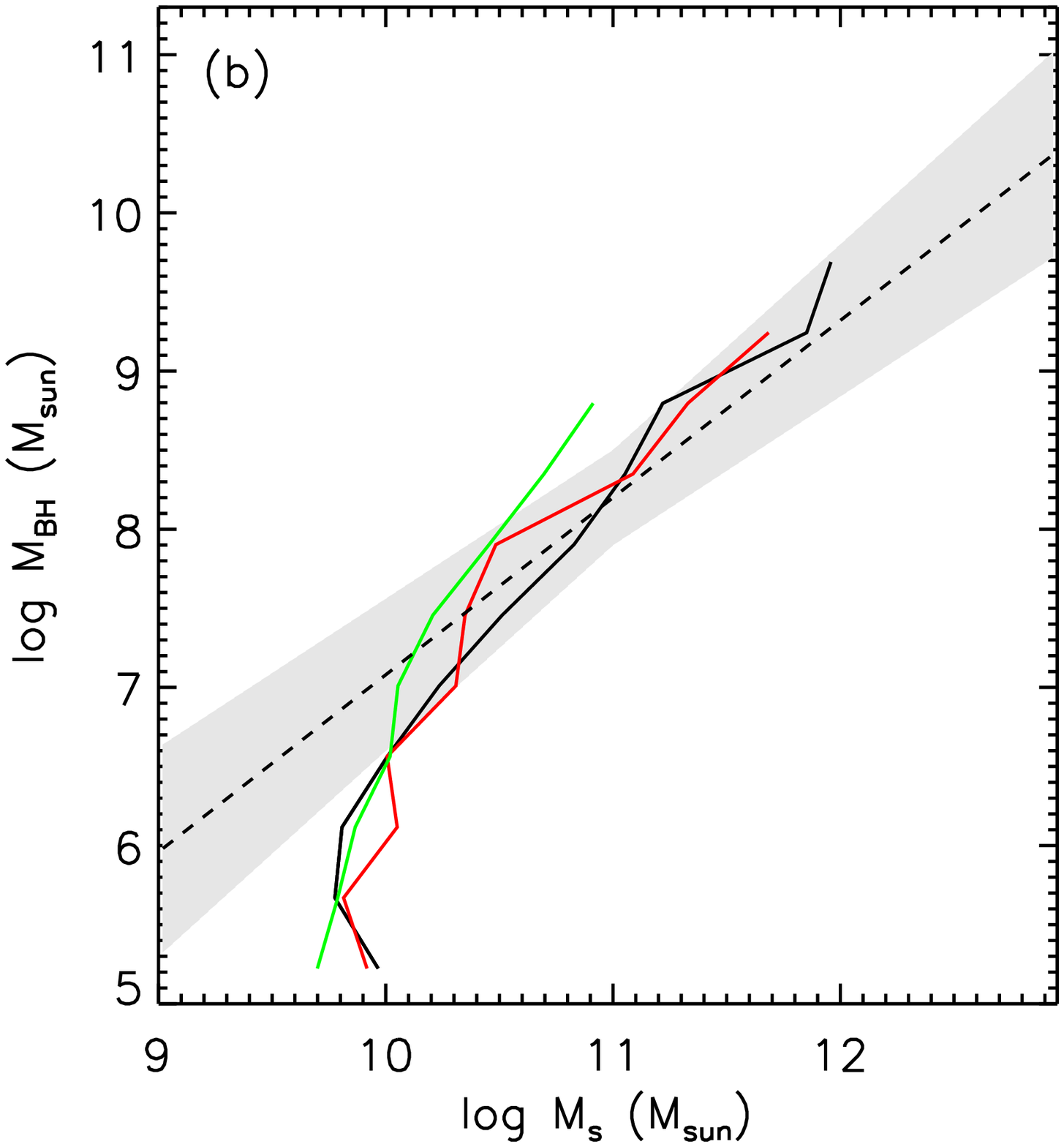}}\hspace{-1.54cm}}
  \centering{\resizebox*{!}{5.cm}{\includegraphics{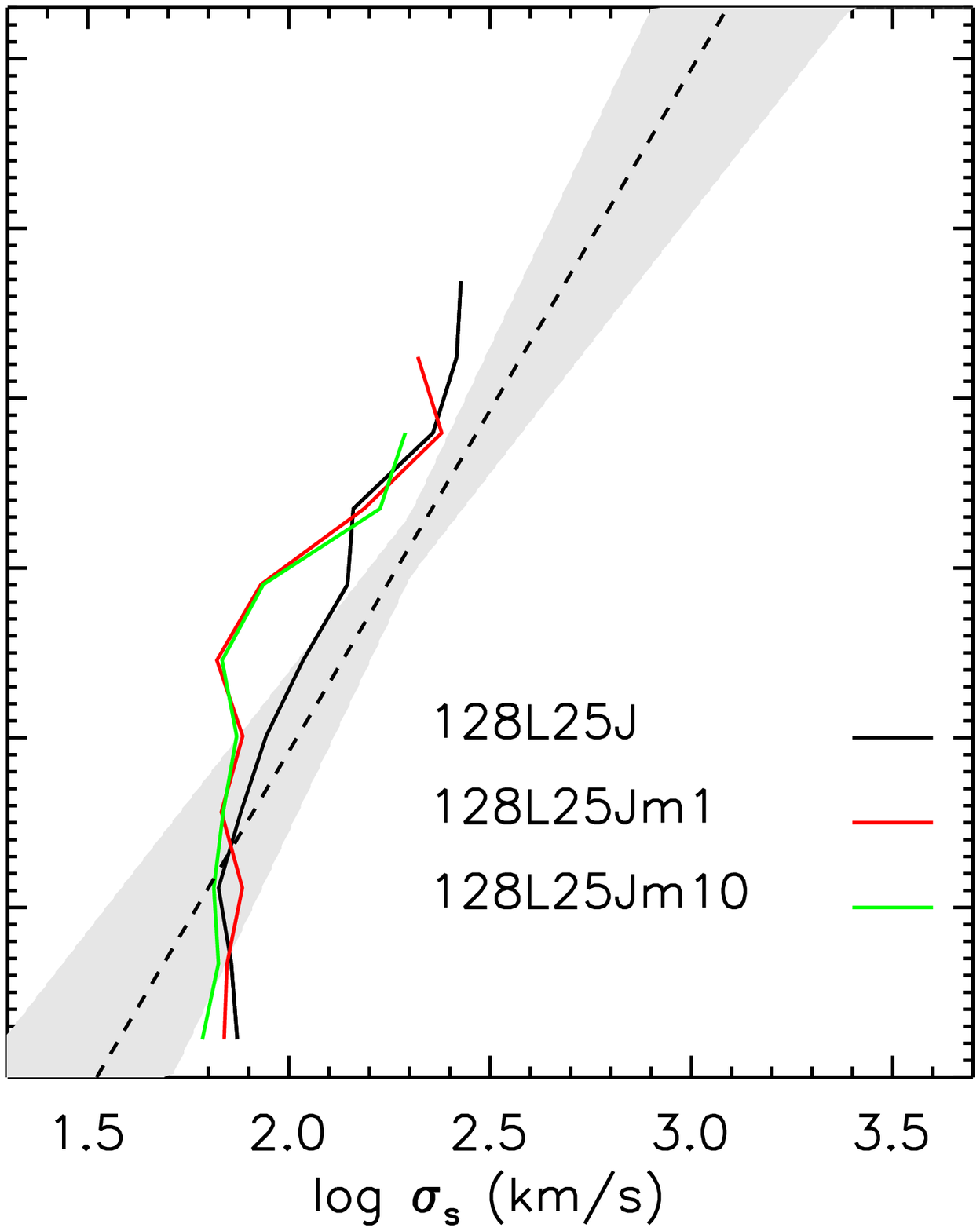}}}
  \centering{\resizebox*{!}{5.cm}{\includegraphics{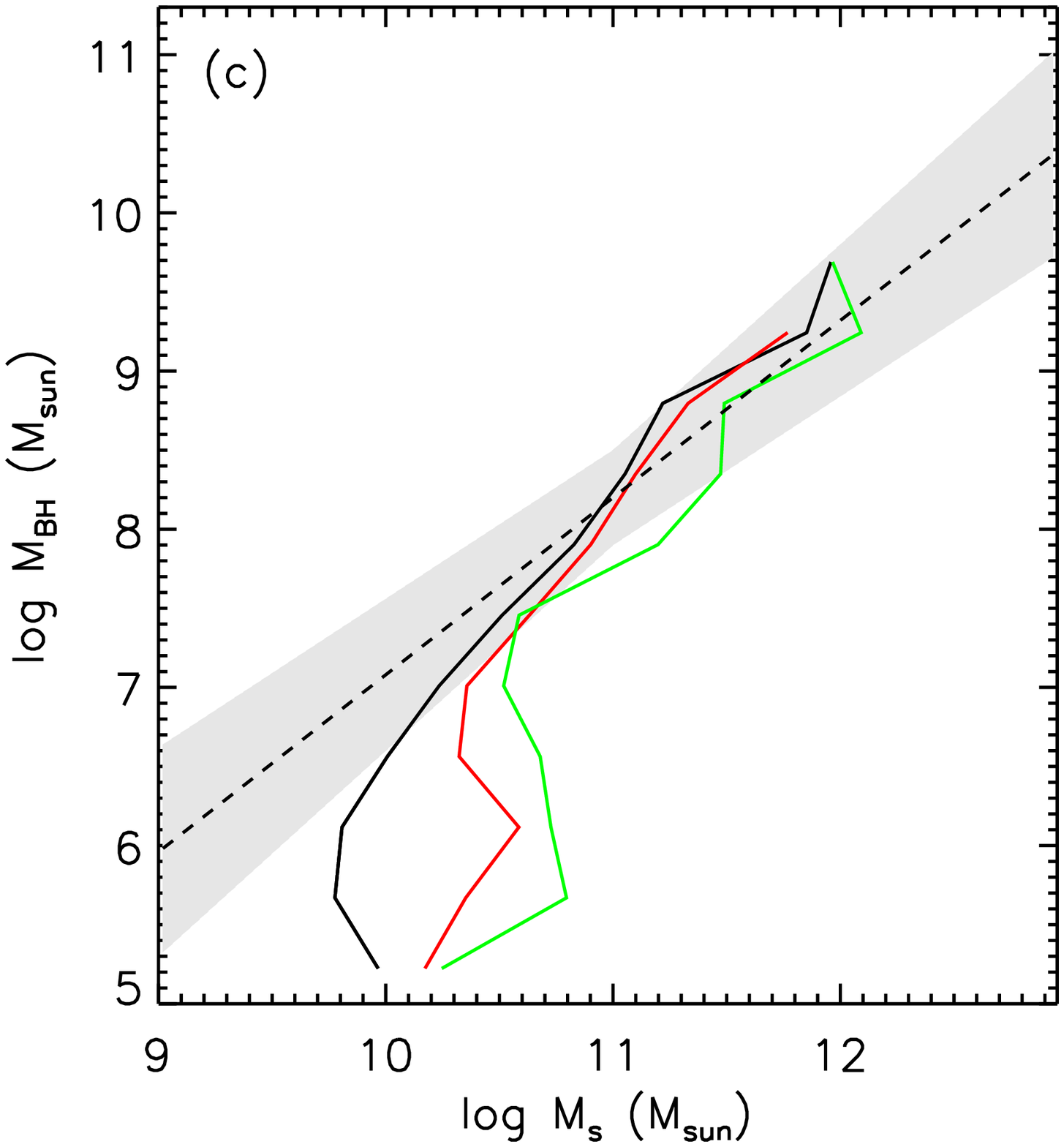}}\hspace{-1.54cm}}
  \centering{\resizebox*{!}{5.cm}{\includegraphics{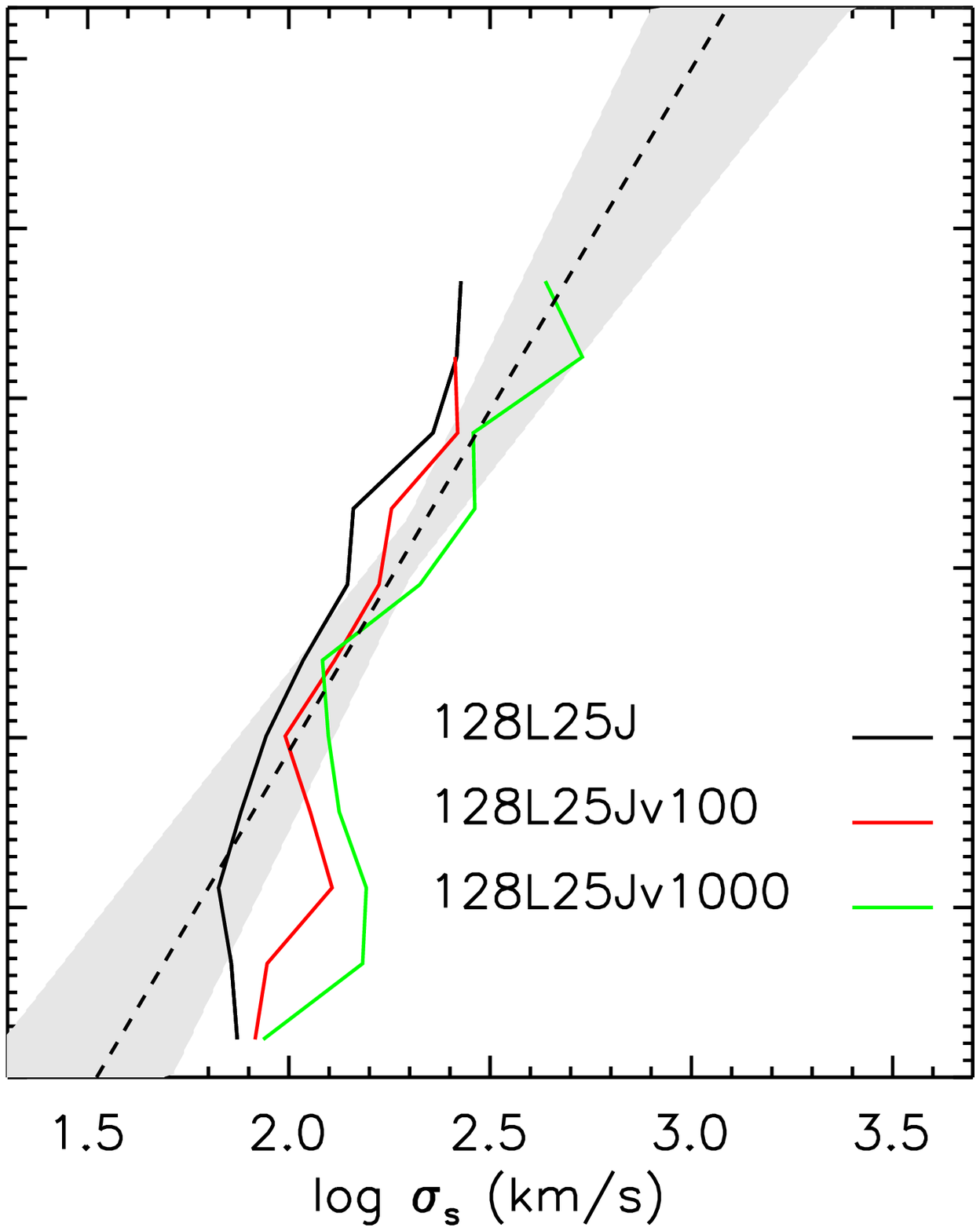}}}
  \centering{\resizebox*{!}{5.cm}{\includegraphics{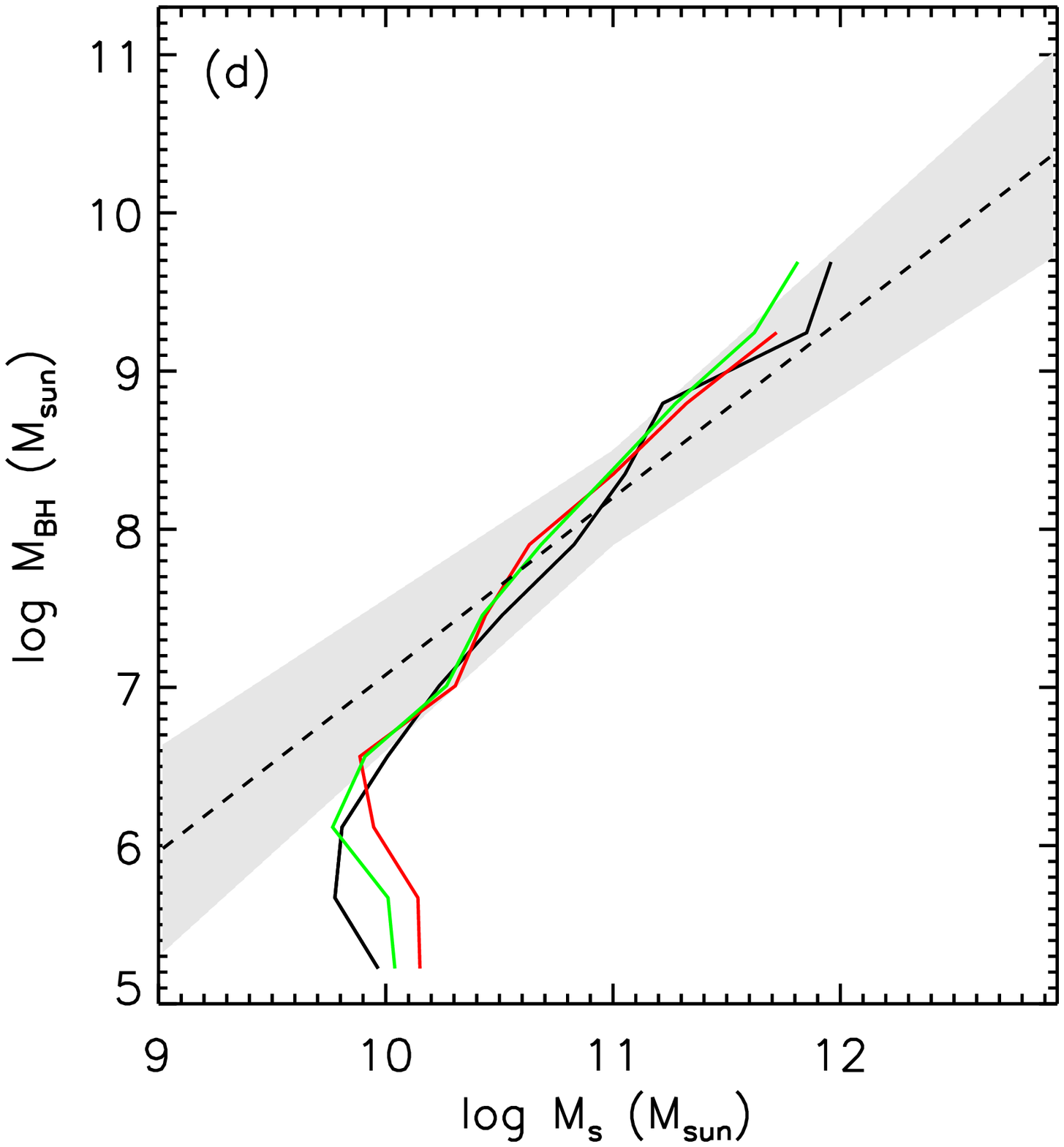}}\hspace{-1.54cm}}
  \centering{\resizebox*{!}{5.cm}{\includegraphics{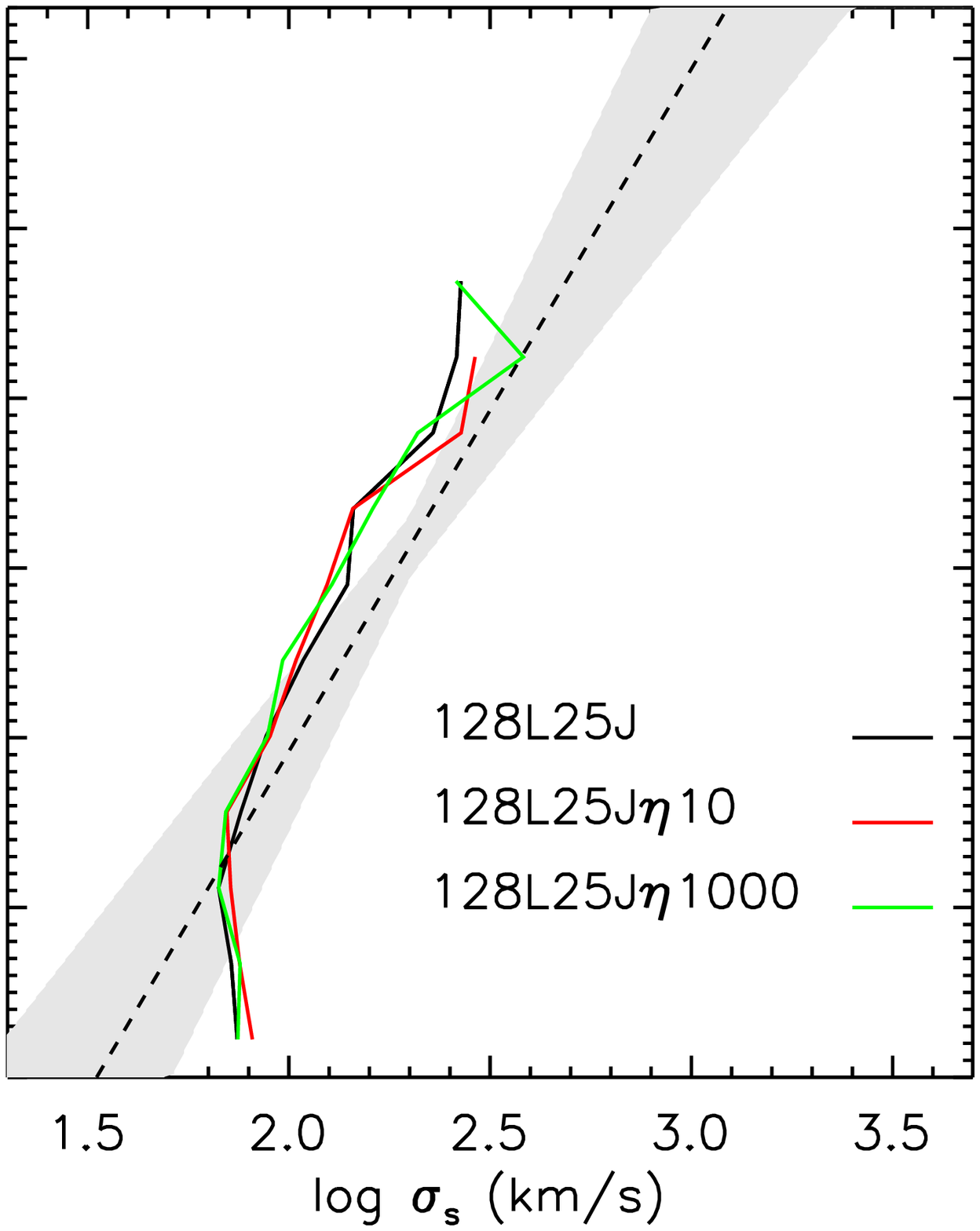}}}
  \centering{\resizebox*{!}{5.cm}{\includegraphics{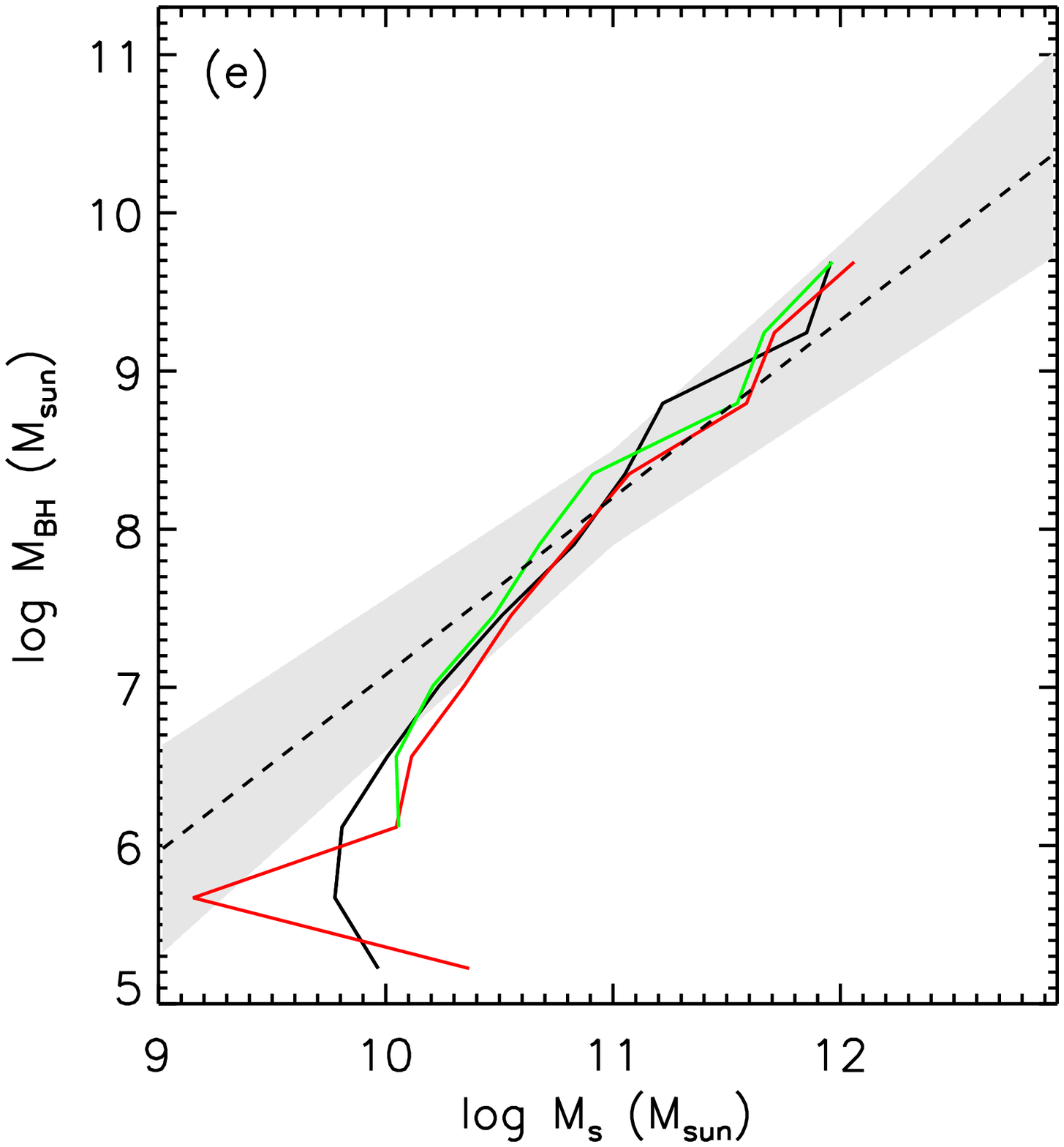}}\hspace{-1.54cm}}
  \centering{\resizebox*{!}{5.cm}{\includegraphics{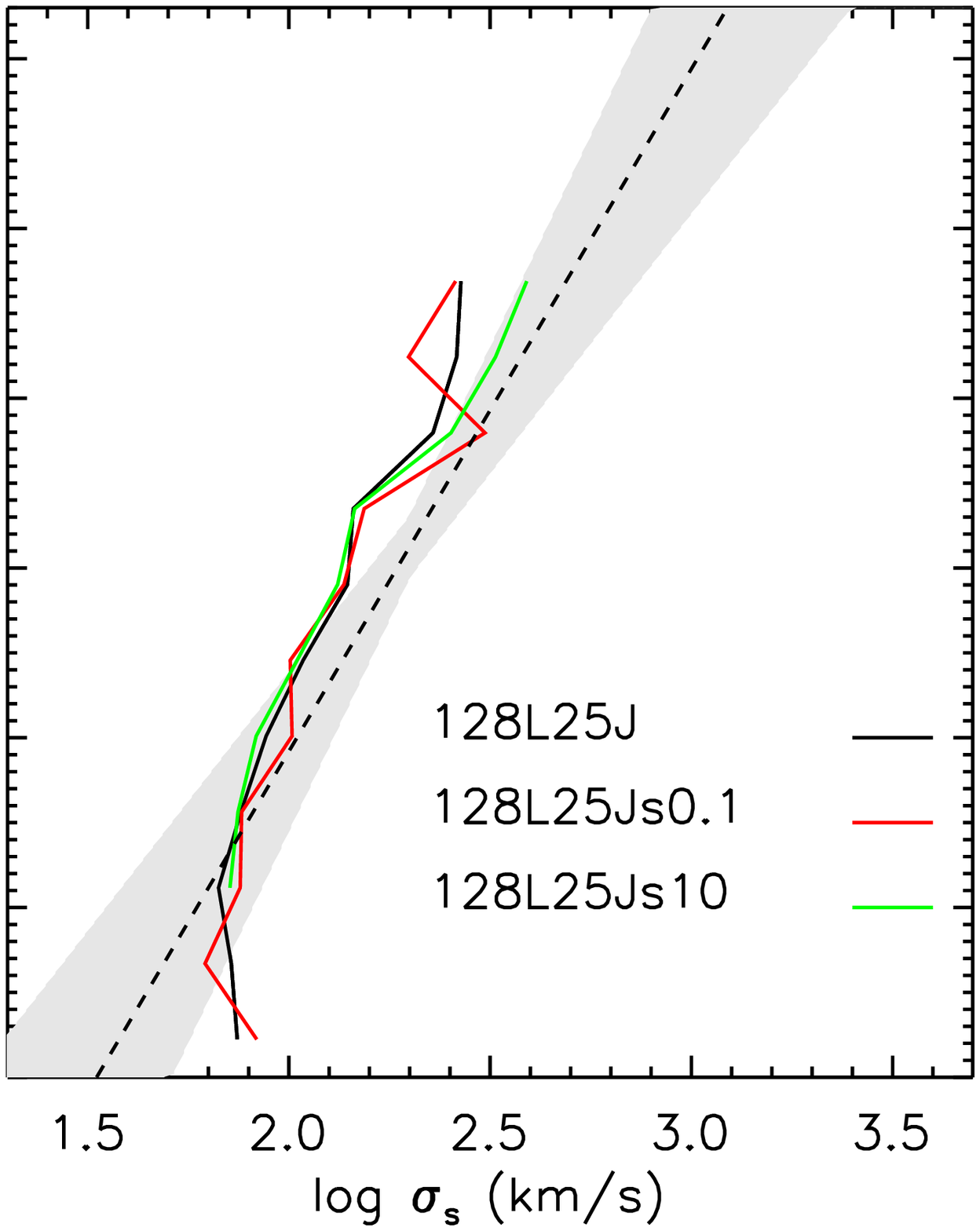}}}
  \centering{\resizebox*{!}{5.cm}{\includegraphics{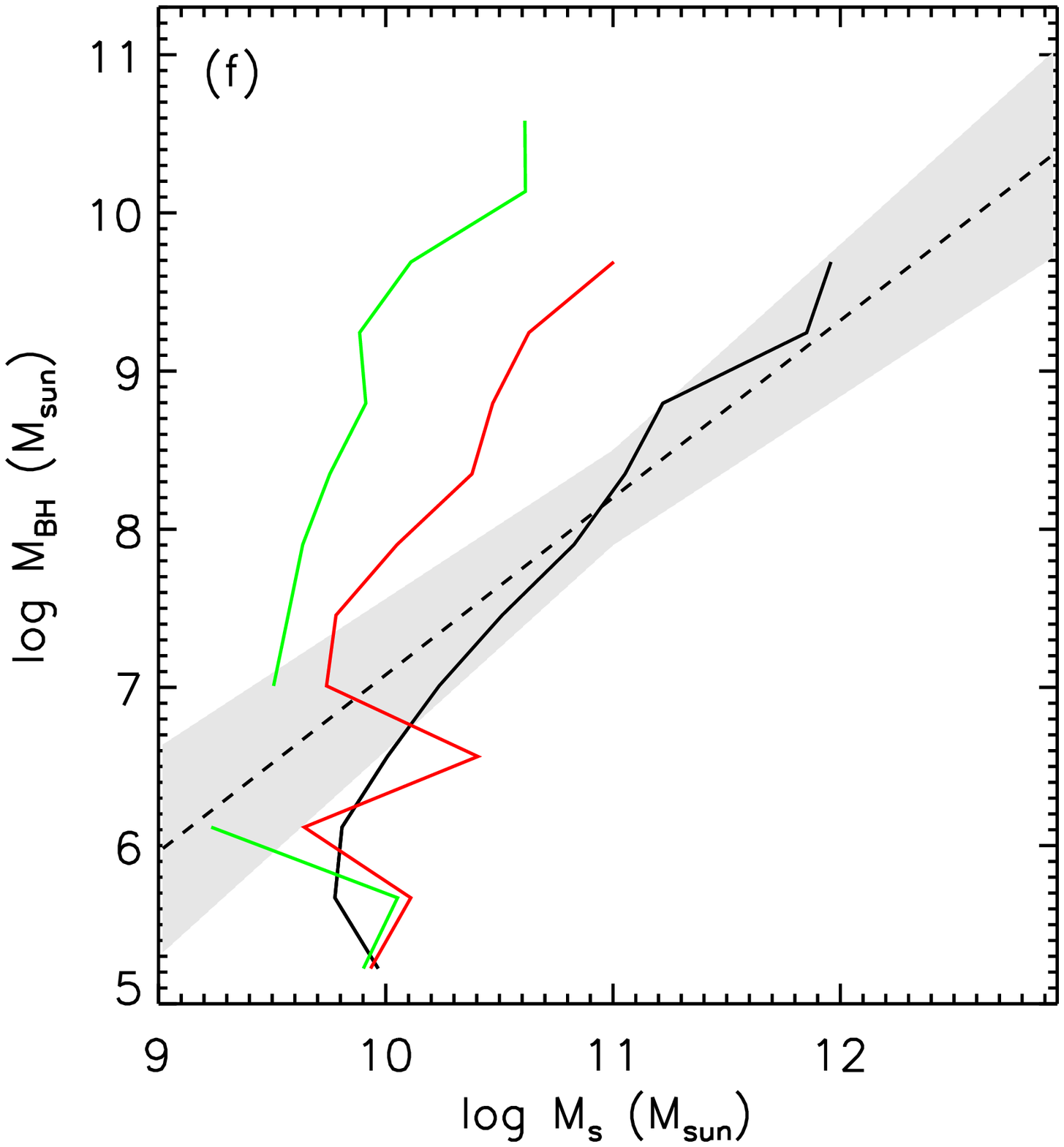}}\hspace{-1.54cm}}
  \centering{\resizebox*{!}{5.cm}{\includegraphics{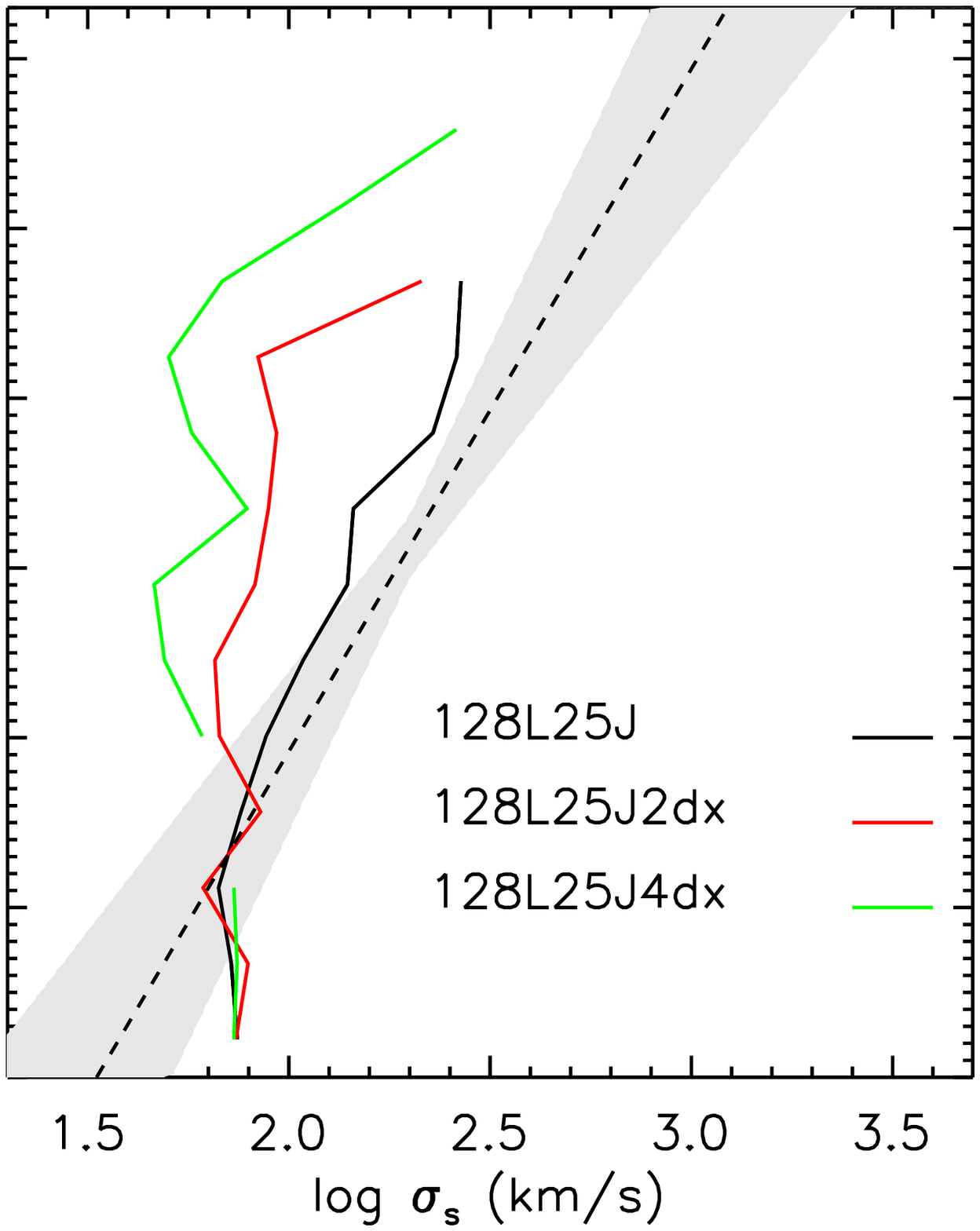}}}
  \centering{\resizebox*{!}{5.cm}{\includegraphics{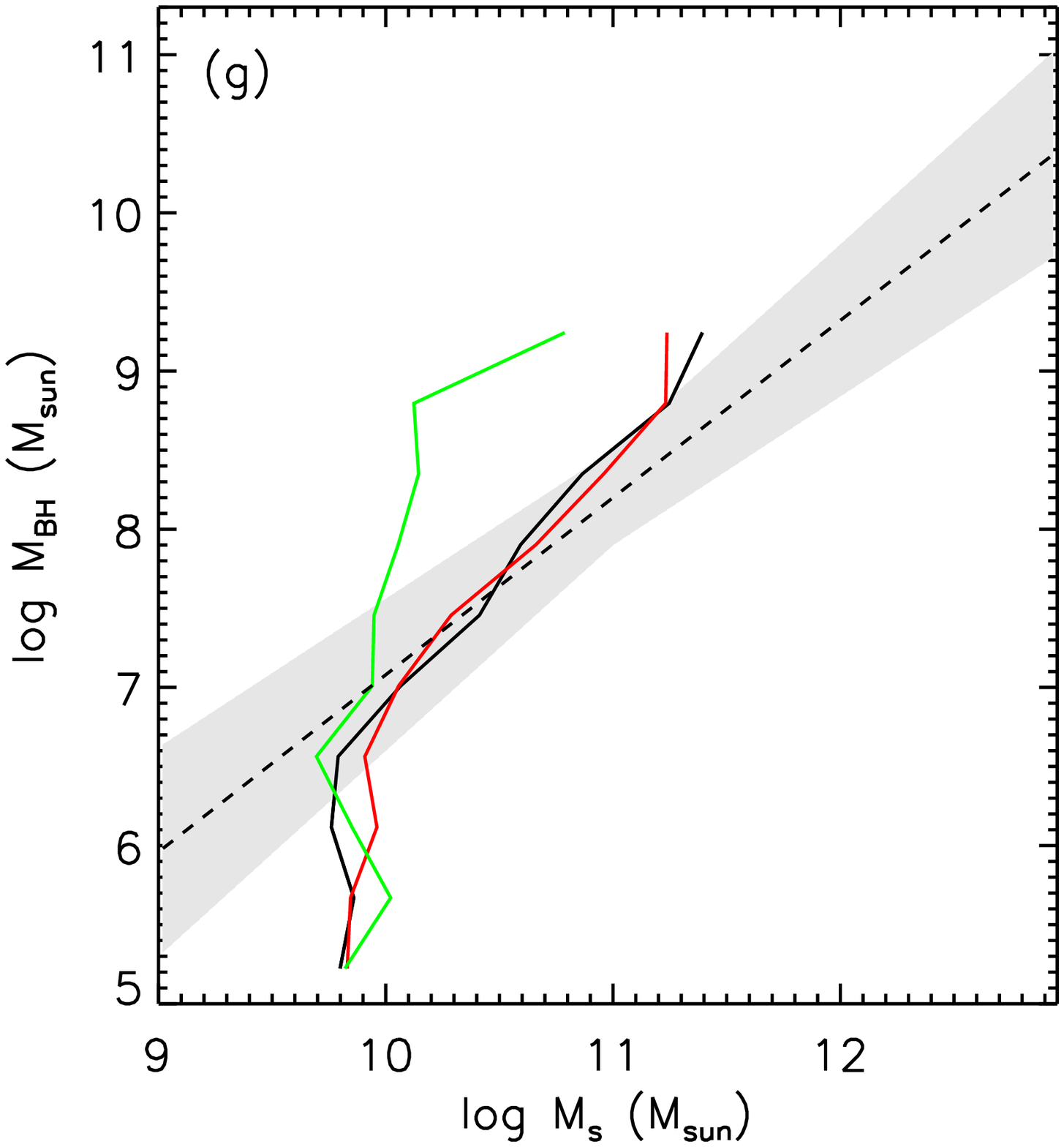}}\hspace{-1.54cm}}
  \centering{\resizebox*{!}{5.cm}{\includegraphics{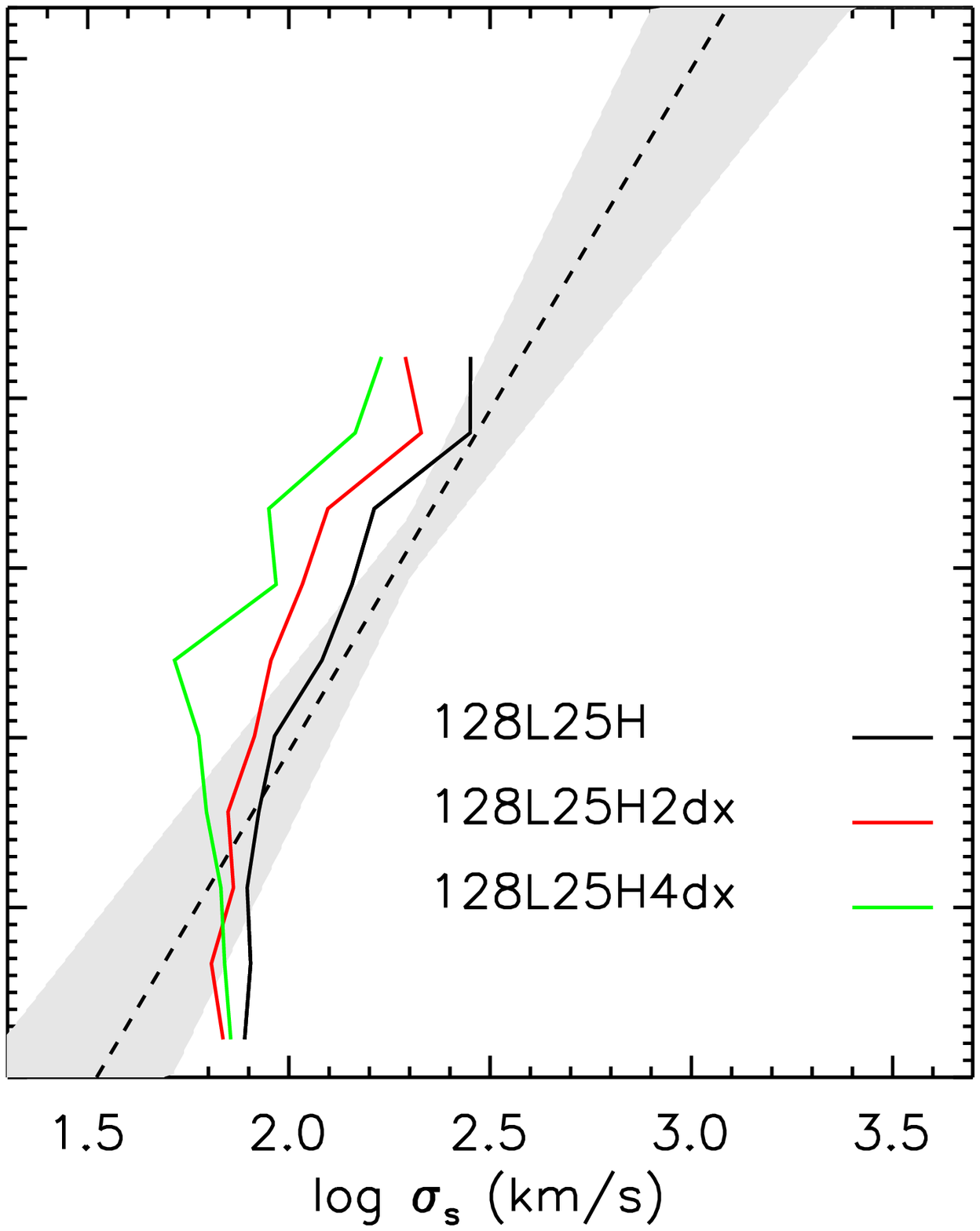}}}
  \centering{\resizebox*{!}{5.cm}{\includegraphics{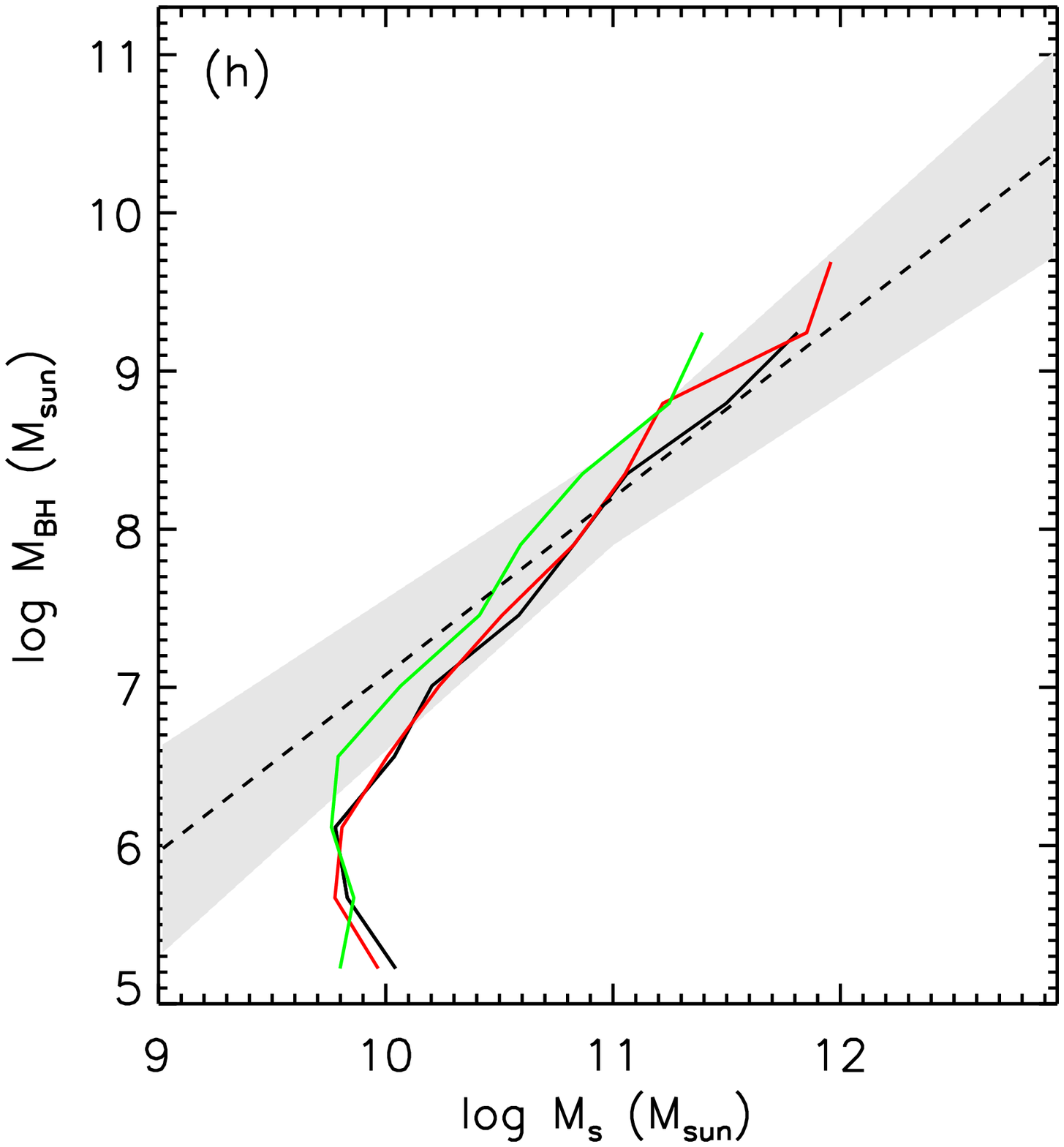}}\hspace{-1.54cm}}
  \centering{\resizebox*{!}{5.cm}{\includegraphics{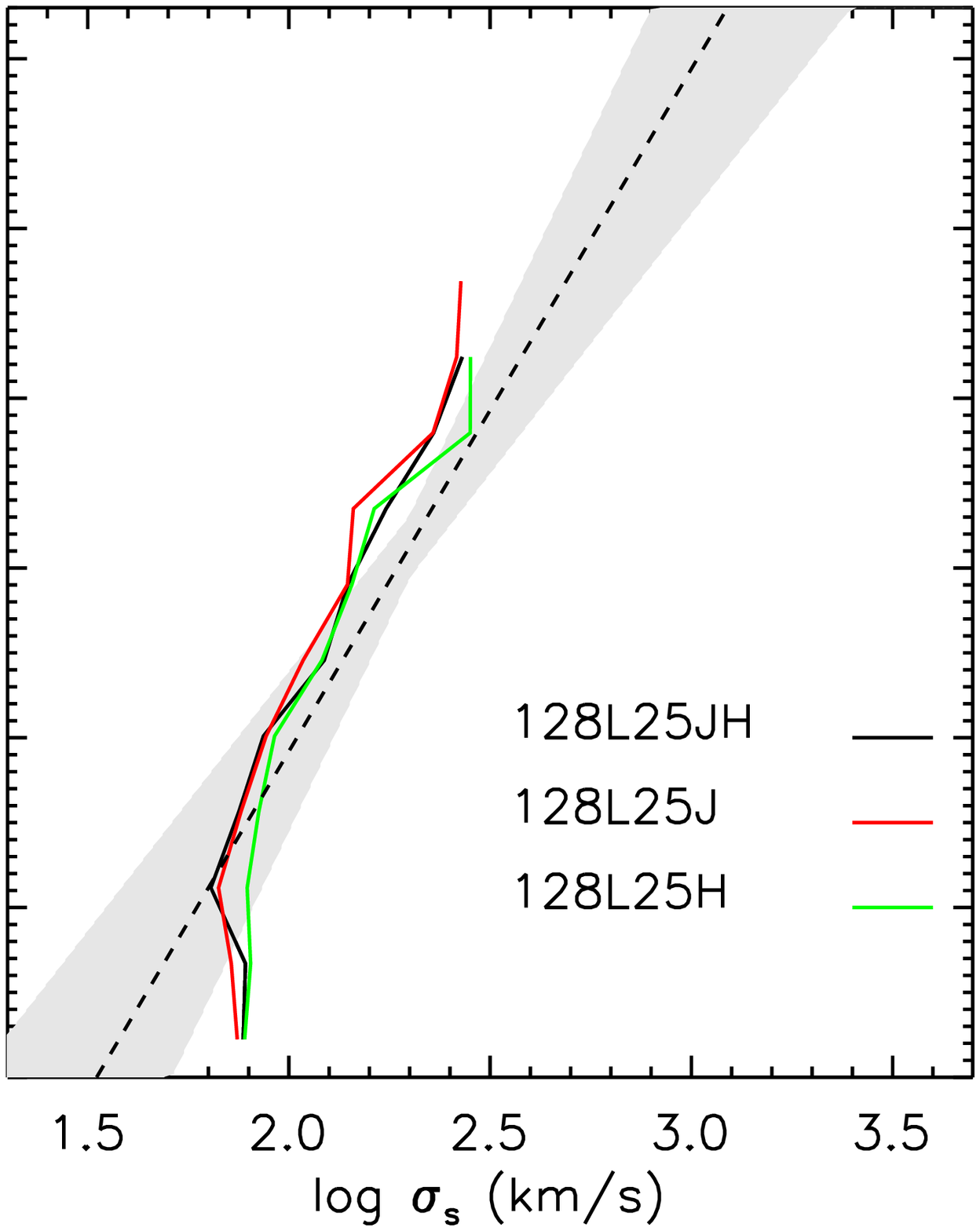}}}
  \caption{Panels (a) through (h) explore a variation of parameters in the same order as listed in the caption
  for fig.~\ref{bhdensity_comp}.  In each panel we plot the  black hole mass as a function of the stellar mass (left), or as a function of stellar velocity dispersion (right). Measurements are done at $z=0$. We overplotted the observational laws as dashed lines from \citet{haring&rix04} for the $M_{\rm BH}$-$M_{\rm s}$ relation and \citet{tremaineetal02} for the $M_{\rm BH}$-$\sigma_{\rm s}$ relation with their 3$\sigma$ uncertainties. The dotted line in the left-hand panel of (a) indicates the relationship between log $M_{\rm BH}$ and log $M_{s}$ when $M_{\rm BH} = M_{\rm s}$.}
    \label{magorrian_comp}
\end{figure*}

\begin{figure*}
  \centering{\resizebox*{!}{5.5cm}{\includegraphics{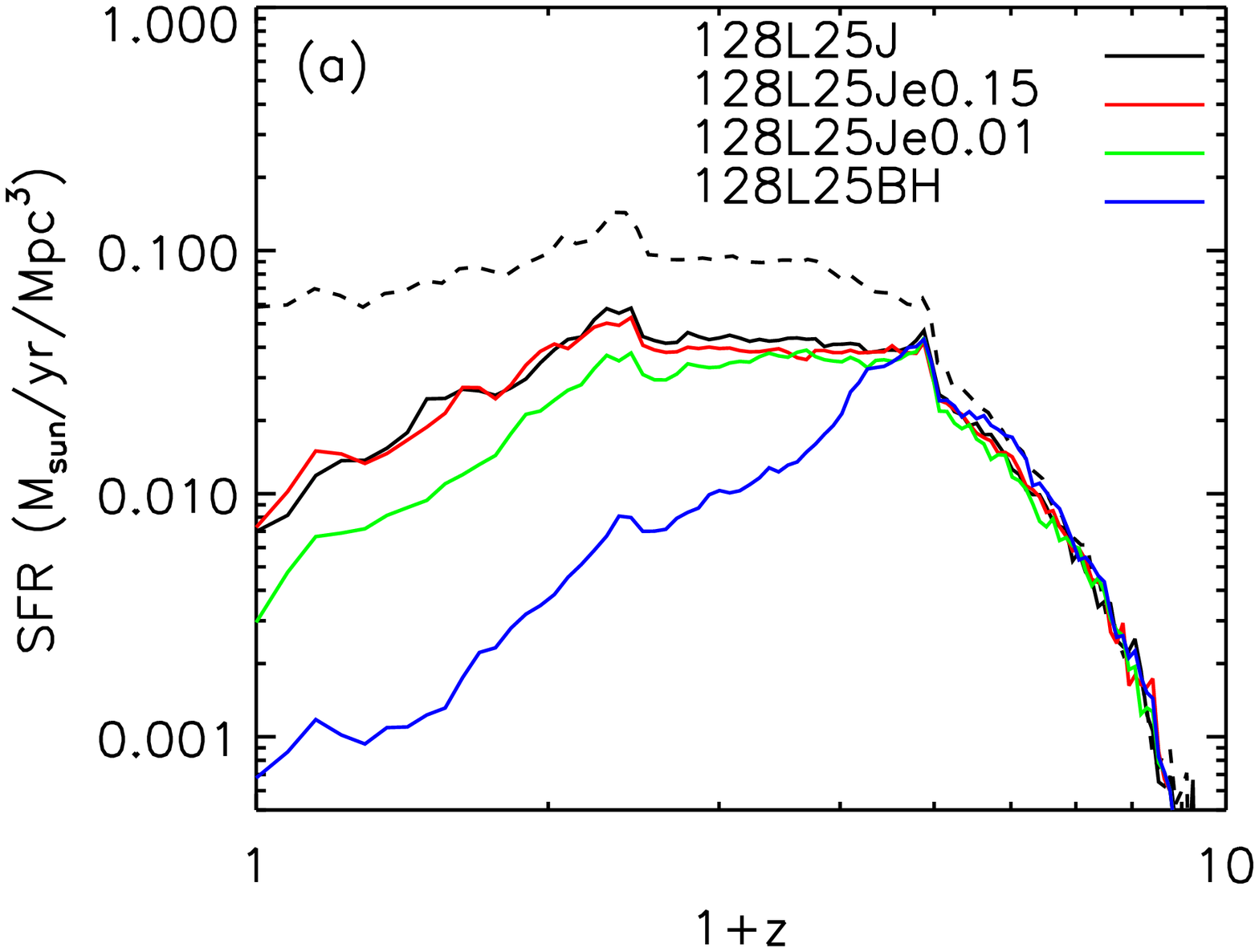}}\hspace{-0cm}}
  \centering{\resizebox*{!}{5.5cm}{\includegraphics{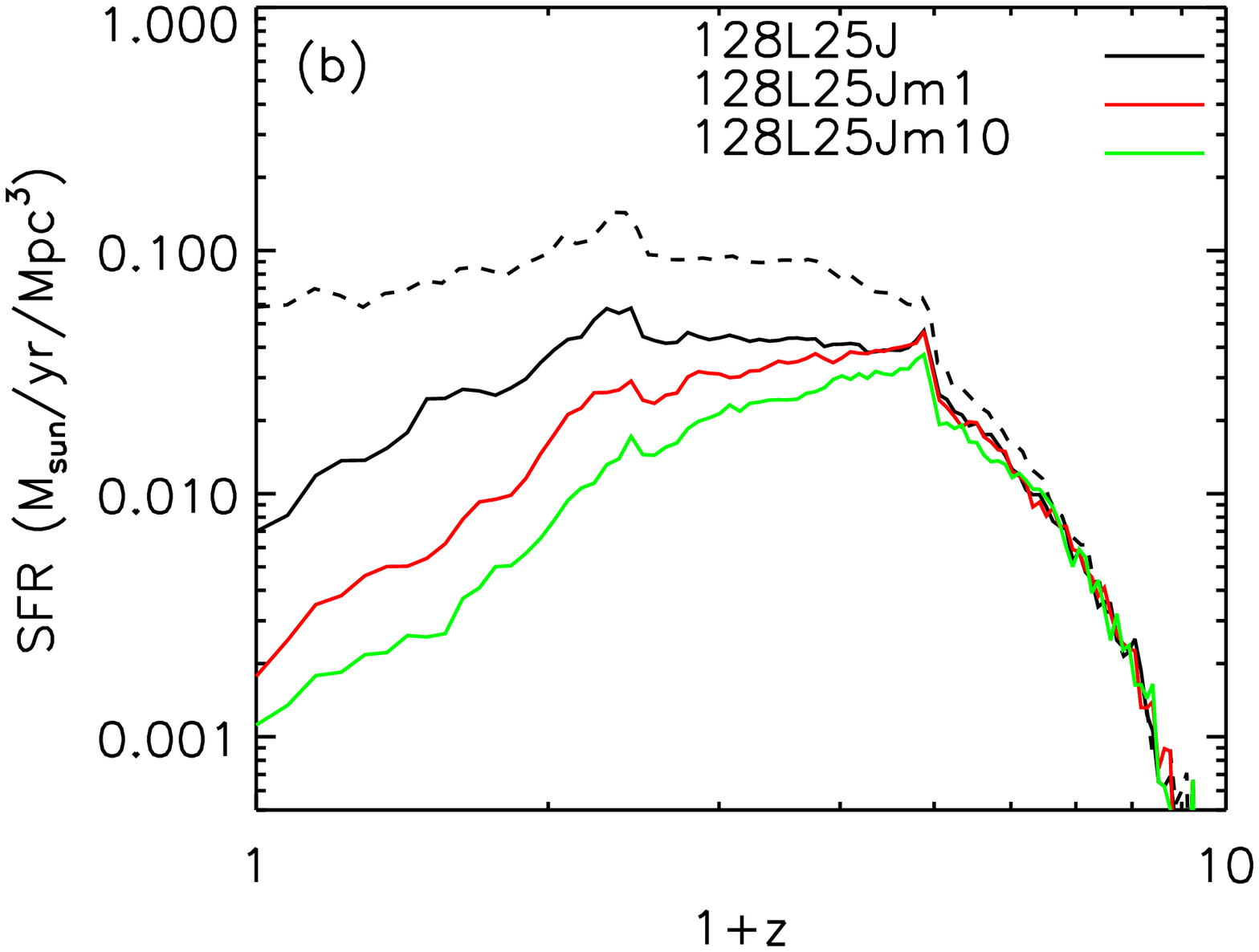}}\hspace{-0cm}}
  \centering{\resizebox*{!}{5.5cm}{\includegraphics{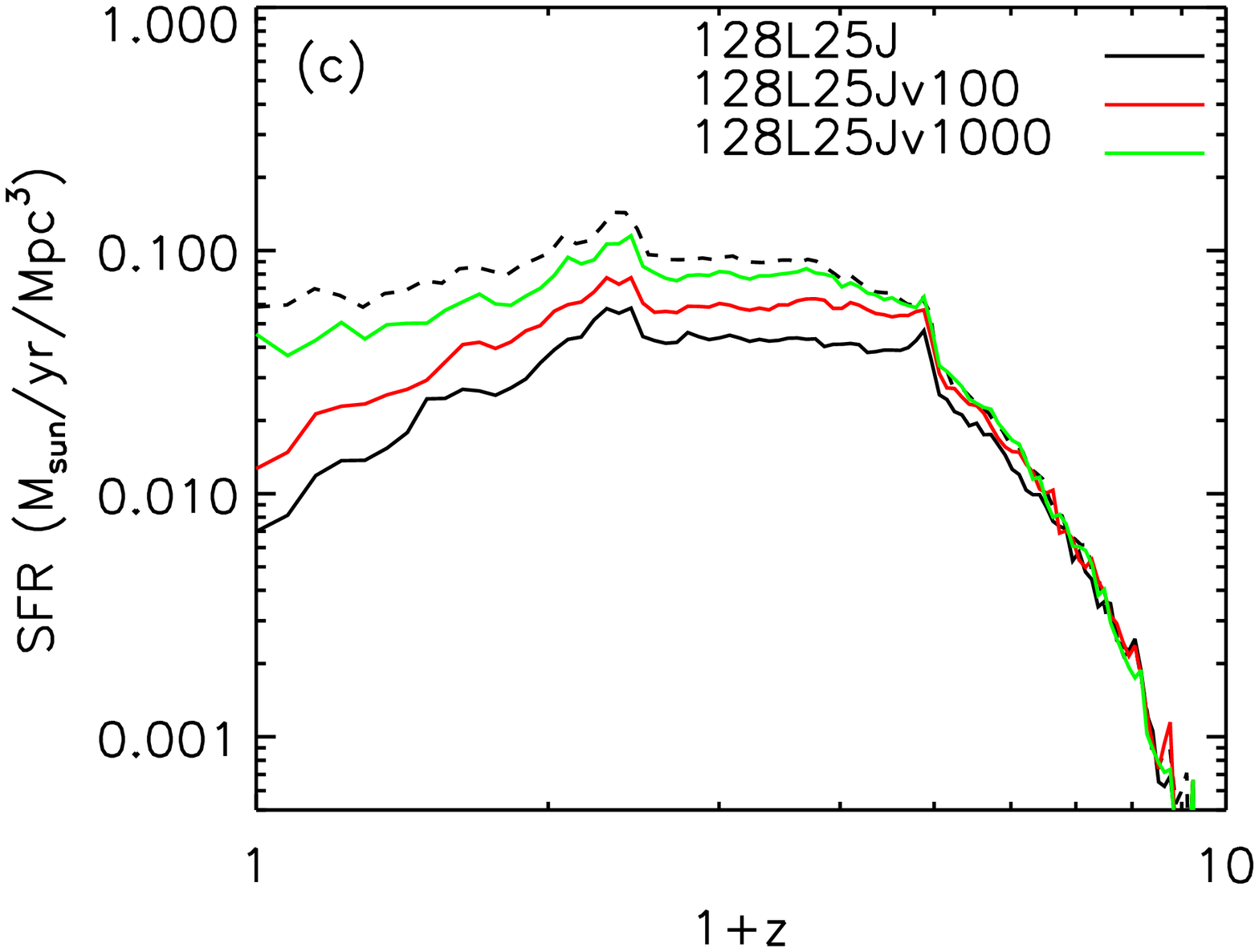}}\hspace{-0cm}}
  \centering{\resizebox*{!}{5.5cm}{\includegraphics{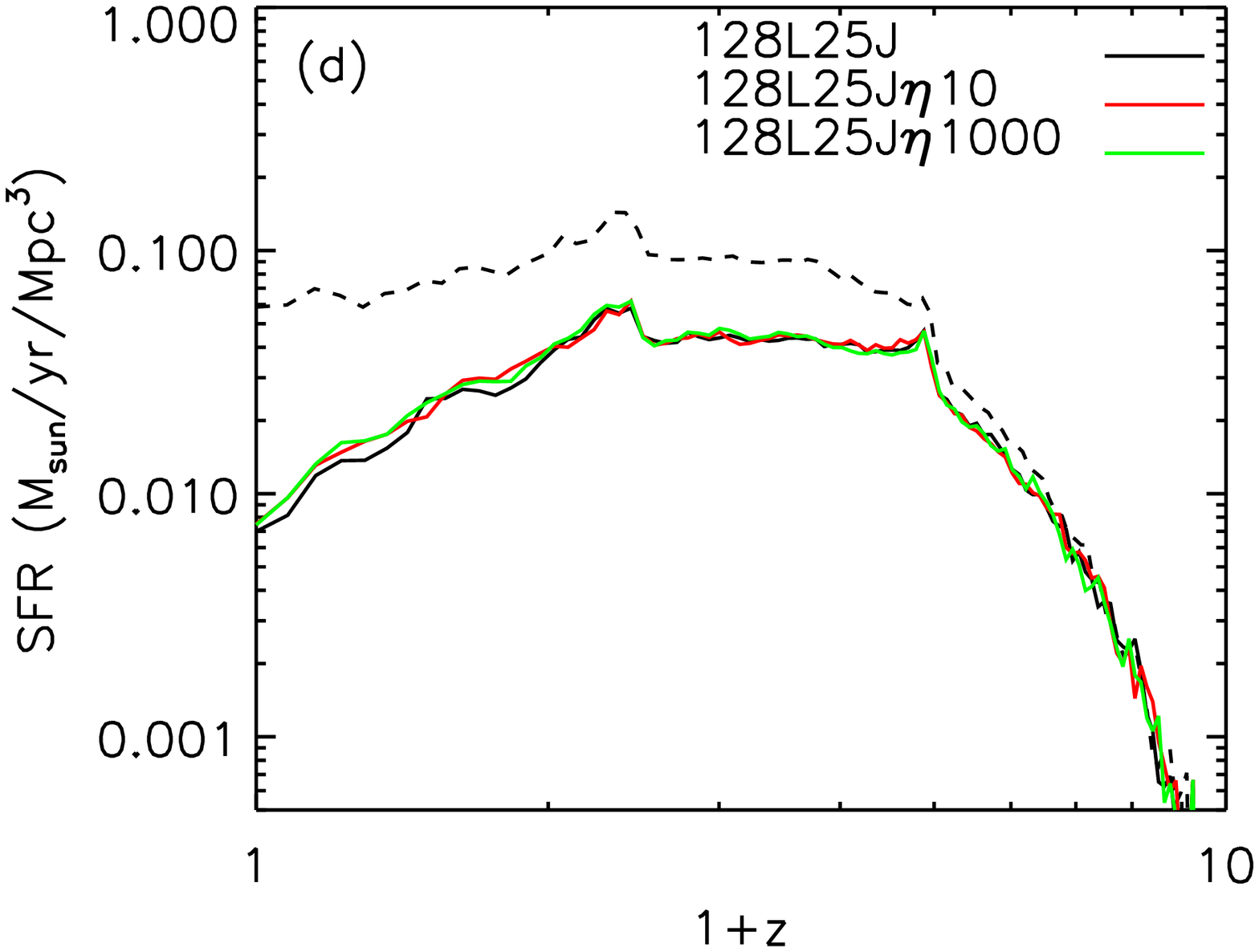}}\hspace{-0cm}}
  \centering{\resizebox*{!}{5.5cm}{\includegraphics{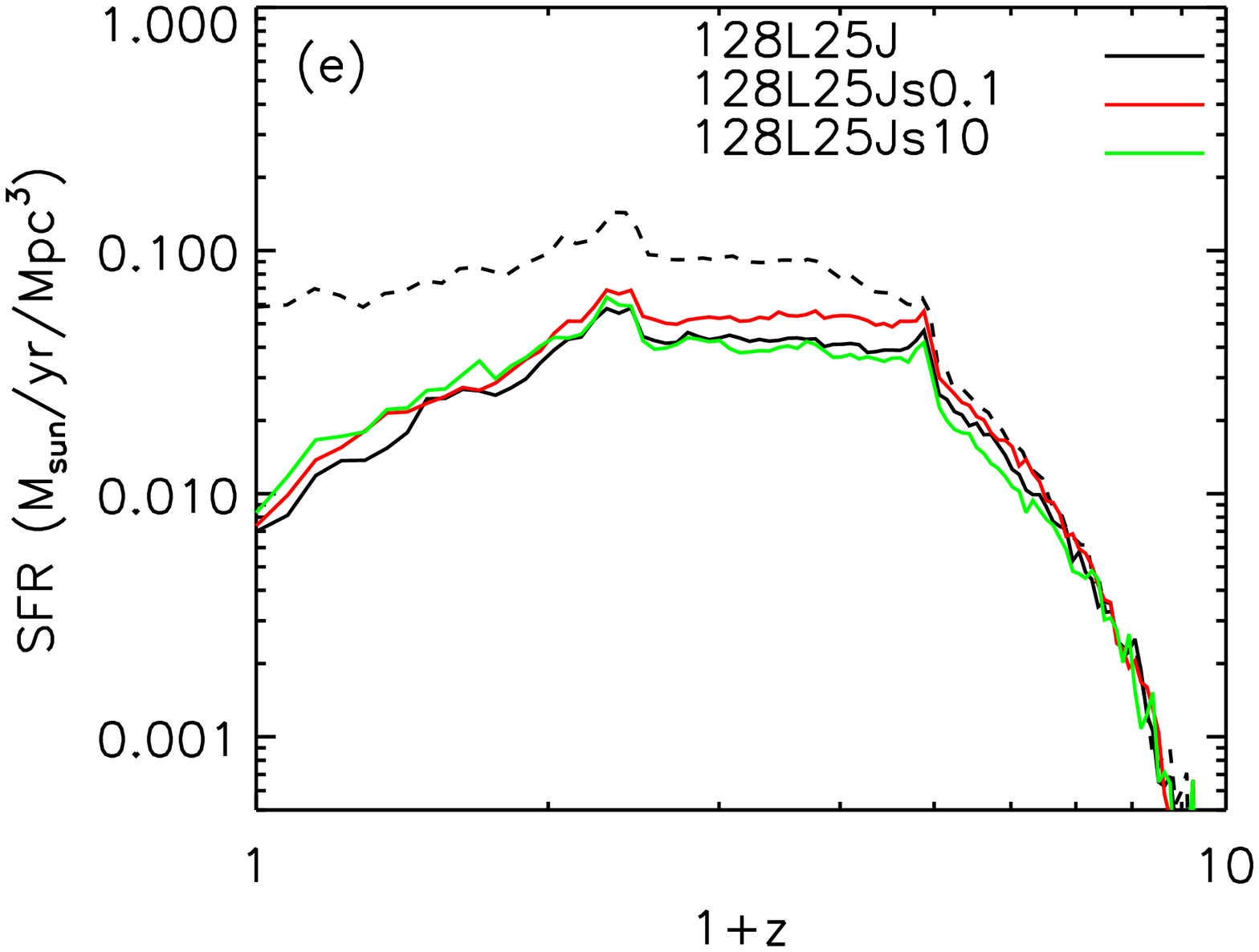}}\hspace{-0cm}}
  \centering{\resizebox*{!}{5.5cm}{\includegraphics{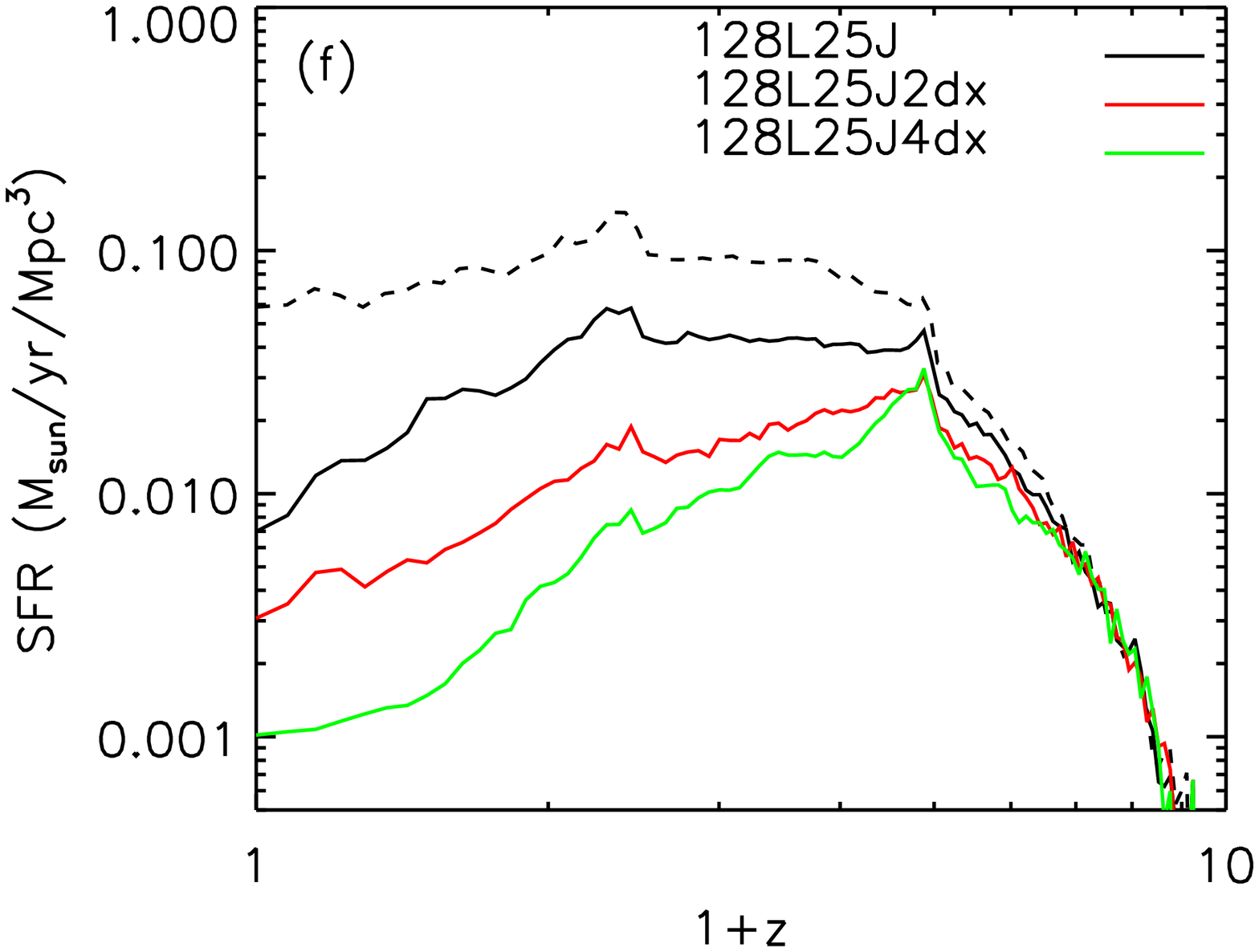}}\hspace{-0cm}}
  \centering{\resizebox*{!}{5.5cm}{\includegraphics{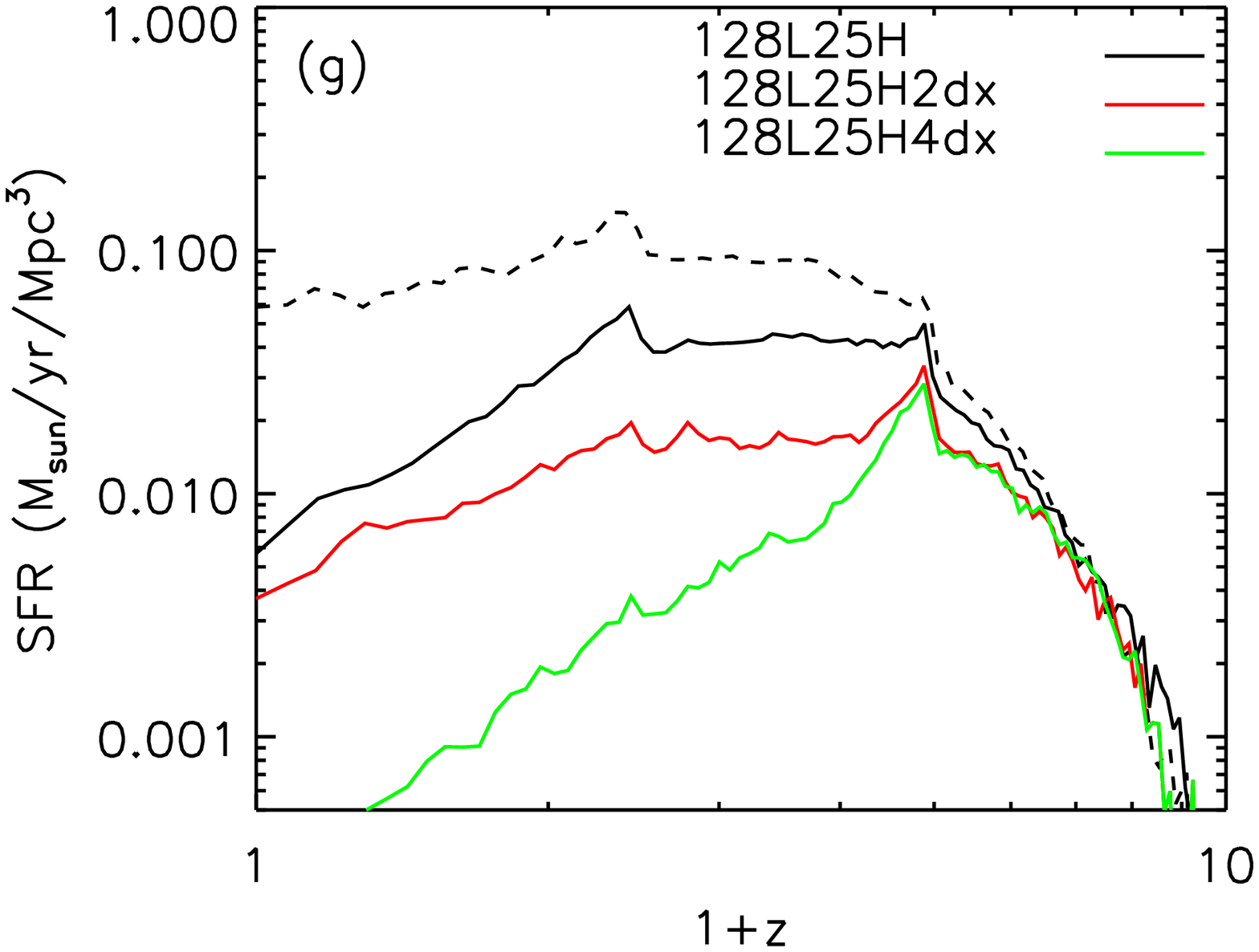}}\hspace{-0cm}}
  \centering{\resizebox*{!}{5.5cm}{\includegraphics{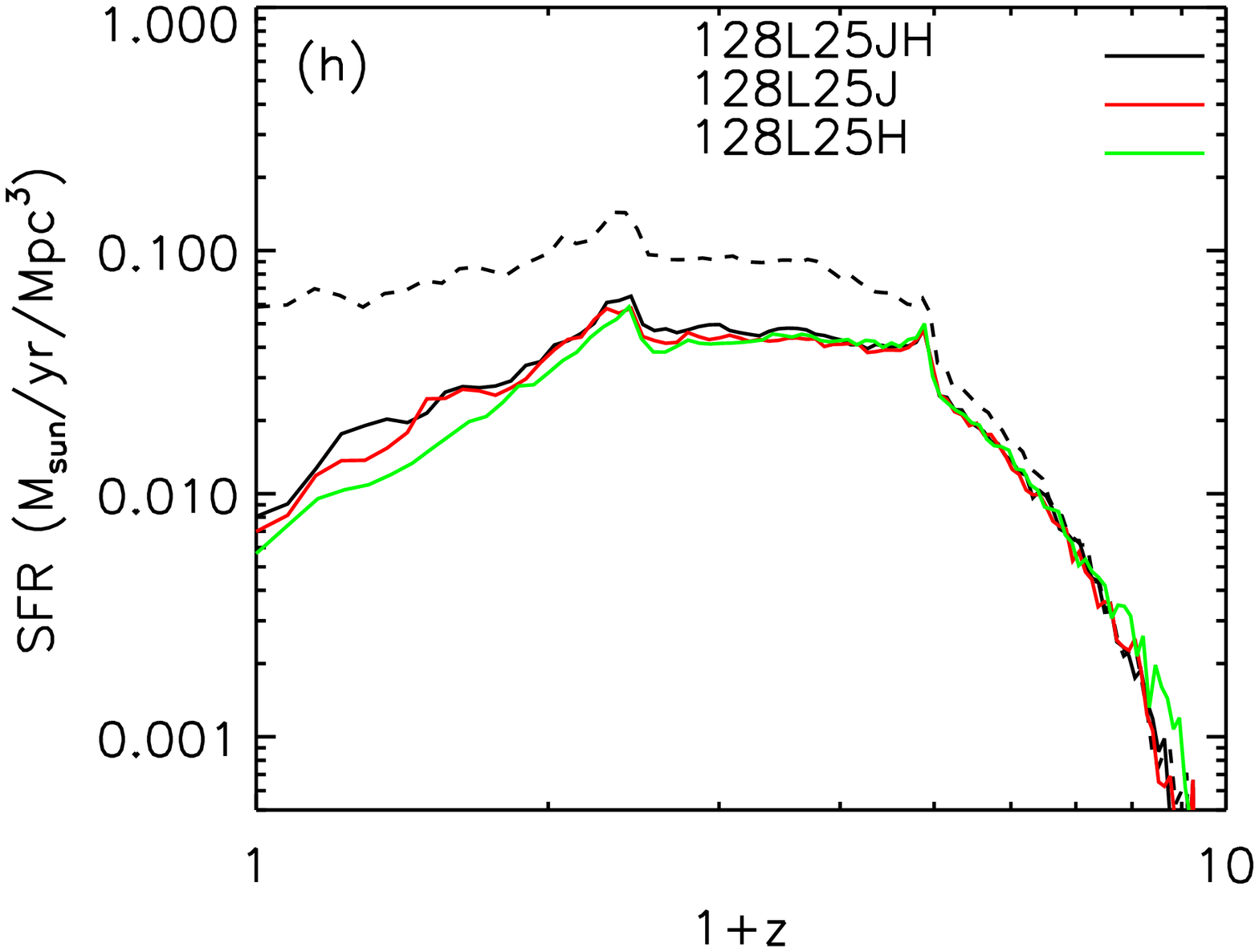}}\hspace{-0cm}}
  \caption{Panels (a) through (h) explore a variation of parameters in the same order as listed in the caption
  for fig.~\ref{bhdensity_comp}. Comoving SFR as a function of redshift. The dashed line corresponds to the 128L25noAGN simulation that does not include AGN feedback. }
    \label{sfr_comp}
\end{figure*}

\subsubsection{Varying efficiency $\epsilon_{\rm f}$}
\label{efficiency_f}

We test the effect of varying the efficiency $\epsilon_{\rm f}$ of the radio mode on the evolution of the cosmic BH density in fig.~\ref{bhdensity_comp}.(a).
We compare three simulations with the implementation of the radio mode identical in every respect expect for the efficiency, $\epsilon_{\rm f}$ which varies by approximately factors of ten from
0.01 (128L25Je0.01) to 0.15 (128L25Je0.15) to 1 (128L25J). We also compare a run with black hole growth but no AGN feedback (128L25BH). 
Efficiency values lower than $\epsilon_{\rm f}=1$ produce larger BH densities at $z=0$ than their observational counterparts.
The case for which no AGN feedback energy is released (128L25BH) produces what we take to be the maximum  attainable BH density.
Even small amounts of energy ($\epsilon_{\rm f}=0.15$ and $\epsilon_{\rm f}=0.01$) prevent BHs from growing to these maximum BH densities.
The maximum possible efficiency $\epsilon_{\rm f}=1$ predicts a slightly larger BH density than the data from \cite{shankaretal04}, but is still within $3\sigma$ error.

\begin{figure}
  \centering{\resizebox*{!}{6.cm}{\includegraphics{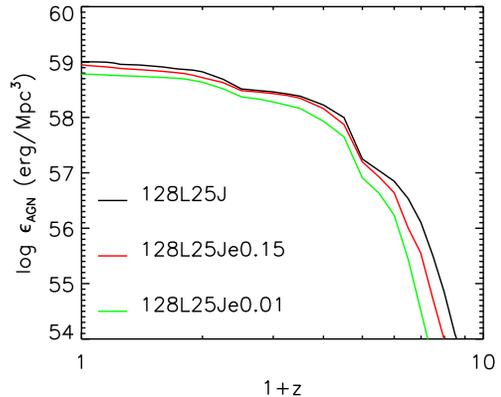}}}
  \caption{Cumulative comoving AGN energy density as a function of redshift for different AGN feedback efficiencies $\epsilon_{\rm f}$. }
    \label{LAGN_density}
\end{figure}

The decrease in efficiency, $\epsilon_{f}$ is compensated by larger accretion rates and more massive BHs, leading to the net result that the total amount of energy released by the AGN feedback is nearly independent of $\epsilon_{f}$ (see also~\citealp{booth&schaye10}).
Fig.~\ref{LAGN_density} substantiates this by showing the comoving cumulative AGN energy density as a function of redshift for different efficiencies.
Indeed, even though AGN efficiency $\epsilon_{\rm f}$ is allowed to vary by two orders of magnitude, the total amount of energy liberated at $z=0$ differs by less than a factor 2.
This suggests that BHs adjust their masses so that the total energy liberated can blow the gas out from the galaxies and stop the accretion of gas. 
We do not present the AGN energy density evolution for the other simulations presented in fig.~\ref{bhdensity_comp} because aside from the simulations presented in panel (h),
they all run with the same $\epsilon_{f}$. Therefore the AGN energy density, $\epsilon_{\rm AGN}$, can be deduced from their $\rho_{\rm BH} $ since 
$\epsilon_{\rm AGN}$ = $\epsilon_{f}\epsilon_{r}c^{2}\rho_{\rm BH}$, where we recall that the radiative efficiency $\epsilon_{r}$ = 0.1 for all simulations and $c$ is the speed of light.

Decreasing the efficiency leads to more massive BHs, but, interestingly, the stellar masses of these galaxies and their stellar velocity dispersions are not significantly impacted by the efficiency (fig.~\ref{magorrian_comp}.(a)).
This is confirmed by the cosmic SFR seen in fig.~\ref{sfr_comp}.(a), which shows little difference in SFR for  different values of the AGN feedback efficiency, especially for $\epsilon_{\rm f} =1$ and $\epsilon_{\rm f}=0.15$.
BHs regulate themselves as well as the gas content of their host galaxy by injecting the same quantity of energy, regardless of $\epsilon_{\rm f}$, to unbind the cold gas component.
The small decrease in the SFR for $\epsilon_{\rm f}=0.01$ occurs because BH masses become comparable to their host galaxy masses so  the BHs accrete gas instead of letting it form stars.
This effect is more obvious when AGN feedback is not allowed but BH growth  is permitted (128L25BH) : the SFR is  suppressed by one order of magnitude because BHs are more massive than the entire stellar content of their host galaxy and they consume all the fresh gas available.

We also plot in fig.~\ref{sfr_comp}, the cosmic SFR for the simulation (128L25noAGN) without AGN feedback nor BHs but including our standard sub-grid physics (cooling, star formation, SN feedback).
In this case, the SFR is systematically higher than any of the simulations including BH growth with or without AGN feedback, and the difference is clearer at low and intermediate redshifts $z=0-4$.
This shows that AGN feedback efficiently suppresses star formation in galaxies, because it prevents gas overcooling and/or ejects large amounts of cold gas back into the Circum-Galactic Medium (CGM).

We remark that the jet velocity depends on the efficiency as $u_{\rm J} \propto \sqrt{\epsilon_{\rm f}/\eta}$, and that the cumulative momentum imparted by all BHs is $Q\propto\sqrt{\eta/\epsilon_{\rm f}}\epsilon_{\rm AGN}$.  Since $\epsilon_{\rm AGN}$ is almost independent of $\epsilon_{\rm f}$ (fig.~\ref{LAGN_density}), $Q$ depends only on $\sqrt{\eta/\epsilon_{\rm f}}$.
 Thus, lower efficiencies produce higher total momentum providing a possible explanation for why self-regulation is weaker for lower efficiency, $\epsilon_{f}$. However by varying the mass loading parameter, $\eta$,  in section~\ref{mass_loading}, we show that self-regulation is controlled by the AGN feedback energy rather than momentum.

This first set of experiments exploring variations in $\epsilon_{f}$  suggests that AGN feedback is a necessary element for the self-regulation of the growth of BHs, and that a high value of the AGN feedback efficiency in the radio mode $\epsilon_{\rm f}=1$ is required to obtain realistic results on the coevolution of BHs and  galaxies.

\subsubsection{Varying AGN energy delay: $\Delta M_{\rm d}$}

We allow for a time delay, represented by the $\Delta M_{\rm d}$ parameter, before releasing AGN energy in the radio mode.
This parameter prescribes that the BH must grow by more than a $\Delta M_{\rm d}$ fraction of its mass before releasing energy into a bipolar kinetic jet.
We test three values for the time delay parameter for the radio mode:  $\Delta M_{\rm d}$ = 0 (128L25J), $1\%$ (128L25Jm1), and $10\%$ (128L25Jm10), 
At high redshift, varying the time delay has a negligible impact on the evolution of the BH density (see fig.~\ref{bhdensity_comp}.(b)) because BHs grow close to the Eddington accretion rate $\dot M_{\rm Edd}$  (see section~\ref{BHevolution}).
Thus, typical growth timescales of BHs are extremely short and even with a non-zero $\Delta M_{\rm d}$, energy is released almost continuously. 
The BH growth timescale can be defined by $t_{\rm BH}=M_{\rm BH}/\dot M_{\rm BH}$, and since $\dot M_{\rm Edd}\propto M_{\rm BH}$, it simplifies to
\begin{equation}
t_{\rm BH}={\epsilon_r \sigma_T c \over 4 \pi G m_{\rm p}} \chi^{-1} \simeq 45.5\, \chi^{-1} \, \rm Myr\, , 
\end{equation}
depending only on the Eddington accretion ratio $\chi$.
As a  consequence, BHs accreting gas at high Eddington ratios have a more continuous and hence immediate impact on the surrounding gas than BHs in a low accretion mode.

However, at low redshift, results from simulations with different time delays diverge. When a delay is permitted (128L25Jm1, 128L25Jm10), the final BH density at $z=0$ is smaller than when energy is continuously deposited (128L25J). This is linked to easier BH self-regulation with larger $\Delta M_{\rm d}$.
Since BHs accrete gas at low redshift in a low-Eddington accretion regime, with non-zero   $\Delta M_{\rm d}$  there are larger stretches of time before a significant amount of energy is  released.
As a result of this accumulation of energy, BHs release a comparable amount of energy to the $\Delta M_{\rm d}$ = 0 case (128L25J) but in a shorter amount of time, allowing for fewer but stronger AGN luminosity bursts.
Thus, BHs can more easily self-regulate with smaller duty cycles~\citep{pope11}. 

The time delay parameterized by  $\Delta M_{\rm d}$ also modifies the relationships between BH masses and their host galaxy properties (fig.~\ref{magorrian_comp}.(b)): BHs with masses $M_{\rm BH}>10^7 M_{\odot}$, sit in lower mass galaxies.
This effect is stronger for $\Delta M_{\rm d}=10\%$ (128L25Jm10) for which stellar masses are reduced by an order of magnitude for the most massive galaxies compared to the 
simulation with a continuous ($\Delta M_{\rm d}$ = 0) injection rate (128L25J).
The maximum stellar velocity dispersions of the host galaxies of the most massive BHs are reduced as a direct consequence of the reduction of the stellar mass in them.
The effect is even more apparent for the intermediate BHs ($10^7<M_{\rm BH}<10^8 M_{\odot}$) in 128L25Jm10, where a clear deviation from the observational fit is observed.
This suggests that the stellar velocity dispersion in massive galaxies is essentially controlled by the cold baryon content, and not by the total halo mass, which is hardly modified by the AGN feedback.

Finally, in fig.~\ref{sfr_comp}.(b), we see that the intermediate and low redshift SFR depends a lot on the  $\Delta M_{\rm d}$ parameter, which is not surprising since we saw that is also influences the $M_{\rm BH}$-$M_{\rm s}$ relationships.
Large $\Delta M_{\rm d}$ more efficiently suppresses the total SFR and has more impact on the gas content in galaxies because they undergo shorter and stronger episodes of AGN feedback.
This parameter study teaches us that not only is the amount of energy deposited important, but that the duration of the energy release plays a key role in unbinding the gas content of galaxies.
Galactic gas exposed to a small deposit of energy can efficiently return to equilibrium by the gas dynamics and by the short cooling times involved in high density gas.

\subsubsection{Varying maximum relative velocity: $u_{\rm max}$}

We measure the effect of varying the maximum allowed velocity $u_{\rm max}$ of the gas relative to the BH in the Bondi formula (equation~\ref{dMBH}).
Increasing our fiducial value $u_{\rm max}=10$ km/s (128L25J) up to $u_{\rm max}=1000$ km/s (128L25Jv1000) decreases the overall BH densities (see fig.~\ref{bhdensity_comp}.(c)), but values are still consistent with the observations.
As we would expect, larger values of the relative velocity inhibit the growth of BHs.
This effect already comes into play at high redshift, when mergers between galaxies are numerous
sometimes resulting in violent excursions of BHs in their host galaxies, leading to large BH velocities relative to the dense gas component.

This spurious effect comes from our inability to resolve the very small scales of the ISM within which BHs should be embedded.
Some authors have circumvented this problem by correcting the positions of BHs when they move too far from the gravitational potential well (private communication with Volker Springel).
Here we prefer to adopt a more straight-forward approach by limiting the maximum gas velocity relative to the BH in the formula for gas accretion (equation~\ref{dMBH}) rather than changing the position of the BH. This also allows us to follow the BHs that are ejected from their galaxies by strong tidal effects and gravitational friction, as material is stripped from galaxy satellites falling into massive halos.
 
Fig.~\ref{magorrian_comp}.(c) shows that different $u_{\rm max}$ produce different BH masses and host galaxy stellar properties. 
When large relative velocities are permitted (128L25Jv100, 128L25v1000), galaxies tend to be more massive, and, as a result, have larger velocity dispersions.
Even though the total BH density at z = 0 is hardly changed for different $u_{\rm max}$, the SFR is very sensitive to $u_{\rm max}$ (fig.~\ref{sfr_comp}.(c)).
Large $u_{\rm max}$ values tend to nullify the effect of the AGN feedback on the total SFR, and the SFR converges to the case where feedback from AGN is not allowed.
As a  consequence $u_{\rm max}=10$ km/s corresponds to a choice that allows for a non-spurious quenching of the accretion rate while keeping the dynamics of the BH particles completely self-consistent. 

\subsubsection{Varying jet mass loading factor: $\eta$}
\label{mass_loading}

The mass loading factor $\eta$ is a free parameter of the radio mode for AGN feedback that controls the velocity  the jet would have if it was propagating into a void.
We compare three simulations with $\eta=10$ (128L25J$\eta$10), $\eta=100$ (128L25J), and $\eta=1000$ (128L25J$\eta$1000).
As can be seen from fig.~\ref{bhdensity_comp}.(d), fig.~\ref{magorrian_comp}.(d), and fig.~\ref{sfr_comp}.(d), the BH and galaxy properties are very insensitive to the adopted values of the mass loading factor of the jet.
The reason is that the jet couples to the gas in its surroundings and AGN feedback becomes ineffective when the energy of the jet becomes comparable to the binding energy of the gas.
The only difference introduced by $\eta$ is that depending on its value, the jet will go more or less quickly into equilibrium with the gas, but as the liberated energy is the same regardless of $\eta$, the same amount of gas is impacted by the jet.

We insist on the fact that the total imparted momentum $Q\propto \sqrt{\eta/\epsilon_{\rm f}} \epsilon_{\rm AGN}$ loses its $\epsilon_{\rm AGN}$ dependence because the latter is constant
given that BH densities are constant for different $\eta$ (fig.~\ref{bhdensity_comp}.(d)) and $\epsilon_{\rm f}$ is the same ($\epsilon_{\rm f}$=1) for the three simulations we are comparing.
As a result, $Q\propto \sqrt{\eta/\epsilon_{\rm f}}$ as it was for the case with varying efficiencies (section~\ref{efficiency_f}).
However since $\epsilon_{\rm f}$ is a constant, momentum, $Q$, only depends on $\eta$ for the set of simulations compared in this section. Figs.~\ref{bhdensity_comp}.(d), ~\ref{magorrian_comp}.(d), and ~\ref{sfr_comp}.(d) show that BHs and their host galaxy properties do not depend on $\eta$, or equivalently on the momentum $Q$. Therefore we
conclude that BHs and their host galaxy properties are only sensitive to jet energies, not their momenta.

\subsubsection{Varying initial BH mass: $M_{\rm seed}$}

We vary the BH initial seed mass by choosing values as small as  $10^4\, \rm M_{\odot}$ (128L25Js0.1) and as large as $10^6\, \rm M_{\odot}$ (128L25Js10).
Simulation 128L25J has $M_{\rm seed}$ = $10^5\, \rm M_{\odot}$.
As seen on fig.~\ref{bhdensity_comp}.(e), this has little effect on the final ($z=0$) BH density.
Differences appear at high redshift, when most of the contribution to the BH density comes from BHs with mass close to their initial seed mass.
Thus, the choice for the initial seed BH mass has an important impact on the BH density at high redshift but memory of it is rapidly erased when BHs grow to values larger than their initial mass.

The Magorrian relationships are almost unchanged by different choices for $M_{\rm seed}$ (fig.~\ref{magorrian_comp}.(e)).
But a careful inspection of the cosmic SFR (fig.~\ref{sfr_comp}.(e)) shows that the SFR is slightly different at high redshift $z=2-4$: lower seed BH masses result in a larger SFR because it takes more time for BHs to reach a self-regulated equilibrium with their environment.

\subsubsection{Varying AGN input size: $r_{\rm AGN}$}
\label{AGN_input_size}
\begin{figure*}
  \centering{\resizebox*{!}{5.5cm}{\includegraphics{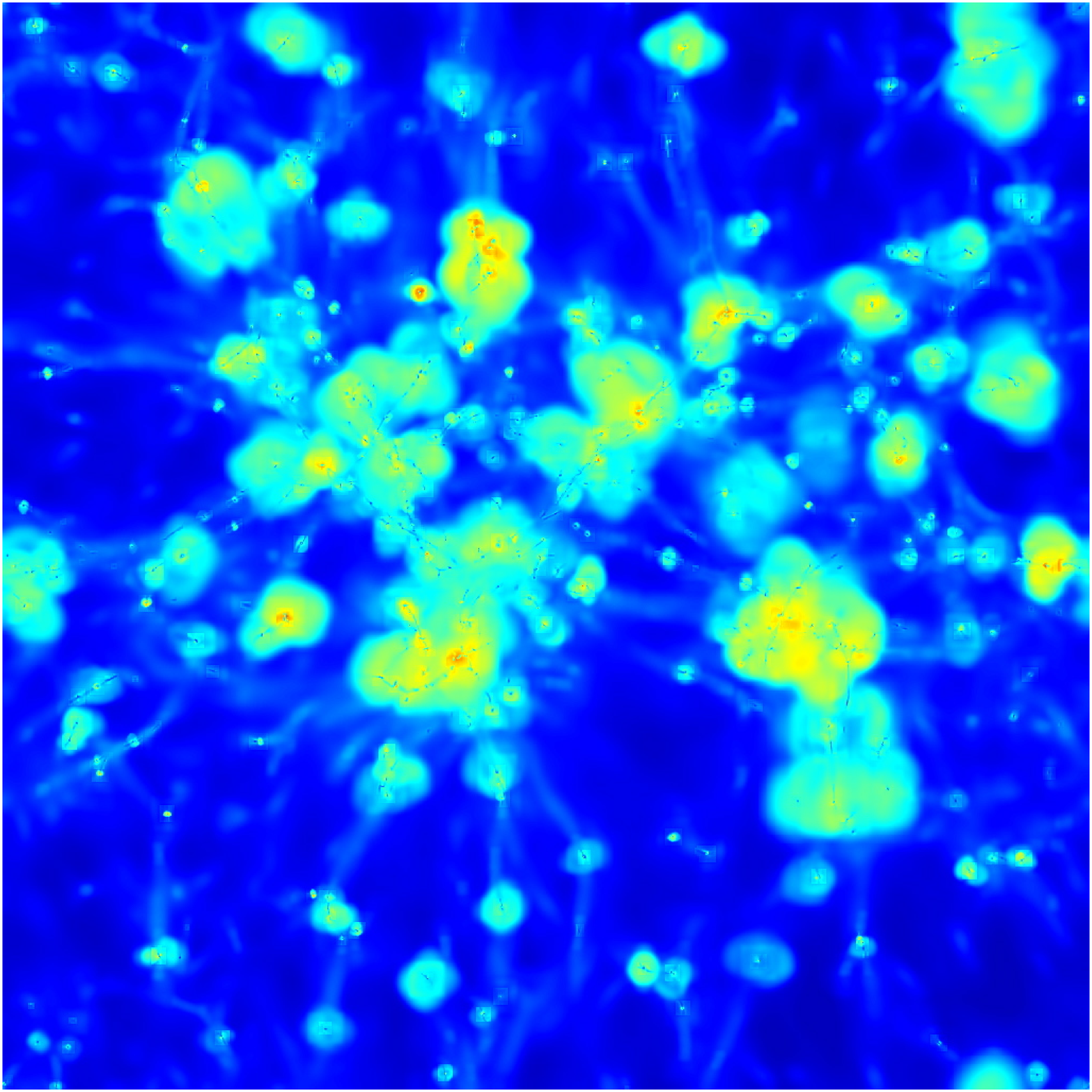}}}
  \centering{\resizebox*{!}{5.5cm}{\includegraphics{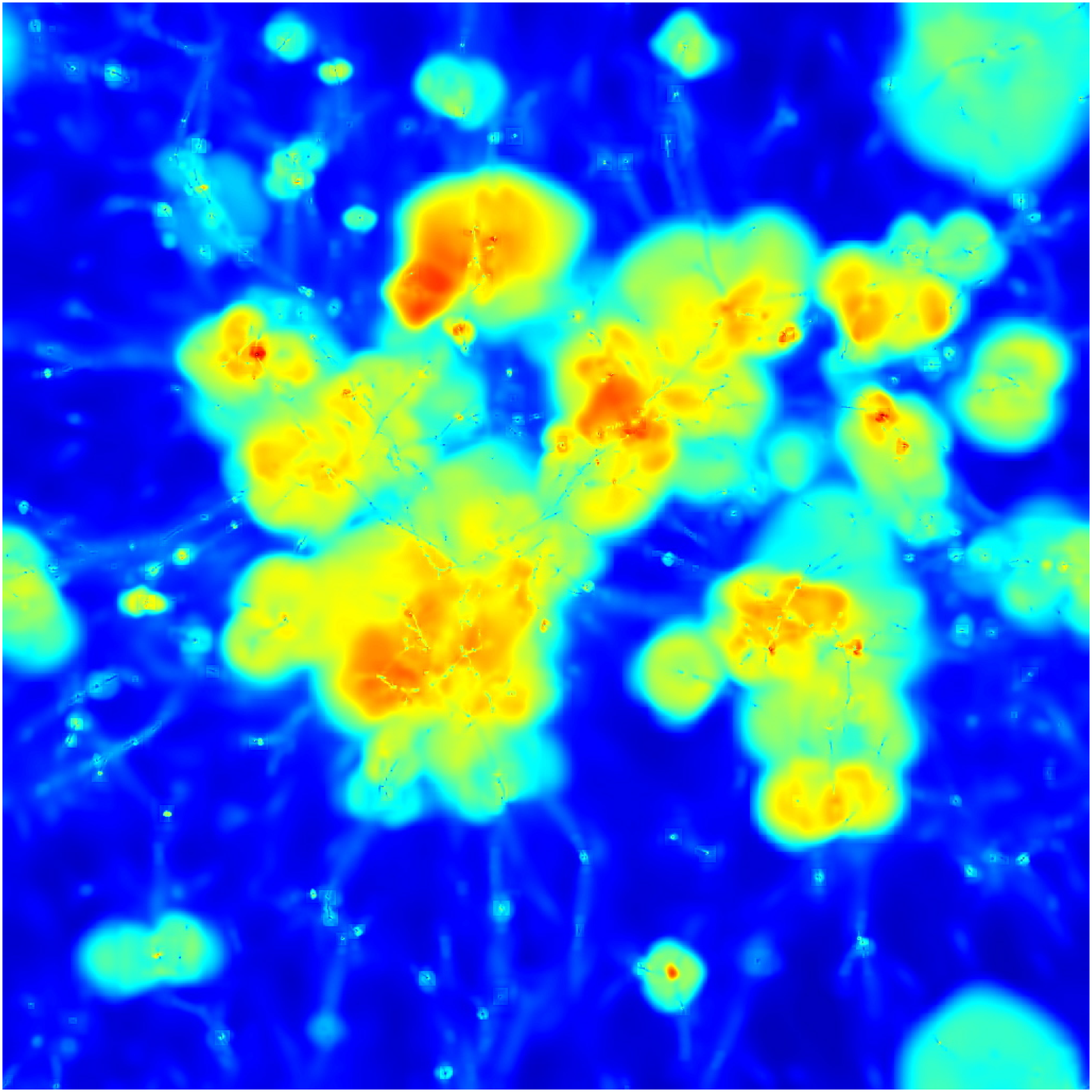}}}
  \centering{\resizebox*{!}{5.5cm}{\includegraphics{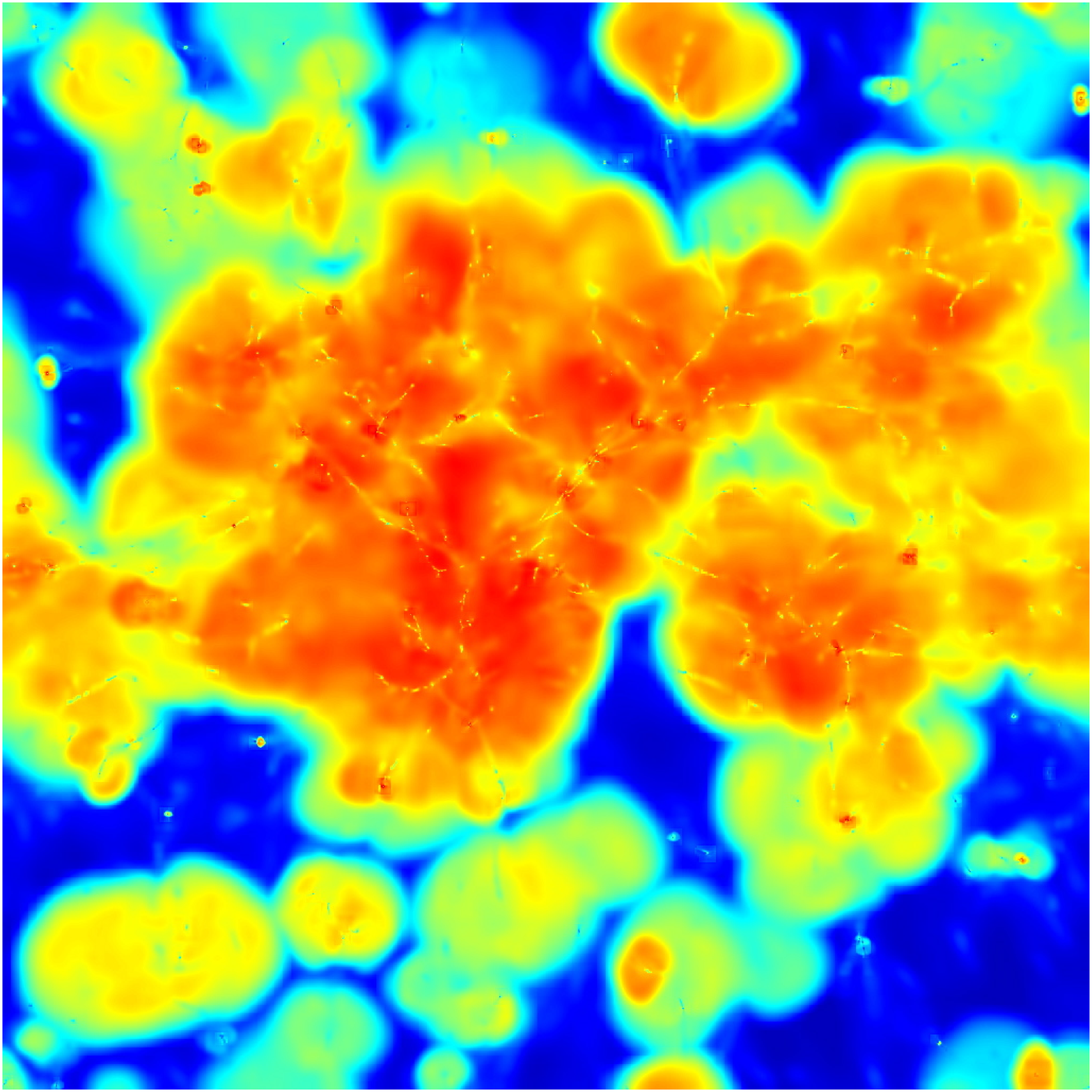}}}
  \centering{\resizebox*{!}{5.5cm}{\includegraphics{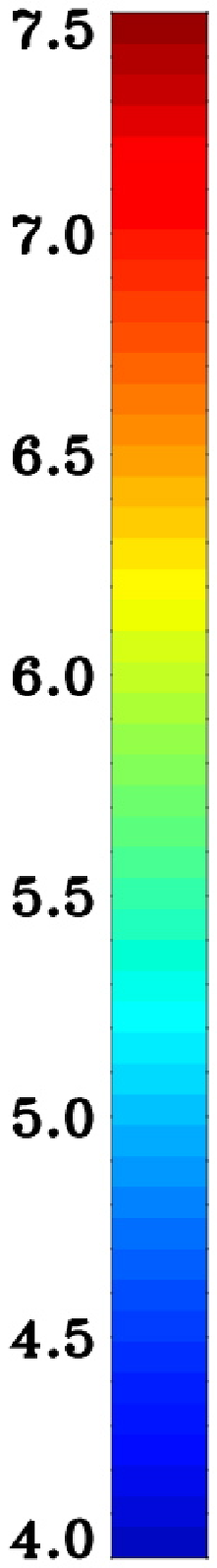}}}
  \caption{Projected temperatures of the 128L25JH (left), 128L25J4dx (middle), 128L25H4dx (right) simulations at $z=3$. The color code is in log(K) units. Images are 6.25 $\rm h^{-1}.Mpc$ physical on a side.}
    \label{temp_dx}
\end{figure*}

We vary the size of the AGN energy input region for both the radio and the quasar mode to deduce its effect on BH self-regulated growth.

First we explore the impact of varying the extent of the jet in the radio mode through a series of three simulations where the AGN input
region varies from $r_{\rm AGN}=\Delta x$ (128L25J), to $r_{\rm AGN}= 2\Delta x$ (128L25J2dx), to $r_{\rm AGN}=4\Delta x$ (128L25J4dx).
Fig.~\ref{bhdensity_comp}.(f) shows that large jet sizes for the radio mode lead to large BH densities.
Since the energy is spread over larger regions when  we increase the jet size, the gas close to the BH
is impacted less by larger jets, thereby accreting more easily onto the BH, hindering its self-regulation, and producing larger BH densities.

Fig.~\ref{magorrian_comp}.(f) shows that jets with sizes larger than $\Delta x$ lead to larger BH masses and lower host galaxy stellar masses, with unrealistic values compared to observations.
Also, the SFR is strongly suppressed when larger jet sizes are chosen (fig.~\ref{sfr_comp}.(f)). 
Thus, large energy injection regions for the radio AGN feedback mode have more impact on galaxy formation because BHs become more massive and as a result  inject more energy to the surrounding gas.
However, as the BHs are much more massive than what is predicted from the $M_{\rm BH}$-$M_{\rm s}$ relationship, they must release more energy to self-regulate.

For the quasar mode the behavior is different. Again, we ran three simulations to explore the effect of changing the size of the energy input region:
from $r_{\rm AGN}=\Delta x$ (128L25H), to $r_{\rm AGN}= 2\Delta x$ (128L25H2dx), to $r_{\rm AGN}=4\Delta x$ (128L25H4dx).
Fig.~\ref{bhdensity_comp}.(g) shows that the trend of BH density with bubble size is non-linear.
Doubling the radius  ($r_{\rm AGN}=2\Delta x$) gives a similar BH density evolution as obtained with $r_{\rm AGN}=\Delta x$ down to redshift $z=1.5$ but shows a drop in BH density below this redshift.
Increasing the radius by four ($r_{\rm AGN}=4\Delta x$), on the other hand, leads to a larger BH density at high redshift, which converges to the same value as the $r_{\rm AGN}=\Delta x$ case at $z=0$.

For the Magorrian relations, choosing  $r_{\rm AGN}=2\Delta x$ rather than  $r_{\rm AGN}=\Delta x$ for the quasar mode leads to smaller stellar velocity dispersions 
but very similar stellar masses.
It seems that for the  $r_{\rm AGN}=2\Delta x$ case, even though the SFR is significantly decreased compared to the $r_{\rm AGN}=\Delta x$ case (see fig.~\ref{sfr_comp}.(g)), the decrease in the BH density keeps the BH mass versus host galaxy stellar mass relation unchanged.
However a larger energy injection region ($r_{\rm AGN}=4\Delta x$) has a dramatic impact on the final galaxy stellar masses and the evolution of the SFR. Both are significantly diminished.
Similar behavior has been found by~\cite{booth&schaye09}, where increasing the number of neighboring SPH particles affected by the AGN bubble  decreases the SFR and increases the BH density.

Both the radio and quasar modes experience a decline in SFR as the size of the injection region (figs.~\ref{sfr_comp}.(f) and (g)) increases because large energy injection regions extend
to less dense regions which are easier to impact. By blowing out the reservoir of hot gas, accretion onto galaxies and hence SFR is suppressed. 
However the self-regulation of BHs is somehow very different for the two modes.
The difference resides in the very nature of energy deposit.
Jets put momentum and kinetic energy into the gas and eventually some of this energy is transformed into thermal energy through shocks, but the more extended the jet, the weaker  the shock.
Fig.~\ref{temp_dx} illustrates this effect in the high redshift Universe: the quasar mode with large $r_{\rm AGN }=4 \Delta x$ (128L25H4dx) inflates larger and hotter bubbles than the radio mode with the same initial jet extent (128L25J4dx).
As the accretion rate is very sensitive to the temperature of the gas $\dot M_{\rm BH}\propto T^{-1.5}$, it is more difficult for jets than for thermal bubbles to self-regulate the growth of BHs. 
This is why BH densities for the jet mode with $r_{\rm AGN}=4 \, \Delta x$ are larger than for the heating mode.

As a final remark, these particular numerical experiments demonstrate that the injection of energy through AGN feedback (but it is true for any type of feedback, see~\citealp{dallavecchia&schaye08} for a similar discussion about SN feedback) is a delicate process that cannot be naively decoupled from the gas dynamics up to large distances, and must be handled with great care. 

\subsubsection{Comparing radio mode and quasar mode}

We compare the choice of using the radio mode (jet mode) to the quasar mode (heating mode), as well as to a combination of both modes.
We found a set of parameters ($\epsilon_{f}=1$ (radio) 0.15 (quasar); $\Delta M_{d} = 0$; $u_{\rm max}=10$ km/s; $\eta=100$; M$_{\rm seed} = 10^{5}$ M$_{\odot}$; r$_{\rm AGN} = \Delta x$) 
consistent with observations for the radio mode and quasar mode used individually
and use this same set of parameters for the dual radio/quasar mode (128L25JH).
With the dual radio/quasar mode of AGN feedback, the feedback from BHs switches to the radio mode when the Eddington accretion ratio $\chi\le \chi_{\rm radio}$ and to the quasar mode when $\chi > \chi_{\rm radio}$.

Fig.~\ref{bhdensity_comp}.(h) shows that the radio (jet) mode (128L25J) produces a larger BH density than the quasar (heating) mode (128L2H), suggesting that the quasar (heating) mode is slightly more efficient  at self-regulating BH growth (although a slightly smaller AGN efficiency, $\epsilon_f$,  of the quasar mode would reduce this difference).
As expected, the dual mode feedback (128L25JH) gives a BH density value which is between the value for individual modes.

There are some `jumps' in the evolution of the BH density at $z=4$ and $z=1.5$ for the radio (jet) mode (fig.~\ref{bhdensity_comp}.(h)), which are not present for the quasar (heating) mode.
This effect is also seen in the SFR evolution (fig.~\ref{sfr_comp}.(h)), both for the radio (jet) and quasar (heating) mode.
It comes from the triggering of a new maximum level of refinement at these redshifts.
With one more level of refinement, the force on small scales is better resolved and the gravitational potential well is deeper.
The gas can therefore get compressed to higher densities, leading to significant boosts in its star formation rate and the accretion rate onto the BHs.
This effect is less pronounced for the quasar (heating) mode where the heating supplies an additional pressure support to the gas, tending to erase this spurious numerical feature.

Fig.~\ref{magorrian_comp}.(h) shows that the stellar mass in the quasar (heating) mode (128L25H) is slightly reduced compared to the radio (jet) mode (128L25J), which is another signature of a slightly more efficient feedback in
the quasar (heating) mode. The mode of AGN feedback has very little effect on the BH mass versus stellar velocity dispersion.
The combination of the two modes show deviations, especially at low $M_{\rm s}$ and $\sigma_{\rm s}$, from  observed BH mass versus stellar mass relationships and observed  BH mass versus stellar velocity dispersion relations. However in fig.~\ref{magorrian_resolution}
we show that this is a consequence of the limited numerical resolution, and that we converge to the observational measurements with increased resolution.

Fig.~\ref{sfr_comp}.(h) shows that the SFR for the different AGN feedback modes are almost undistinguishable.
A small difference can be seen at low redshift $z=0-1$ where the quasar (heating) mode seems to more efficiently suppress the total SFR than the radio (jet) mode. The latter explains the difference seen in the BH mass versus stellar mass relationships (fig.~\ref{magorrian_comp}.(h)).

Finally, this parameter study has allowed us to choose the best fitting parameters for our dual radio/quasar AGN feedback model compared to observations, namely the parameters used in the 128L25JH simulation (see table~\ref{tabnames}).

\subsection{Resolution study}

\begin{figure}
  \centering{\resizebox*{!}{6.cm}{\includegraphics{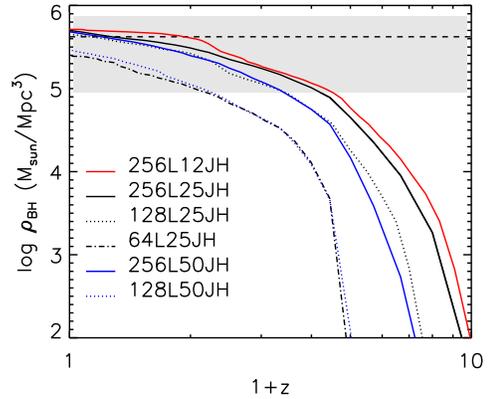}}}
  \caption{Comoving black hole mass density as a function of redshift for different box sizes and resolutions. The grey shaded area is the black hole mass density in our local Universe with its 3$\sigma$ uncertainty from \citet{shankaretal04}.}
    \label{bhdensity_resolution}
\end{figure}

\begin{figure}
  \centering{\resizebox*{!}{6.cm}{\includegraphics{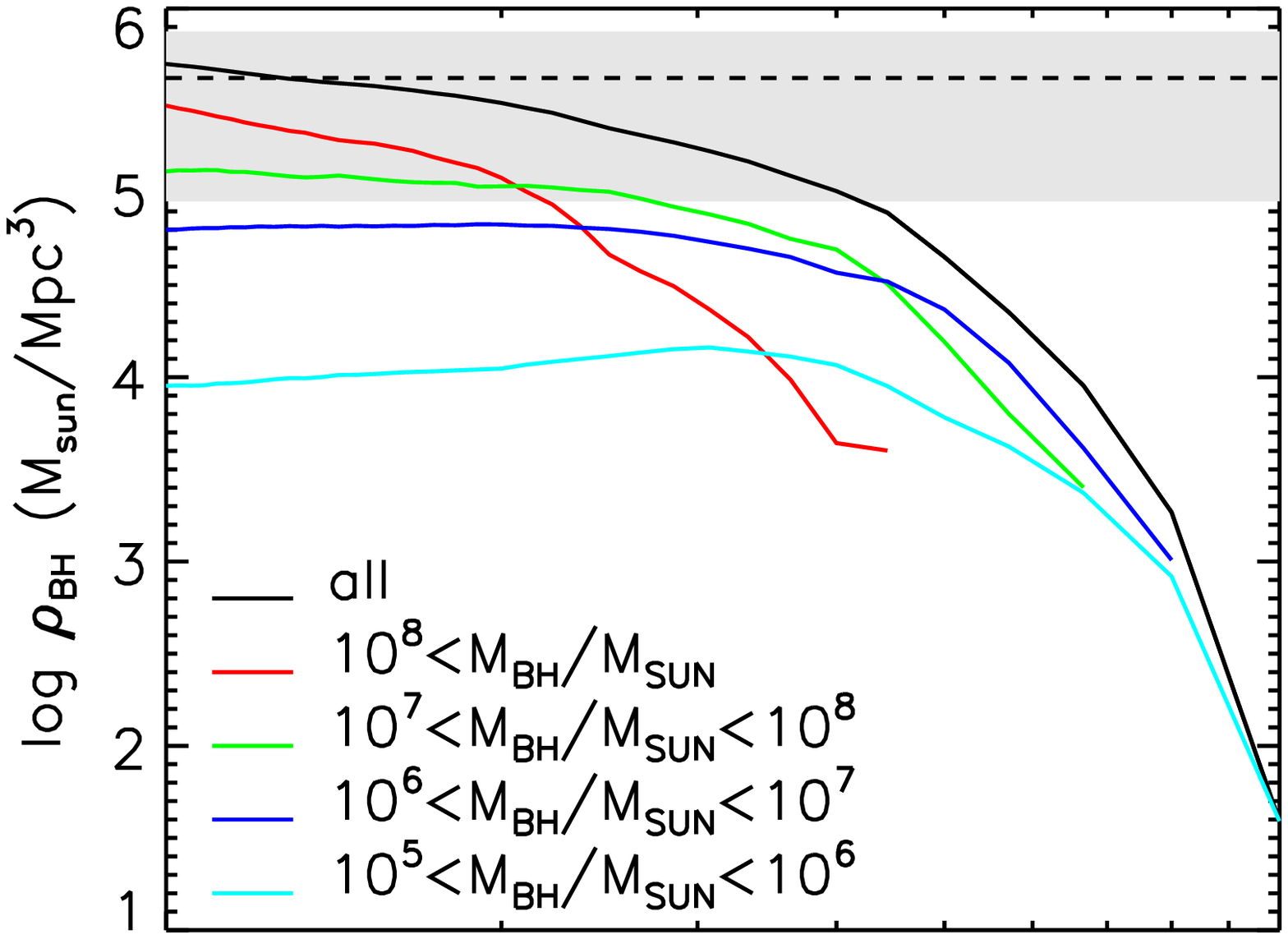}}\vspace{-1.615cm}}
  \centering{\resizebox*{!}{6.cm}{\includegraphics{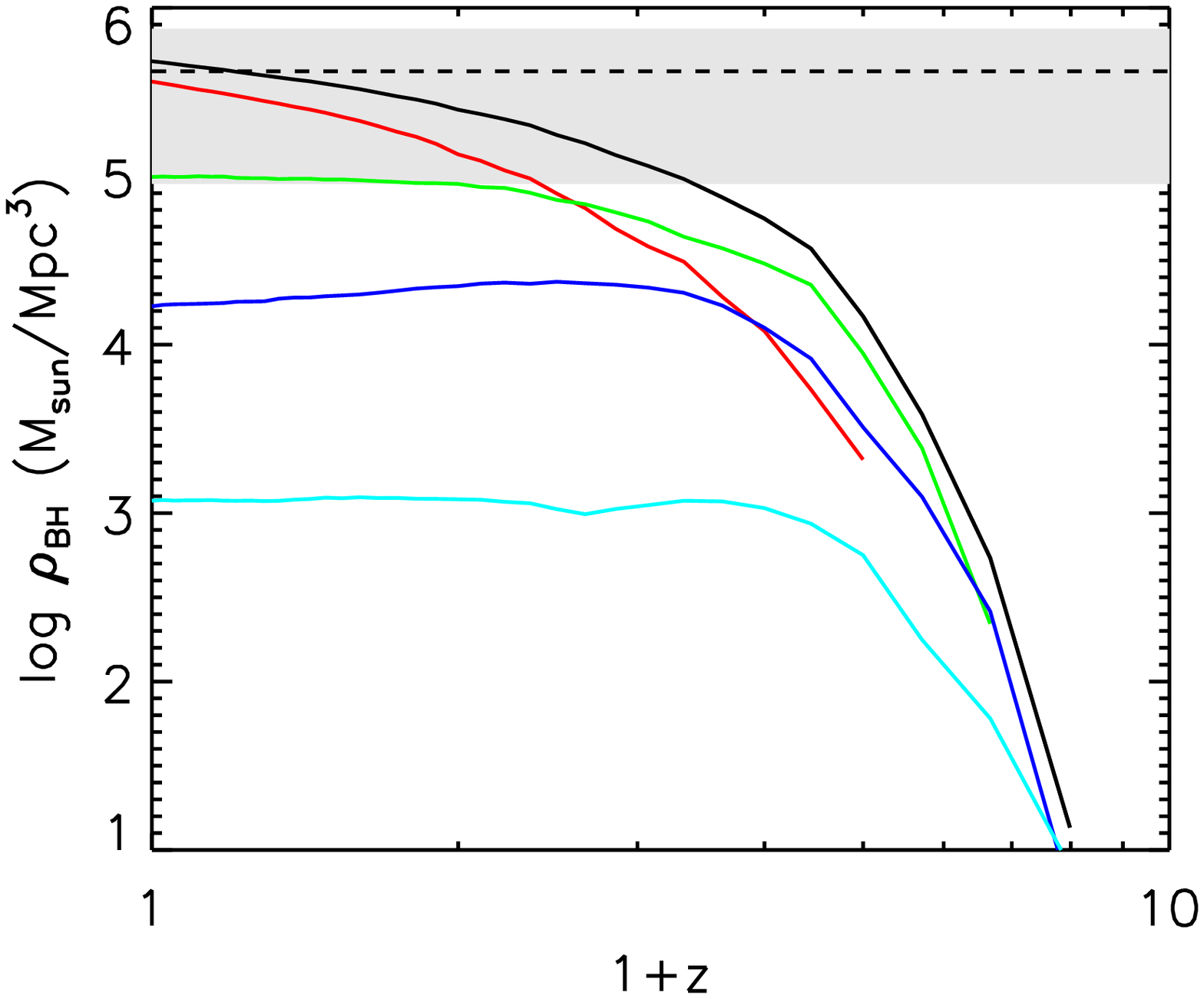}}}
  \caption{Comoving black hole mass density as a function of redshift with contributions from different BH mass ranges. Upper panel is for the 256L25JH simulation, and bottom panel is for the 256L50JH simulation. The grey shaded area is the black hole mass density in our local Universe with its 3$\sigma$ uncertainty from \citet{shankaretal04}.}
    \label{bhdensity_256L25-50}
\end{figure}

In order to test the convergence of our models, we vary the resolution, by changing the  DM mass, cell size, and box size.
We run five
 different simulations with our fiducial model for AGN feedback (256L12JH, 256L25JH, 128L25JH, 64L25JH, 256L50JH, 128L50JH), with three different box sizes $L_{\rm box}=12.5 \, \rm h^{-1}.Mpc$, $L_{\rm box}=25 \, \rm h^{-1}.Mpc$ and $L_{\rm box}=50 \, \rm h^{-1}.Mpc$, and four different resolutions \{$M_{\rm DM}=3.5\, \rm 10^9\, \rm M_{\odot}$, $\Delta x=3.04\, \rm h^{-1}.kpc$\} (64L25JH, 128L50JH), \{$M_{\rm DM}=4.4\, \rm 10^8\, \rm M_{\odot}$, $\Delta x=1.52\, \rm h^{-1}.kpc$\} (128L25JH, 256L50JH), \{$M_{\rm DM}=5.5\, \rm 10^7\, \rm M_{\odot}$, $\Delta x=0.76\, \rm h^{-1}.kpc$\} (256L25JH), and \{$M_{\rm DM}=6.9\, \rm 10^6\, \rm M_{\odot}$, $\Delta x=0.38\, \rm h^{-1}.kpc$\} (256L12JH) (see table~\ref{tabnames} for details).

The BH densities shown in fig.~\ref{bhdensity_resolution} slowly converge to the same value at $z=0$ when the resolution is increased.
Even though low resolution simulations (64L25JH and 128L50JH) are within the observational $3\, \sigma$ error bars, they tend to underestimate the BH density at all redshifts compared to more resolved simulations.
Intermediate resolution simulations with \{$M_{\rm DM}=4.4\, \rm 10^8\, \rm M_{\odot}$, $\Delta x=1.52\, \rm h^{-1}.kpc$\}, which correspond to runs 128L25JH and 256L50JH, have already converged at $z=0$ and differ  only by $\sim$ 10\% at z=0 from the simulation (256L25JH) with one additional refinement level.  
However, at high redshift, the difference is larger because galaxies in the field with intermediate BH masses contribute more to the total BH density (fig.~\ref{bhdensity_256L25-50}), and some of these galaxies are not resolved in simulations 128L25JH and 256L50JH with lower resolution.

\begin{figure}
  \centering{\resizebox*{!}{4.8cm}{\includegraphics{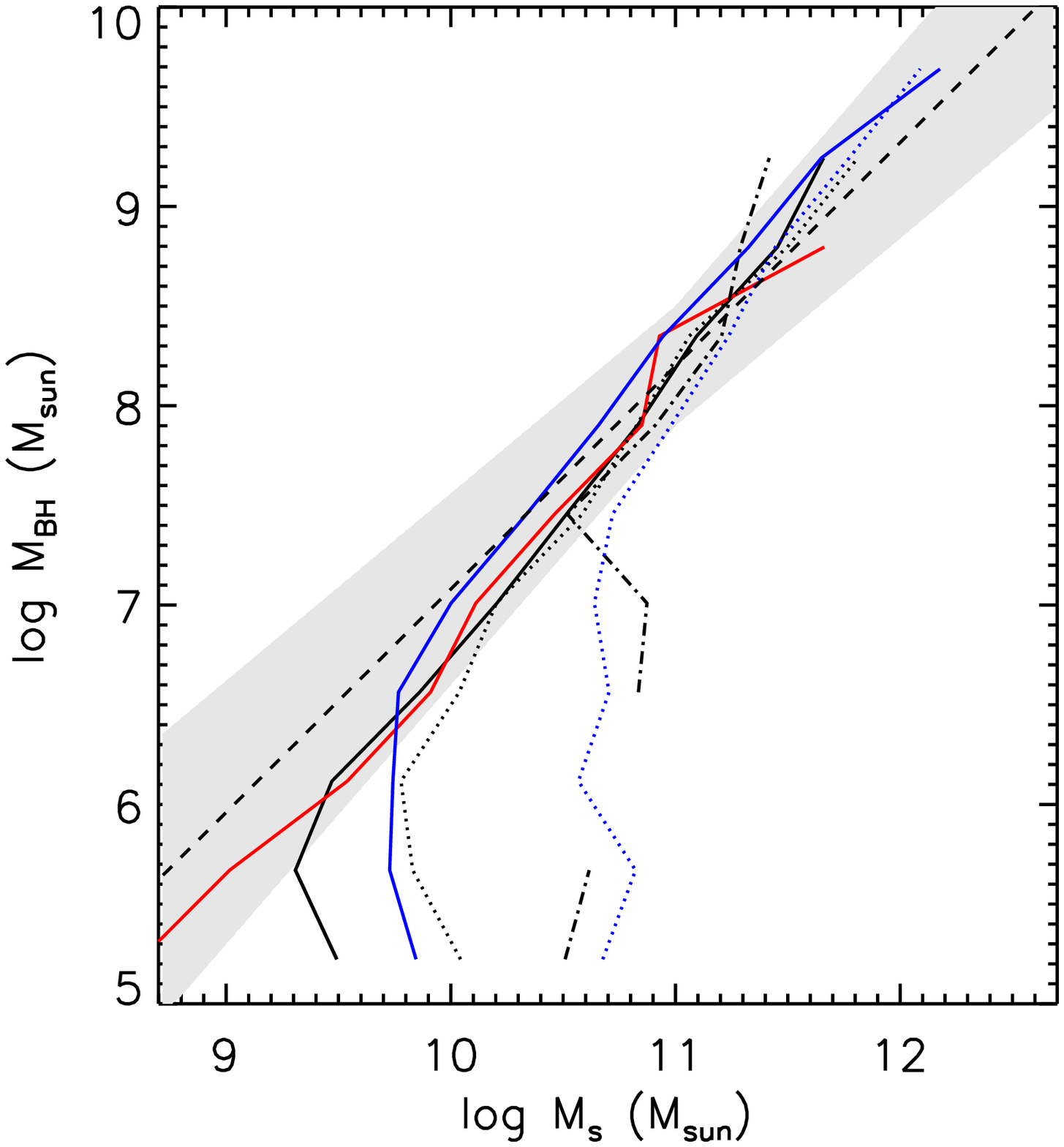}}\hspace{-1.48cm}}
  \centering{\resizebox*{!}{4.8cm}{\includegraphics{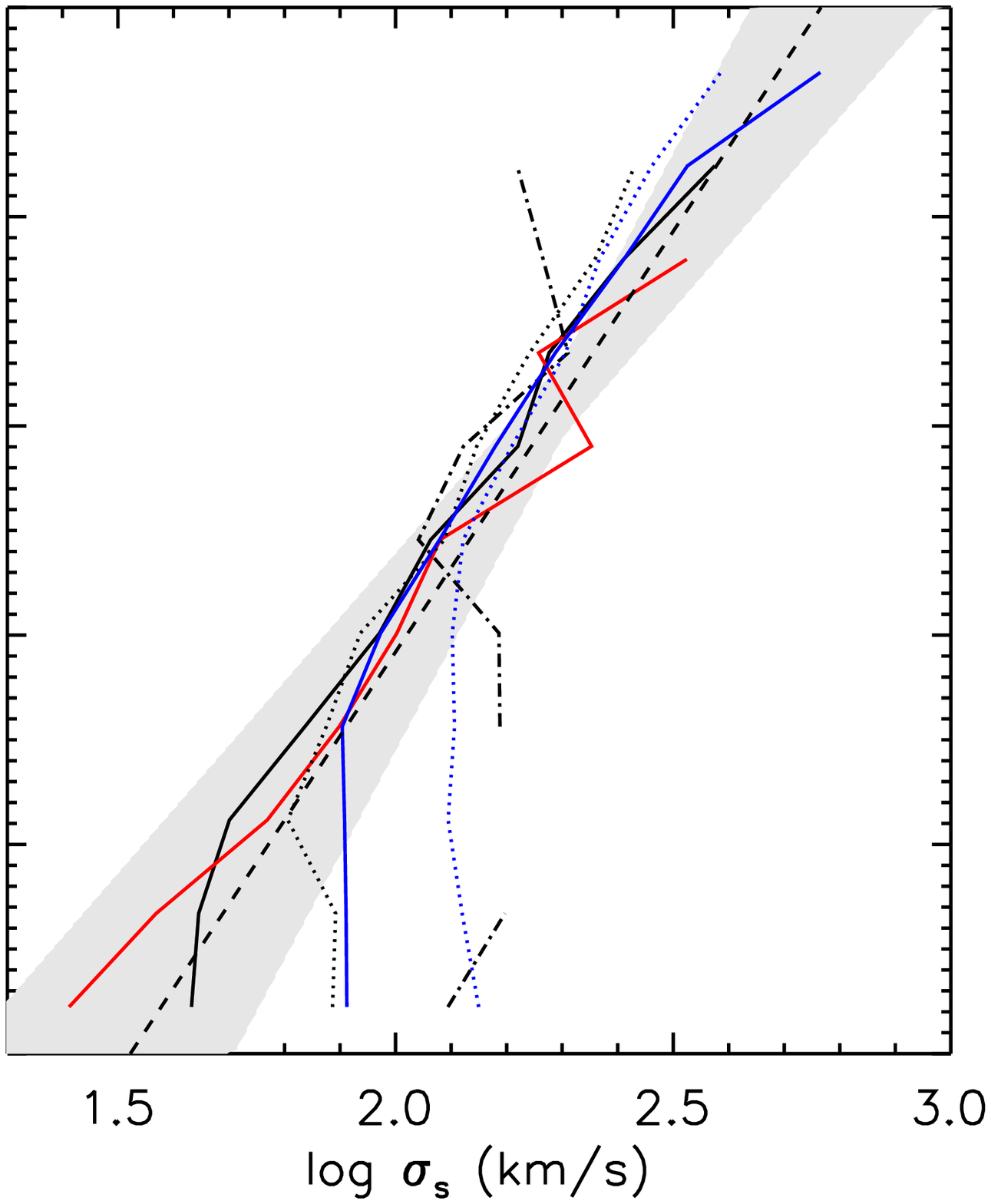}}}
  \caption{Black hole mass as a function of the stellar mass (left panel), or as a function of stellar velocity dispersion (right panel) at $z=0$ for different box sizes and resolutions. The color code is the same as for fig.~\ref{bhdensity_resolution}. We overplotted the observational laws as dashed lines from \citet{haring&rix04} (left) and \citet{tremaineetal02} (right) with their 3$\sigma$ uncertainties. }
    \label{magorrian_resolution}
\end{figure}

The same convergence can be seen in the BH mass versus stellar mass relationships (fig.~\ref{magorrian_resolution}, left panel).
When the resolution is increased, these relationships converge to the same value close to the observations from~\cite{haring&rix04}.
There is a departure from the observational constraint for the least massive galaxies that reside in halos that are barely resolved.
For PM codes, DM halos with more than $\sim 1000$ DM particles are followed with sufficient force resolution \citep{osheaetal05, heitmannetal08}.
Thus, the break observed at the low galaxy mass end corresponds to this low limit.

For example, the 256L25JH simulation has a DM mass resolution $M_{\rm DM}=5.5 \, 10^7 \, \rm h^{-1}.M_{\odot}$, which gives a minimum DM halo mass $M_{\rm h, min}\sim 8 \, 10^{10} \, \rm M_{\odot}$ that corresponds to a total gas content of $f_{b} M_{\rm h, min}\sim 10^{10} \, \rm M_{\odot}$.
A non-negligible fraction of the baryon content is locked into stars.  Assuming 25 \% for this fraction, we obtain $M_{\rm s, min}\sim 2.5 \, 10^9 \, \rm M_{\odot}$, i.e. the value of the bulge stellar mass where the break appears in the $M_{\rm BH}$-$M_{\rm s}$ relationship.
Indeed, the same behavior is seen when relating the BH mass to the stellar velocity dispersion of their host galaxy: low-mass BHs hosted by galaxies with low stellar velocity dispersion show a significant deviation from the constraints of~\cite{tremaineetal02} (fig.~\ref{magorrian_resolution}, right panel).

\begin{figure}
  \centering{\resizebox*{!}{6cm}{\includegraphics{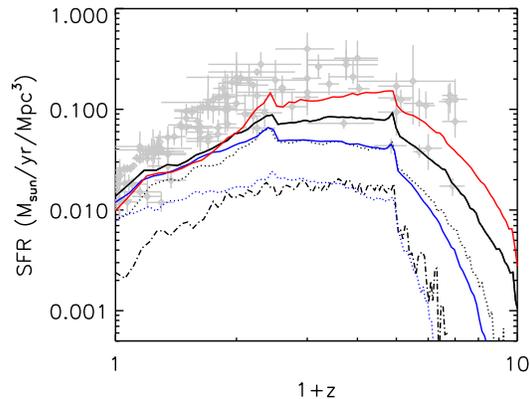}}}
  \caption{Comoving SFR as a function of redshift for different box sizes and resolutions. The color code is the same as for fig.~\ref{bhdensity_resolution}. The light grey circles with error bars correspond to observational points from~\citet{hopkins&beacom06}. }
    \label{sfr_resolution}
\end{figure}

Fig.~\ref{sfr_resolution} shows the cosmic SFR for simulations with our fiducial model for AGN feedback and for different resolutions and box sizes.
It appears that this quantity is strongly dependant on DM mass resolution at high redshift because of two effects.
When increasing the resolution, smaller halos which contribute a lot to the total star formation rate at high redshift are resolved.
Furthermore, structures that were already resolved at lower resolution collapse earlier when resolution is increased, thus, forming stars earlier.
This effect is well constrained and, if one has sufficient resolution, the numerical solution converges to its analytical prediction \citep{rasera&teyssier06}.

We point out that Tree codes or Tree-PM codes, such as the {\sc gadget} code \citep{springel05}, have a better force resolution in the early Universe than PM codes. The former therefore can follow the formation of smaller halos (down to 10 DM particles) with equivalent initial conditions.
This is why, in general, the convergence in the SFR with such codes is more rapidly obtained, even though 10 particles per halo are not sufficient to properly treat the gas dynamics.

Even though the convergence for a given box size is not reached, the SFR for the box size $25 \, \rm  h^{-1}.Mpc$ slowly converges at $z=0$. It increases by a factor 4 from 64L25JH to 128L25JH, and by a factor 2 from 128L25JH to 256L25JH.
Changing box size at constant resolution has also some non-negligible effect on the SFR at low redshift.
The reason is that at low redshift the SFR is essentially dominated by massive galaxies.
Thus, with larger box sizes, very massive clusters of galaxies are more likely to be present \citep{davisetal11}.
Finally, we conclude that getting convergence for the SFR is extremely difficult, because it requires both a large box size and a small DM mass resolution, which can only be achieved with tremendous computational power.
 
\begin{figure}
  \centering{\resizebox*{!}{6cm}{\includegraphics{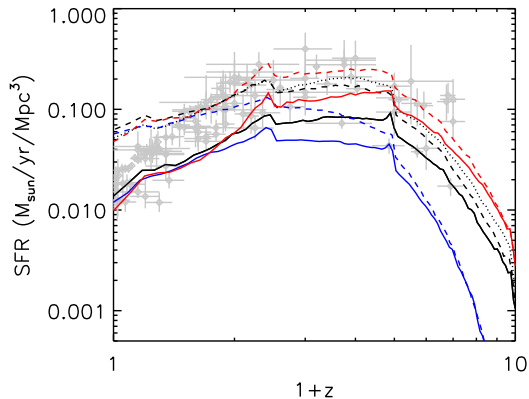}}}
  \caption{Comoving SFR as a function of redshift for different box sizes $L_{\rm box}=12.5 \, \rm h^{-1}.Mpc$ (red), $L_{\rm box}=25 \, \rm h^{-1}.Mpc$ (black), $L_{\rm box}=50 \, \rm h^{-1}.Mpc$ (blue) including SN and AGN feedback (solid), including SN feedback and no AGN feedback (dashed), and without SN or AGN feedback (dotted). The light grey circles with error bars corresponds to observational points from~\citet{hopkins&beacom06}. }
    \label{sfr_AGNvsnoAGN}
\end{figure}

It is interesting to note that the DM resolution effect on SFR is also present for the simulations without AGN feedback (fig.~\ref{sfr_AGNvsnoAGN}), which proves that this effect is uncorrelated to AGN feedback but only to DM mass resolution and box size.
AGN feedback is  most efficient at suppressing the SFR at low redshift when massive structures such as groups and clusters of galaxies are formed.
At high redshift the SFR is dominated by galaxies in the field which are not progenitors of groups or clusters of galaxies, thus, the effect of AGN feedback is less visible.

There is an interesting behavior of the simulation without AGN feedback and without SN feedback (256L25noSNAGN, dotted line in fig.~\ref{sfr_AGNvsnoAGN}) compared to the simulation without AGN feedback but with SN feedback (256L25noAGN, black dashed line in fig.~\ref{sfr_AGNvsnoAGN}).
At high redshift SN feedback reduces the SFR, because galaxies form large-scale galactic winds \citep{springel&hernquist03, dubois&teyssier08winds} that remove some baryons from them and prevent some gas from collapsing into them.
But as time goes by, structures become more massive and the ram-pressure confinement from the halo becomes higher, preventing galactic winds from escaping the discs. As a result they develop galactic fountains \citep{dubois&teyssier08winds}.
The SN feedback also enriches the gas with metals and enhances the gas cooling rates, developing stronger accretion flows and SFRs at late times, when metals are confined in halos~\citep{duboisetal11}.
Thus the feedback of SN has a negative impact on SFR at high redshift but a positive effect at low redshift due to metal enrichment.

\section{Redshift evolution of BH properties}
\label{BHevolution}

\begin{figure}
  \centering{\resizebox*{!}{4.8cm}{\includegraphics{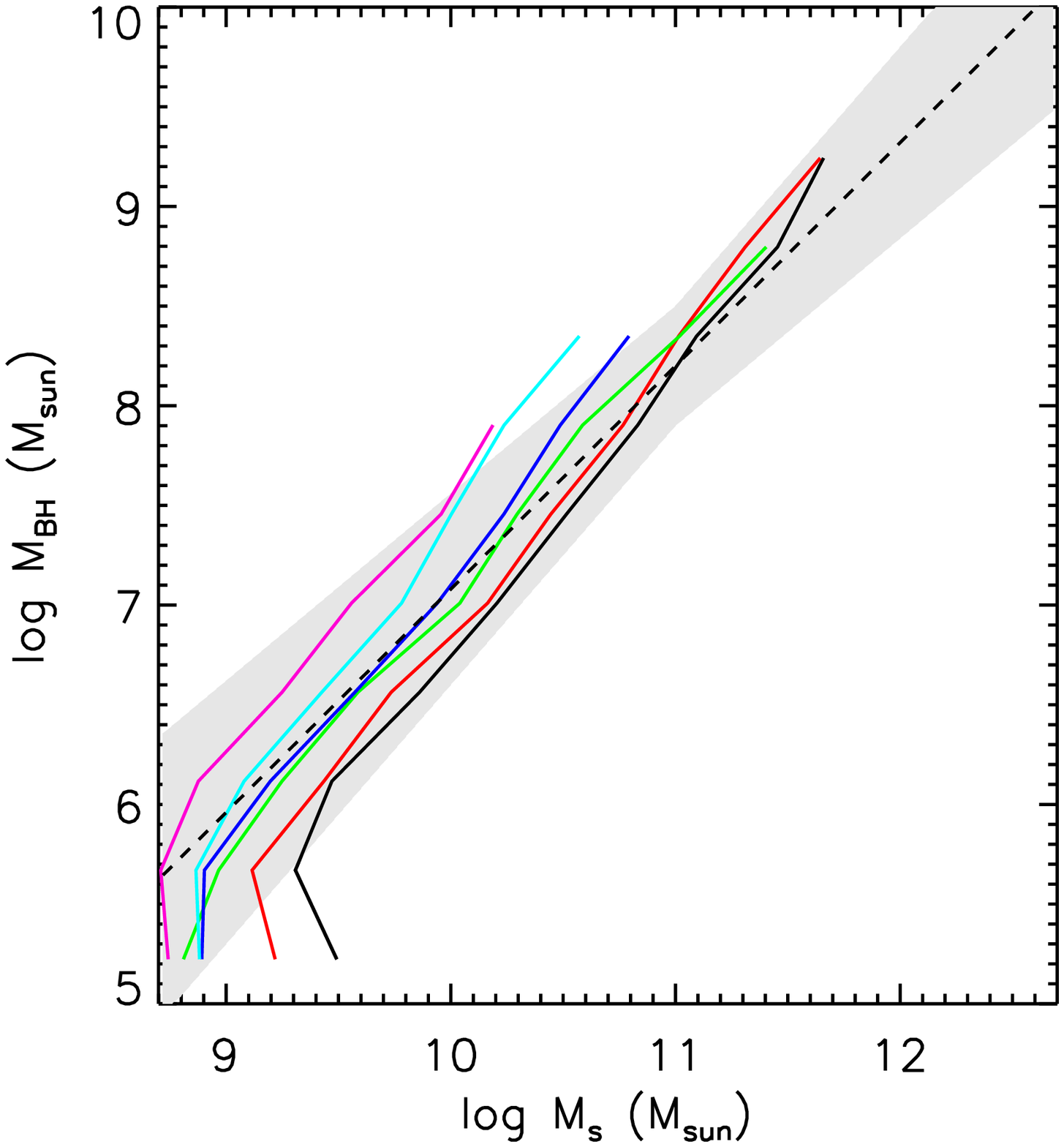}}\hspace{-1.48cm}}
  \centering{\resizebox*{!}{4.8cm}{\includegraphics{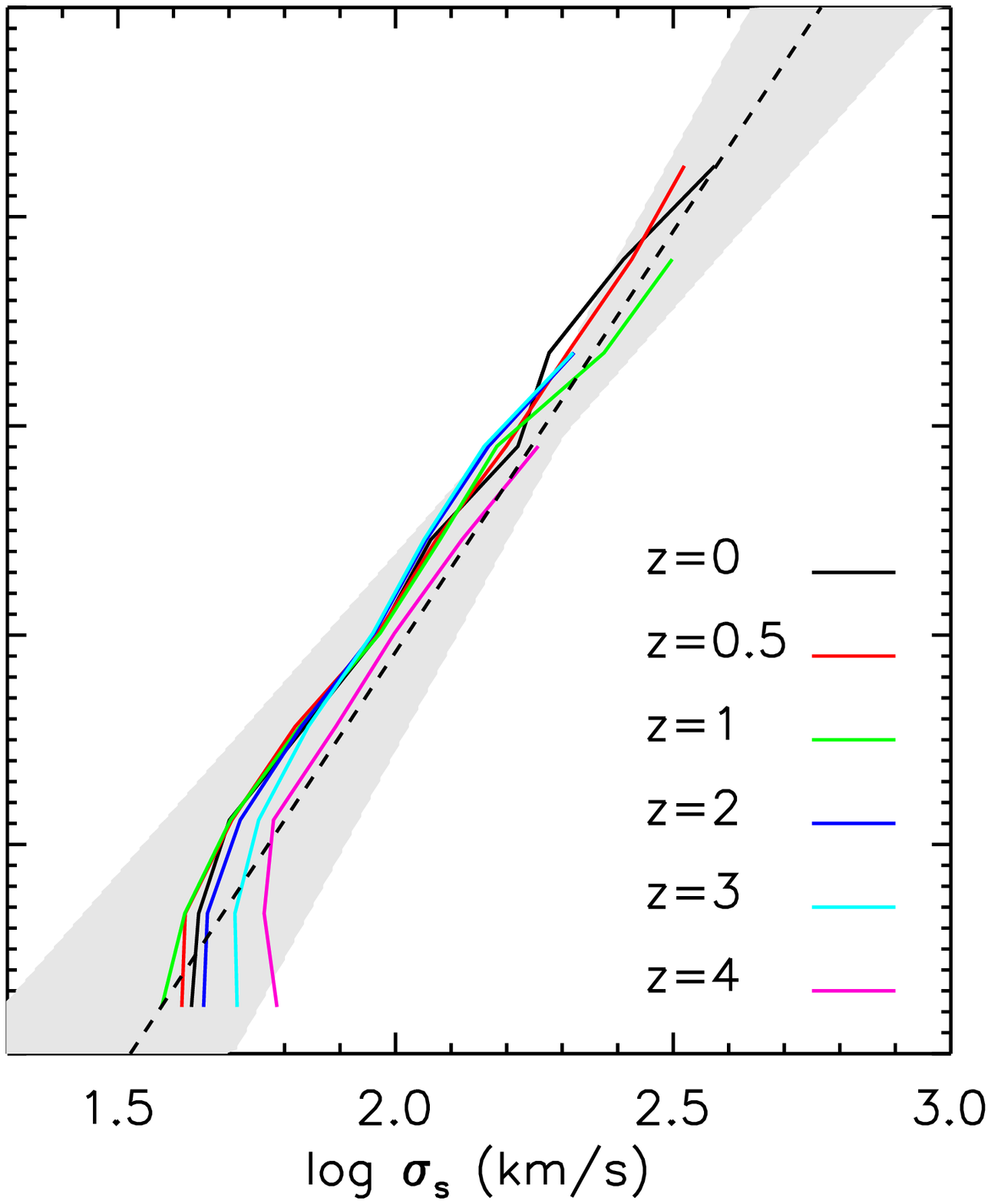}}}
  \centering{\resizebox*{!}{4.8cm}{\includegraphics{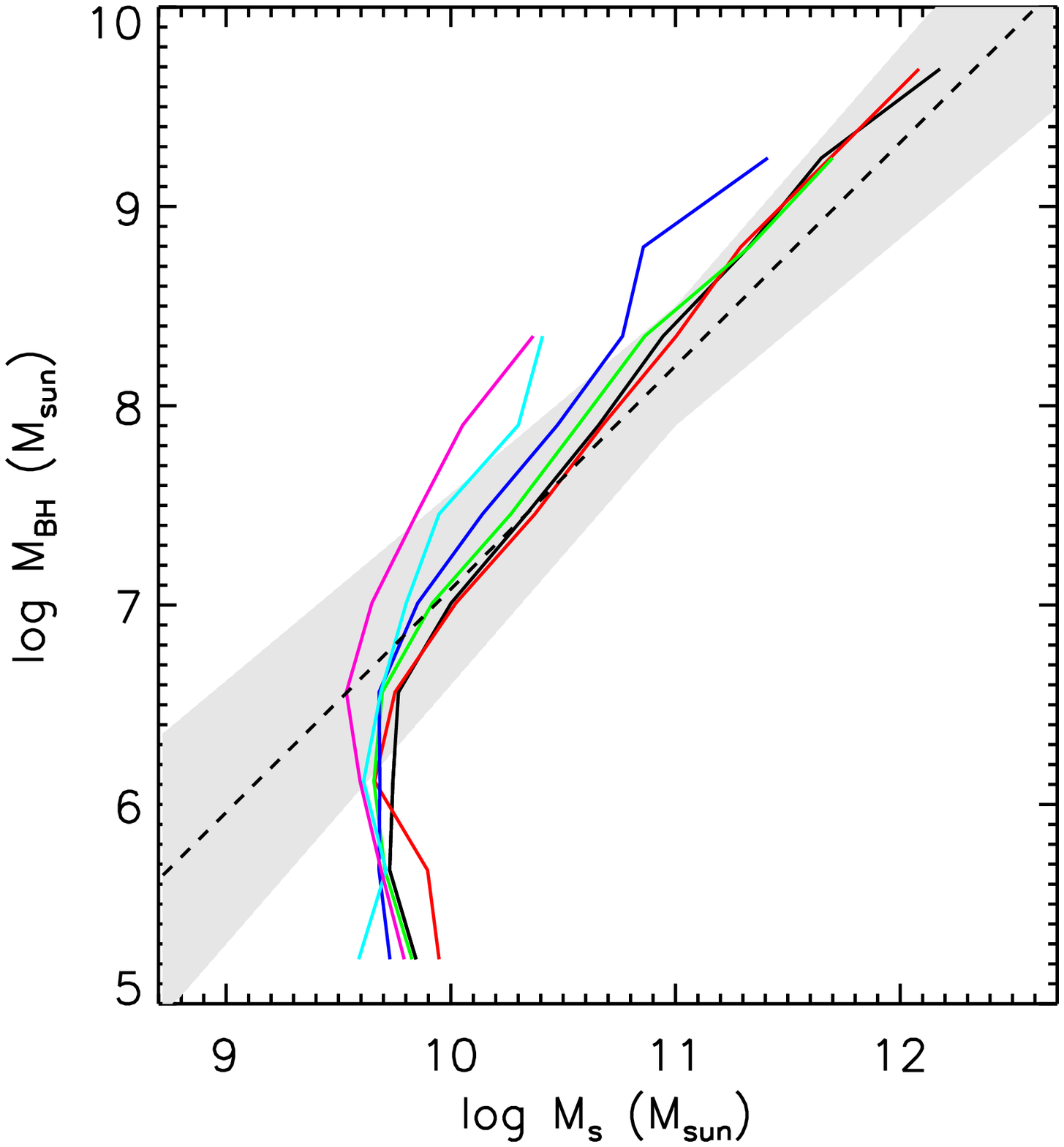}}\hspace{-1.48cm}}
  \centering{\resizebox*{!}{4.8cm}{\includegraphics{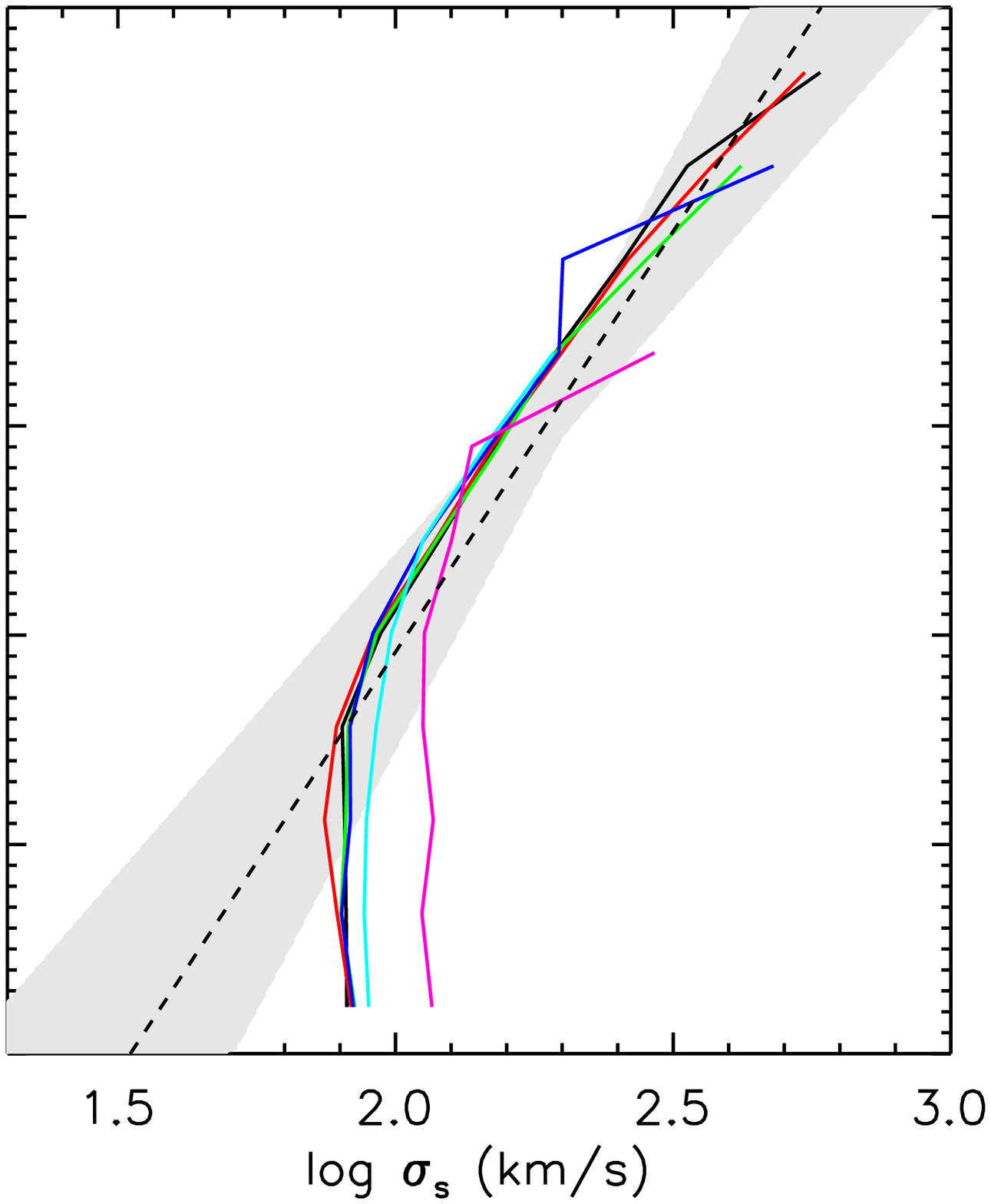}}}
  \caption{For each plot: black hole mass as a function of the stellar mass (left panel), or as a function of stellar velocity dispersion (right panel). Measurements are done at different redshifts as labelled in the upper right panel. Top plot corresponds to the 256L25JH simulation, and bottom plot to the 256L50JH simulation. We overplotted the observational laws as dashed lines from \citet{tremaineetal02} and \citet{haring&rix04} with their 3$\sigma$ uncertainties. }
    \label{magorrian_redshift}
\end{figure}

\begin{figure}
  \centering{\resizebox*{!}{6cm}{\includegraphics{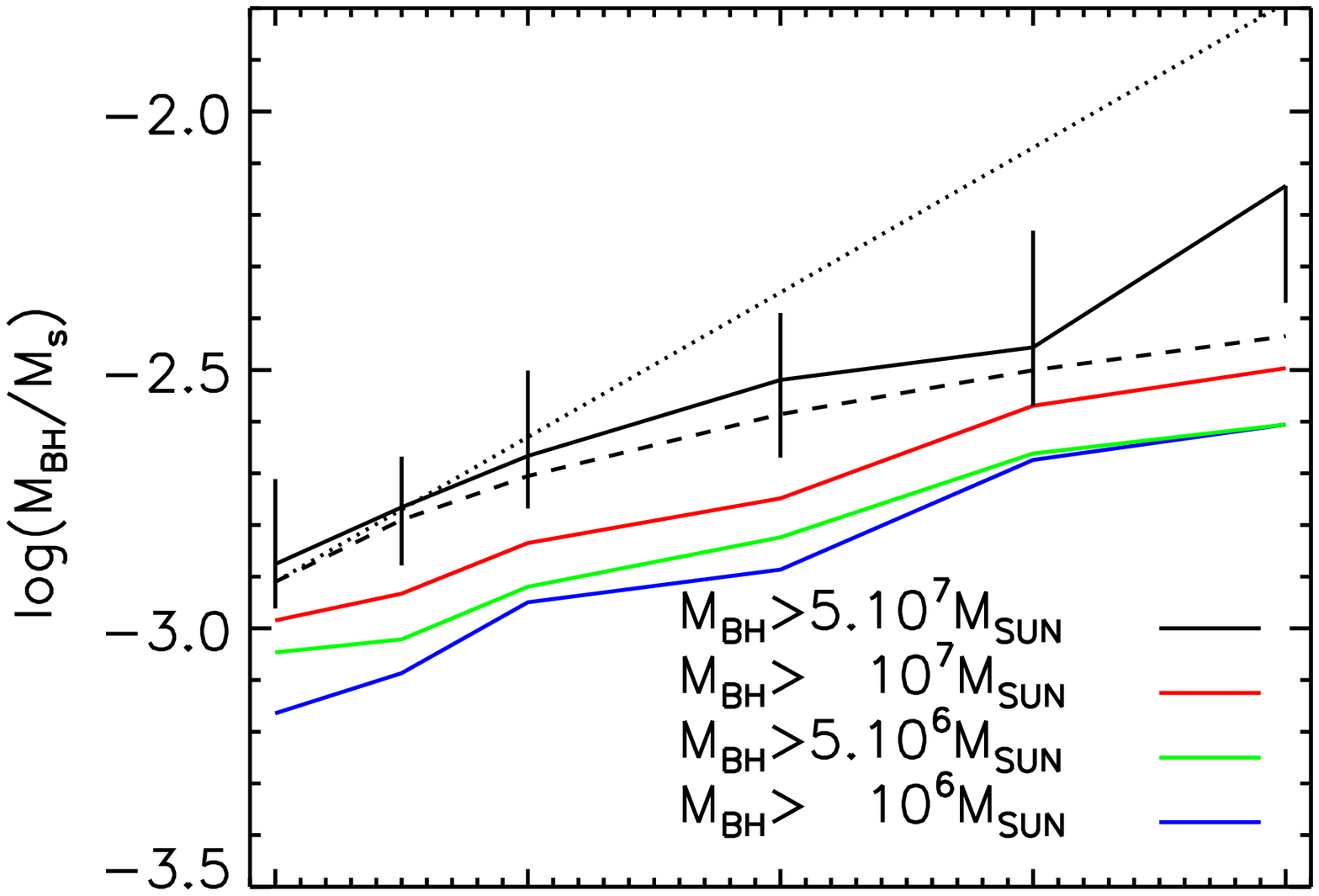}}\vspace{-1.615cm}}
  \centering{\resizebox*{!}{6cm}{\includegraphics{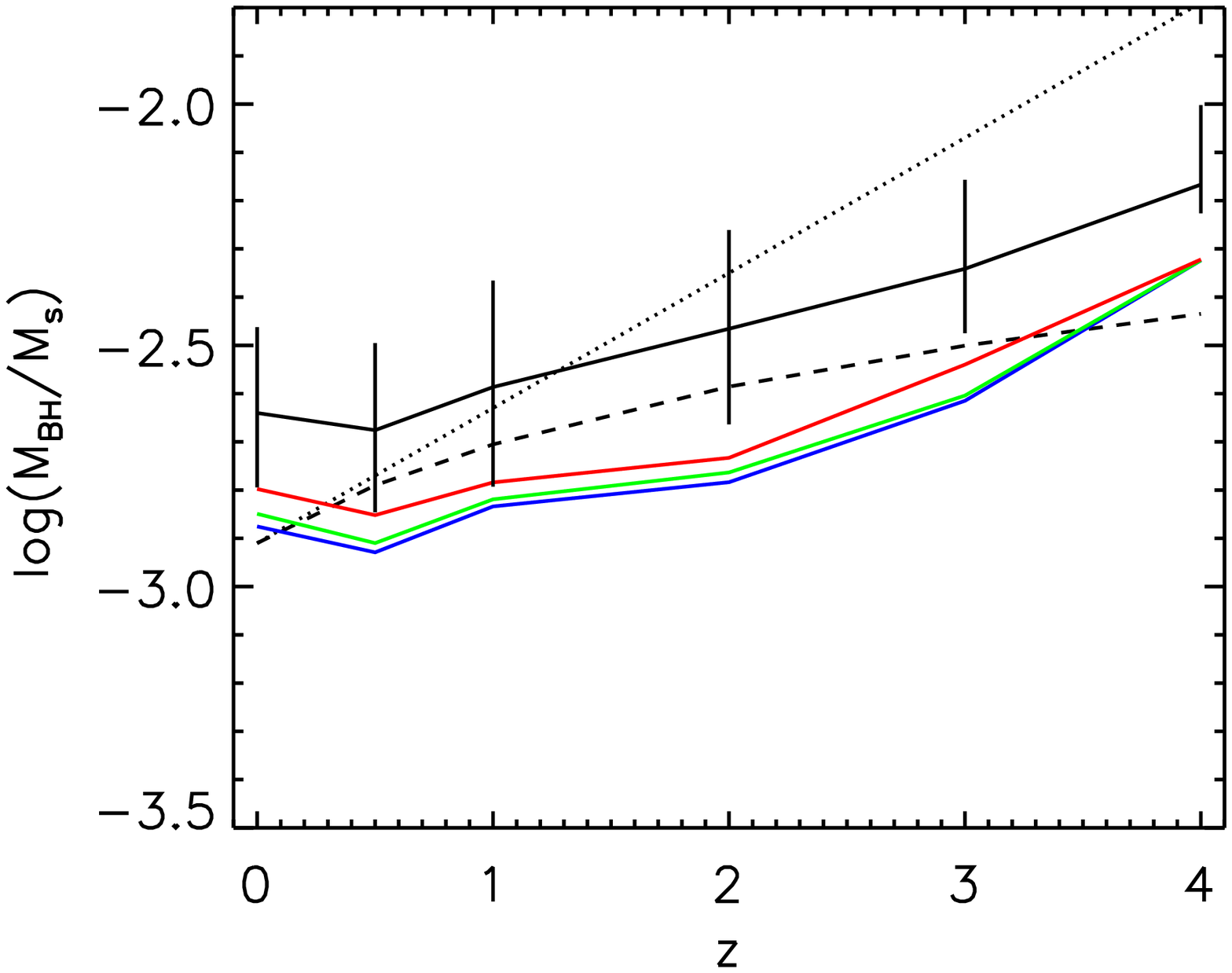}}}
  \caption{Median value of the BH mass ($M_{\rm BH}$) over its host stellar bulge mass ($M_{\rm s}$) as a function of redshift for the 256L25JH simulation (upper panel) and the 256L50JH simulation (bottom panel). Different colors correspond to different BH mass cut-offs used to get the median value of $M_{\rm BH}/M_{\rm s}$. The vertical error bars correspond to first and third quartile limits of the distribution of points. The dotted line is the trend from a fit to observations in~\citet{decarlietal10} and the dashed line is the trend from~\citet{merlonietal10}.  }
    \label{tauvsz}
\end{figure}

\begin{figure}
  \centering{\resizebox*{!}{6cm}{\includegraphics{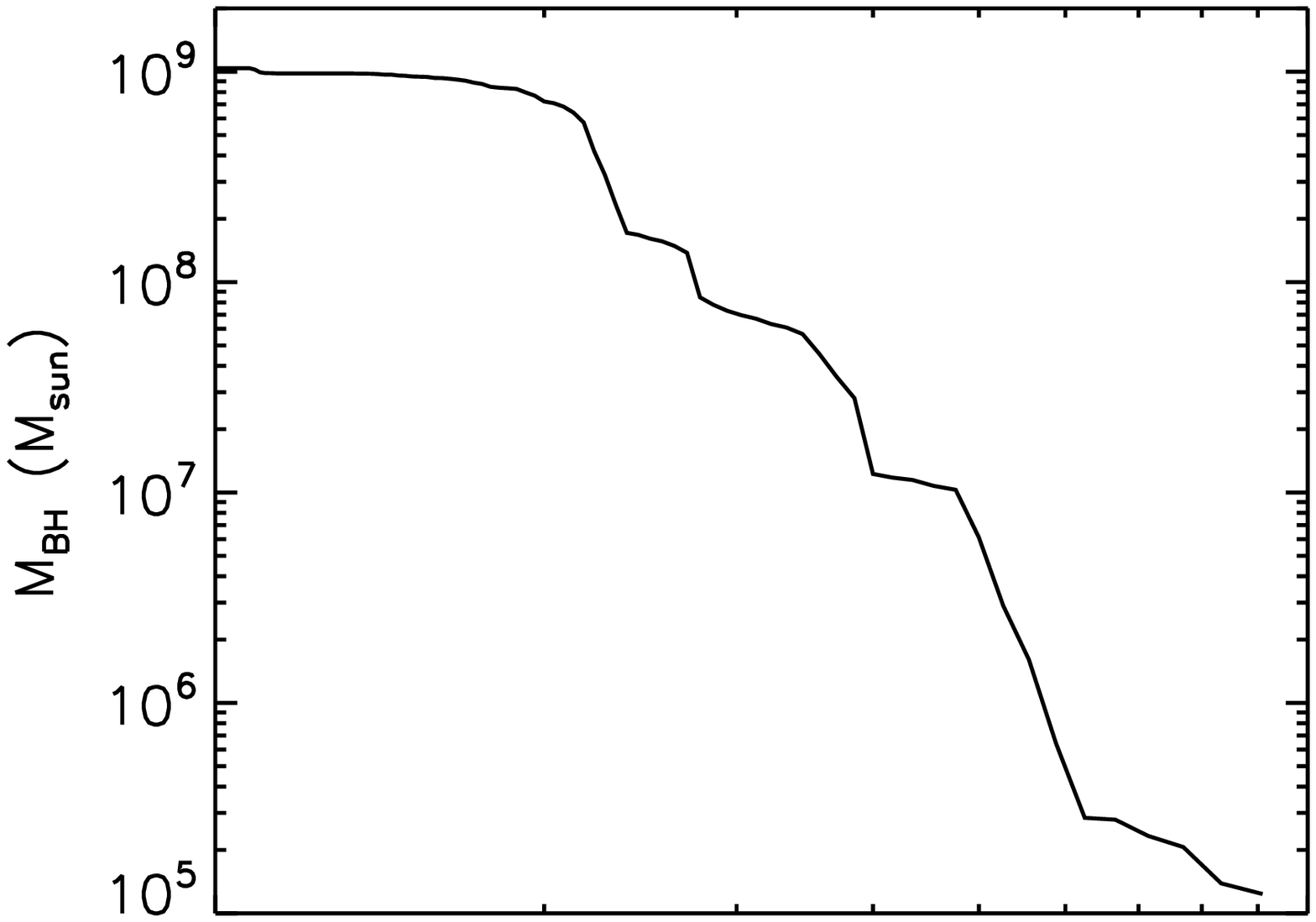}}\vspace{-1.615cm}}
  \centering{\resizebox*{!}{6cm}{\includegraphics{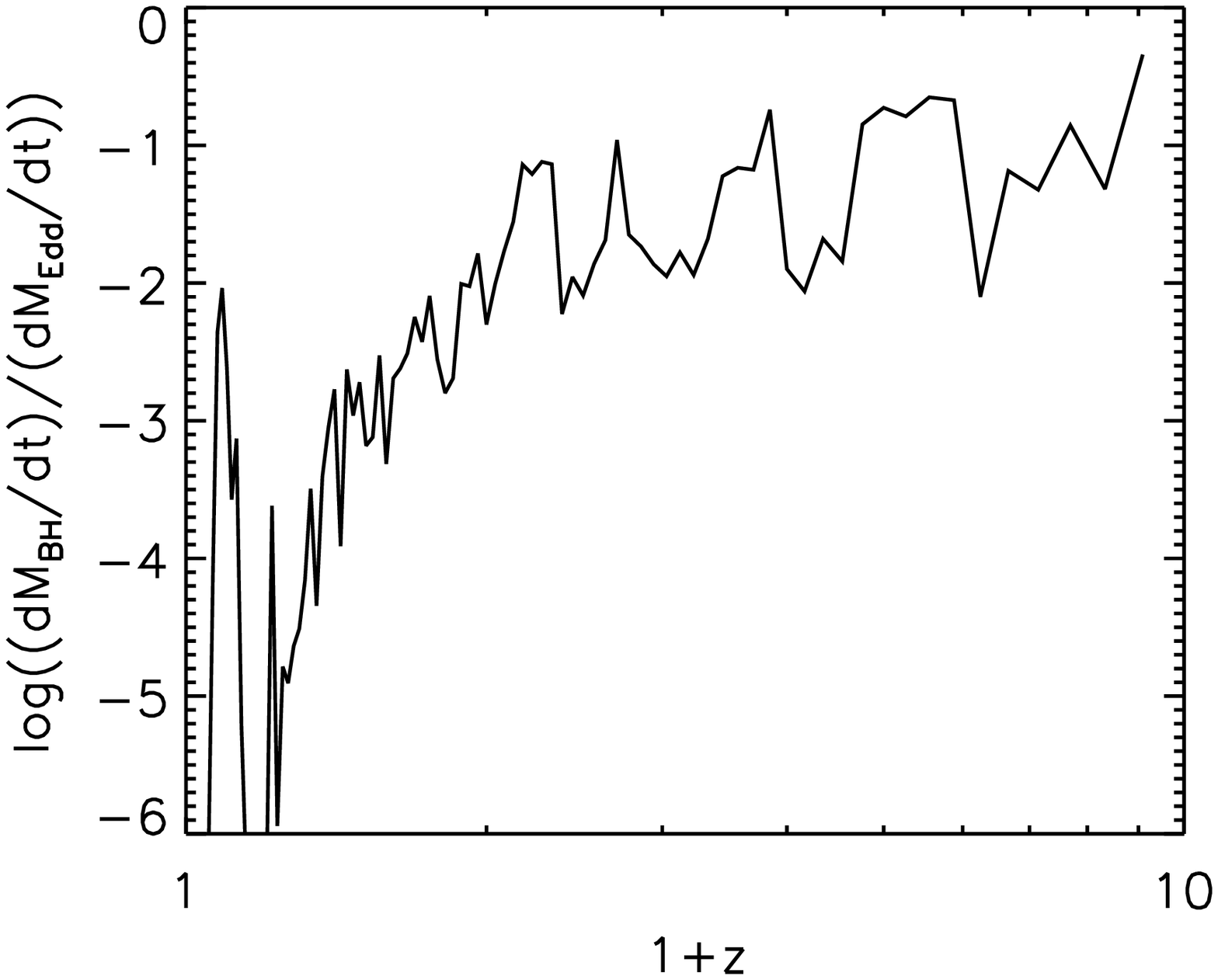}}}
  \caption{BH mass as a function of redshift for the most massive BH in the 256L12JH simulation (upper panel) and the logarithm of the ratio of its accretion rate to the Eddington accretion limit (bottom panel).}
    \label{mbhvstime}
\end{figure}

The way BHs acquire their mass and how they liberate energy to the gas is important for constraining their co-evolution with their host galaxy.
Observations of the relationships between BH masses and galaxy masses at high redshift are extremely difficult because the luminosity of host galaxies of massive BHs is dominated by the AGN component. 
Despite these difficulties, an increasing amount of data is suggesting that the $M_{\rm BH}/M_{\rm s}$ relationship shows some positive evolution with redshift \citep{mclureetal06, pengetal06, shieldsetal06, salvianderetal07, bennertetal10, decarlietal10, merlonietal10}.
Simulations can provide insight on the time evolution of these relationships.

Fig.~\ref{magorrian_redshift} shows the BH mass versus bulge stellar mass at different redshifts for the 25 $\rm h^{-1}.Mpc$ (256L25JH) and the 50 $\rm h^{-1}.Mpc$ (256L50JH) simulations.
At high redshift, the BHs are more massive than the $z=0$ Magorrian relation would predict given the stellar mass of their host galaxy. This is supported by observations (see \citealp{merlonietal10}).
We evaluate this deviation from the z = 0 relationships by measuring the median of the distribution of the $M_{\rm BH}/M_{\rm s}$ measurements for the 256L25JH and the 256L50JH simulations in fig.~\ref{tauvsz}.
We explore the effect of removing from the sample BHs with masses smaller than a mass threshold to see if there is a BH mass for which the deviation is most pronounced. Like the observations~\citep{decarlietal10, merlonietal10} for which BH masses are larger than a few $10^7\, \rm M_{\odot}$, we observe a positive trend with redshift of the $M_{\rm BH}/M_{\rm s}$ ratio.
In our simulations, the ratio is larger for more massive BHs, but the trend is independent of the BH mass threshold. Quantifying the trend, 
we find $M_{\rm BH}/M_{\rm s}\propto (1+z)^{\alpha_s}$ with $\alpha_s=0.42 \pm 0.09$ for the 256L25JH simulation and $\alpha_s=0.42 \pm 0.06$ for the 256L50JH simulation when fitting our simulation data on BHs with masses larger than $>5\, 10^7\, \rm M_{\odot}$ between redshifts $z=0$ to $z=3$ .
This is in relatively good agreement with the value $\alpha_s=0.68 \pm 0.12$ measured in the observational data by~\cite{merlonietal10}, and with numerical simulations from \cite{dimatteoetal08} ($\alpha_s=0.5$) and \cite{booth&schaye11} ($\alpha_s=0.52 \pm 0.05$).

The increase in $M_{\rm BH}/M_{\rm s}$ reflects the different accretion modes onto the BH and the gas content and properties at different redshifts.
Fig.~\ref{mbhvstime} illustrates how massive BHs grow through time, with very fast accretion of gas at high redshift due to the presence of a cold and dense ISM.
The accretion proceeds by bursts accompanied by large releases of AGN energy that temporarily delay the accretion onto the BH.
At high redshift, two kinds of accretion occur: the accretion of a diffuse component that can eventually shock and virialize the gas in the halo, as well as accretion of dense filaments of gas~\citep{keresetal05, ocvirketal08, brooksetal09, dekeletal09}.
The gaseous filaments feed galaxies so that their BHs can grow to larger masses,  pre-heat their proto-cluster environment  and remove gas thereby halting the adiabatic contraction in the proto-cluster cores.

\begin{figure}
  \centering{\resizebox*{!}{6cm}{\includegraphics{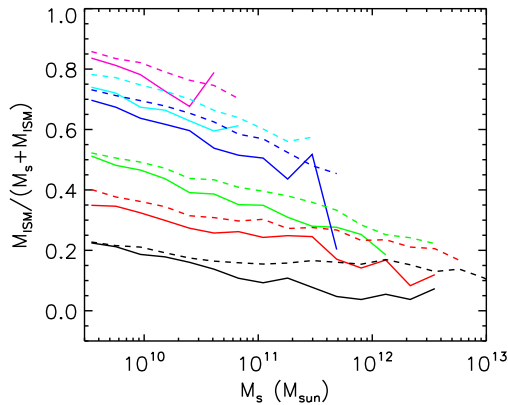}}}
  \caption{Average gas mass in the disc ($M_{\rm ISM}$) over the total mass of gas plus stars ($M_{\rm ISM}+M_{\rm s}$) as a function of the total stellar mass ($M_{\rm s}$) for different redshifts $z=4$, $z=3$, $z=2$, $z=1$, $z=0.5$, and $z=0$ from top to bottom for the 256L50JH simulation (solid lines)  and for the 256L50noAGN simulation (dashed lines). }
    \label{fgasgal}
\end{figure}

For simulations 256L50JH and 256L50noAGN, fig.~\ref{fgasgal} shows the fraction of baryons in galaxies in the form of a cold gas component with gas density larger than $>0.1\, \rm H.cm^{-3}$ for different stellar masses at different redshifts.
It appears that more massive galaxies have lower gas fractions that  decline with time. This can be explained by two effects.
Firstly, galaxies efficiently consume their gas to form stars without replenishing their cold gas content quickly enough to maintain a constant specific star formation rate.
Secondly, AGN feedback reduces the amount of cold gas available in galaxies by ejecting dense material into the CGM. 
Hence AGN feedback coupled to a vigorous consumption of gas via star formation reduces the gas content in galaxies, resulting in a much lower accretion rate at low-redshift, where BHs enter a low-Eddington accretion regime (fig.~\ref{mbhvstime}).

\begin{figure*}
  \centering{\resizebox*{!}{7cm}{\includegraphics{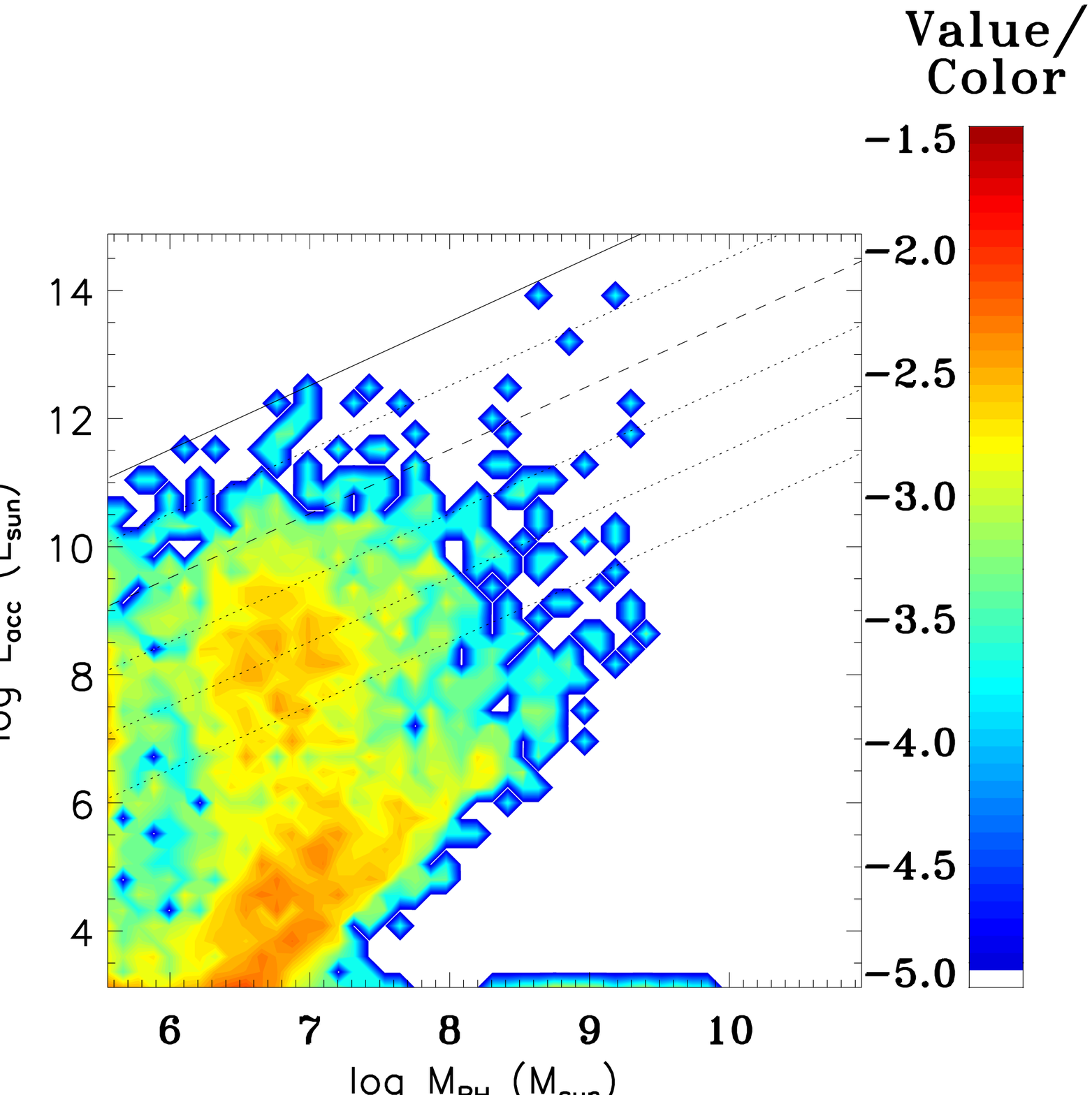}}}
  \centering{\resizebox*{!}{7cm}{\includegraphics{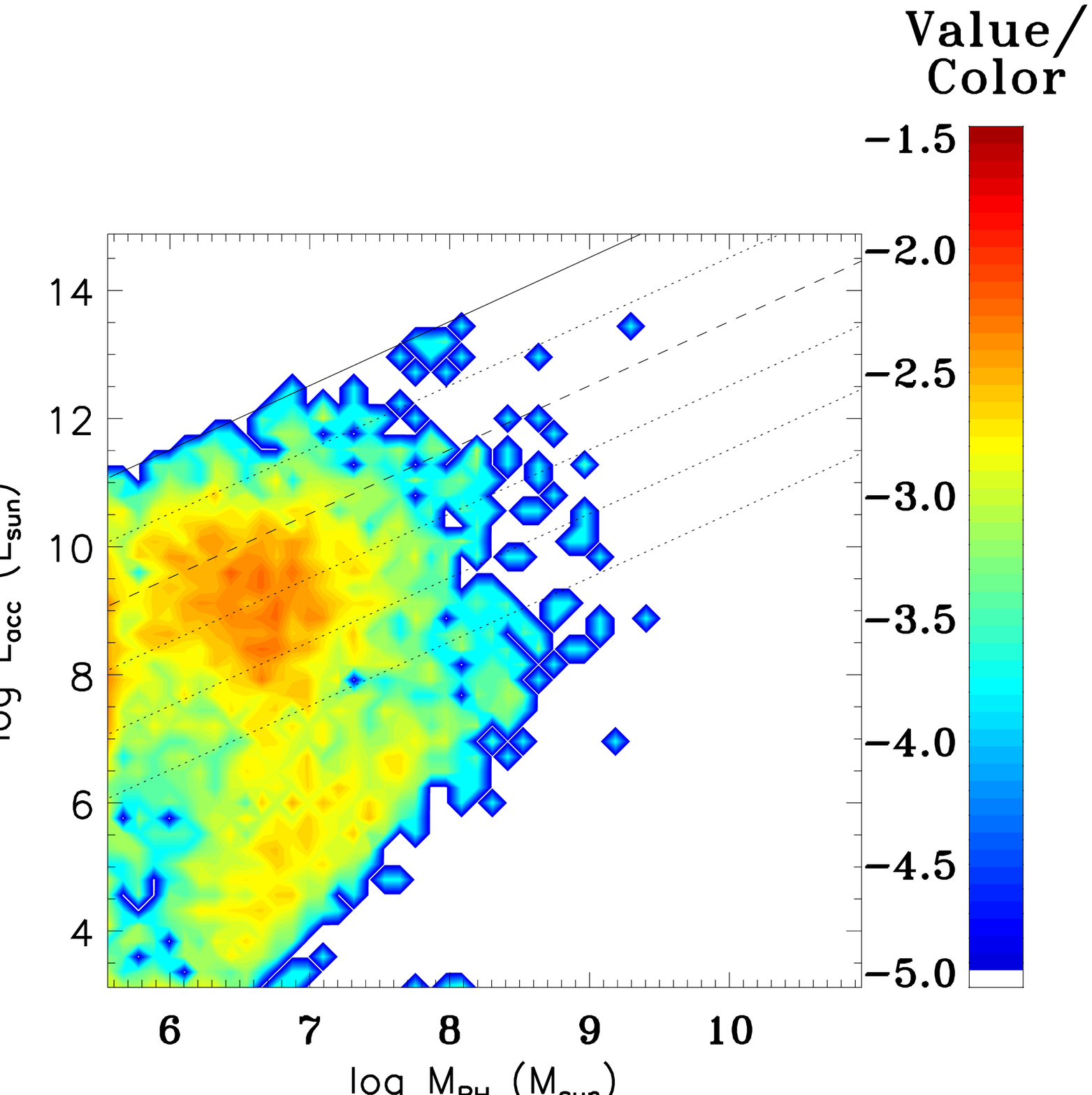}}}
  \centering{\resizebox*{!}{7cm}{\includegraphics{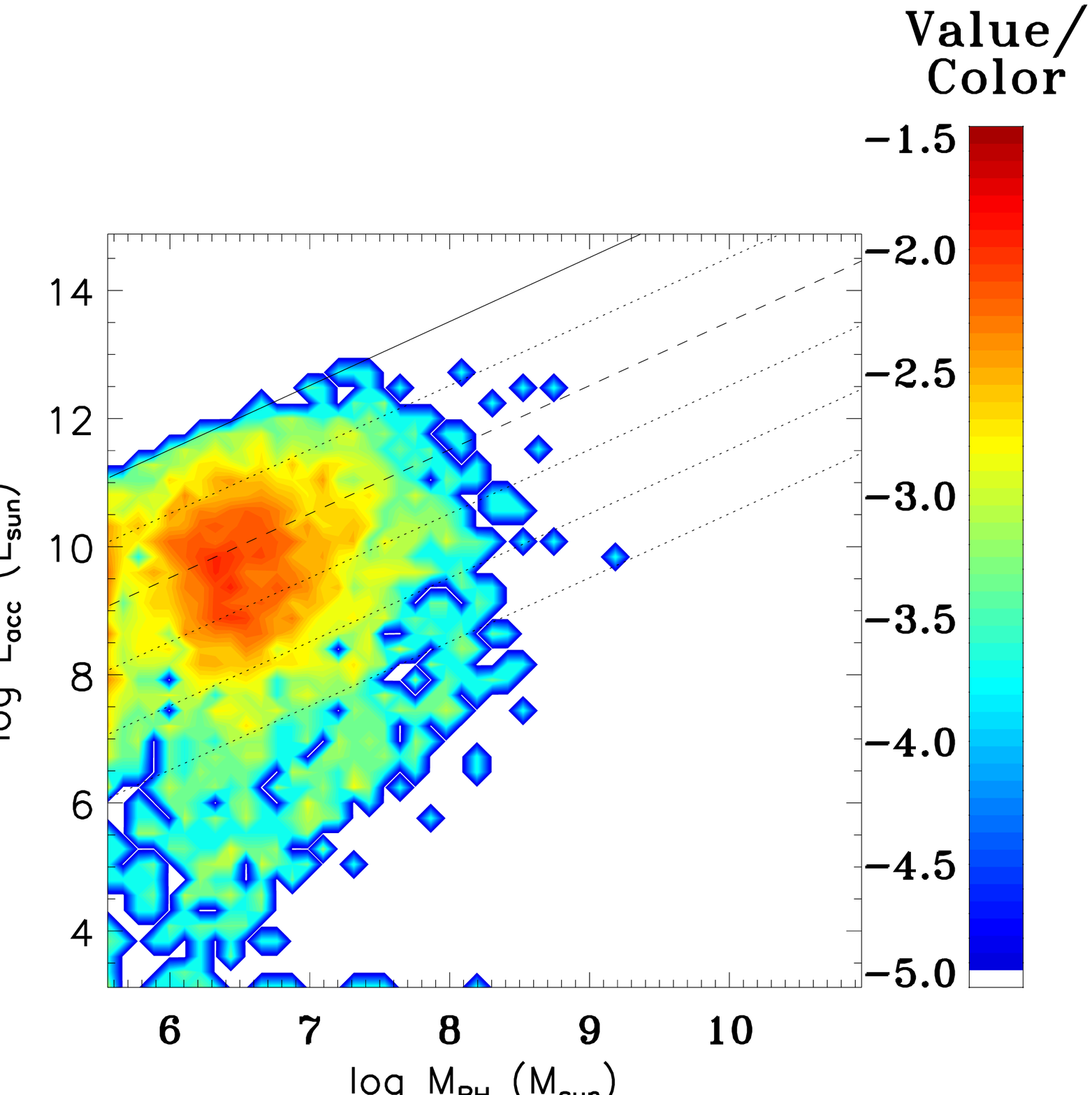}}}
  \centering{\resizebox*{!}{7cm}{\includegraphics{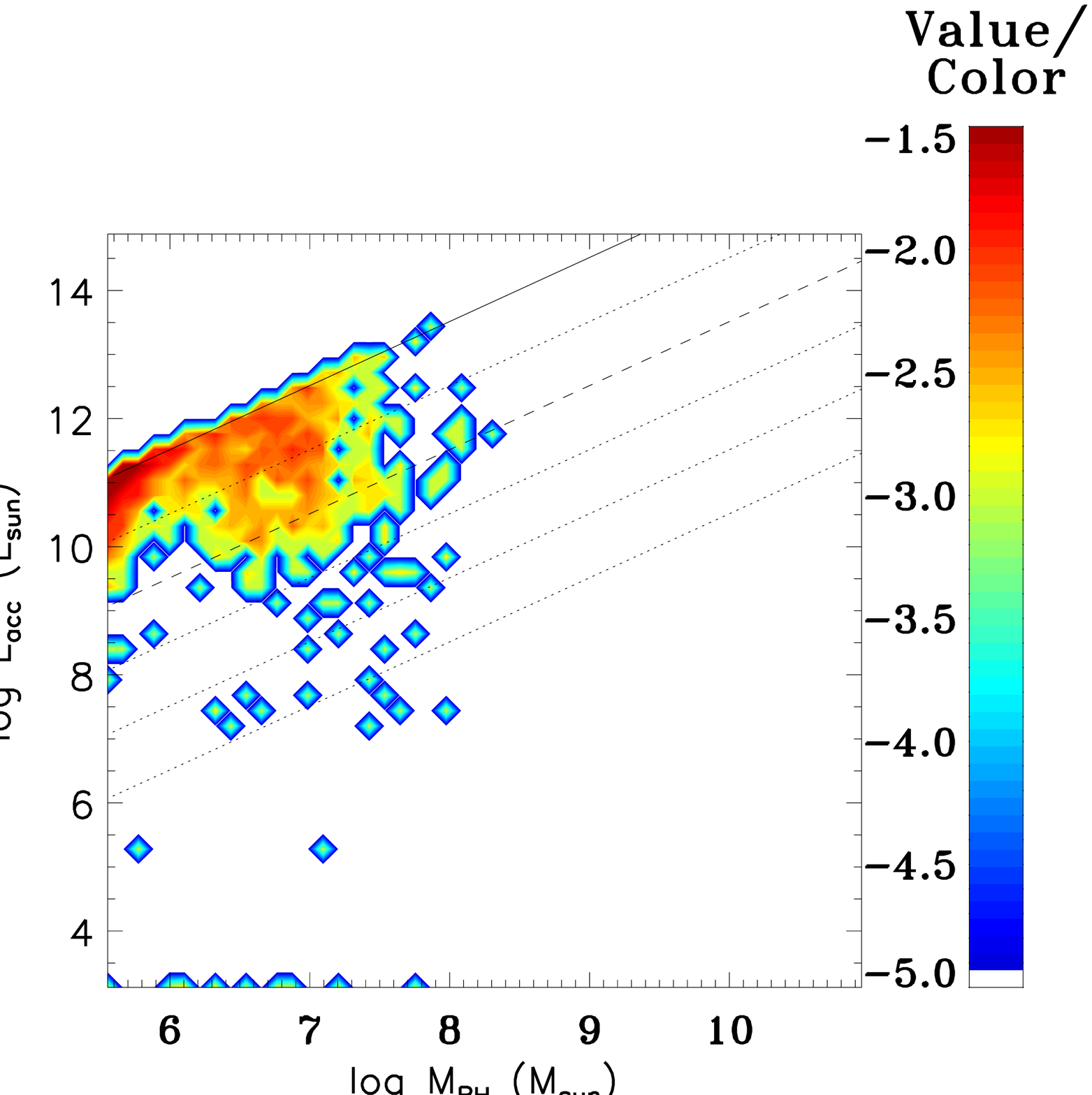}}}
  \caption{Number-weighted histogram as a function of the BH mass and their accretion luminosity for the 256L50JH simulation. The solid line corresponds to the Eddington limit. The dashed line separates our fiducial radio and quasar mode. The dotted lines corresponds to $10^{-1}\dot M_{\rm Edd}$, $10^{-3}\dot M_{\rm Edd}$, $10^{-4}\dot M_{\rm Edd}$, and $10^{-5}\dot M_{\rm Edd}$ from top to bottom. Upper left plot is for $z=0$, upper right plot for $z=1$, bottom left plot for $z=2$, and bottom right plot for $z=4$.}
    \label{bh_lum_func_MBH_256L50JH}
\end{figure*}

The behavior of the accretion rate shown in fig.~\ref{mbhvstime} for a single BH is common to BHs of a large range of  masses.
We represent the number-weighted diagram of the BH accretion luminosity $L_{\rm acc}=\dot M_{\rm BH}c^2$ versus the BH mass at different redshifts in fig.~\ref{bh_lum_func_MBH_256L50JH} for the 50 $\rm h^{-1}.Mpc$ simulation (256L50JH).
At high redshift ($z=4$), the accretion proceeds in a high accretion regime, where most BHs accrete very close to their Eddington accretion rate and release their energy in a `quasar' mode.
Some of the most massive BHs have already entered a low-accretion regime suggesting that the most massive objects are the first to self-regulate their gas content.
Then, as the simulation evolves, more and more BHs enter the low-accretion regime providing a `radio'  mode of AGN feedback.
The core of the distribution which is located at $\sim 10^{-1}$-$1\, \dot M_{\rm Edd}$ at $z=4$ is at $\sim 10^{-2}\, \dot M_{\rm Edd}$ at $z=2$, $\sim 10^{-3}\, \dot M_{\rm Edd}$ at $z=1$, and $\sim 10^{-4}\, \dot M_{\rm Edd}$ at $z=0$, with very fewer and fewer Eddington-limited BHs at lower redshifts.
We observe a lower-limit trend that goes like $M_{\rm BH}^2$ produced by a combination of the minimum density and maximum temperature reached in the CGM very close to the galaxy hosting the BH.
This lower bound evolves with time because of both a rarefaction of the gas and an increase in temperature as haloes get more shock-heated as they grow in mass~\citep{birnboim&dekel03, dekel&birnboim06, birnboimetal07}.

These diagrams suggest that the feedback from AGN at high redshifts is essentially dominated by a quasar mode ($\chi>10^{-2}$), whereas at low redshift a radio mode ($\chi \le 10^{-2}$) prevails within the core of massive structures~\citep{duboisetal10}.

\begin{figure}
  \centering{\resizebox*{!}{4.5cm}{\includegraphics{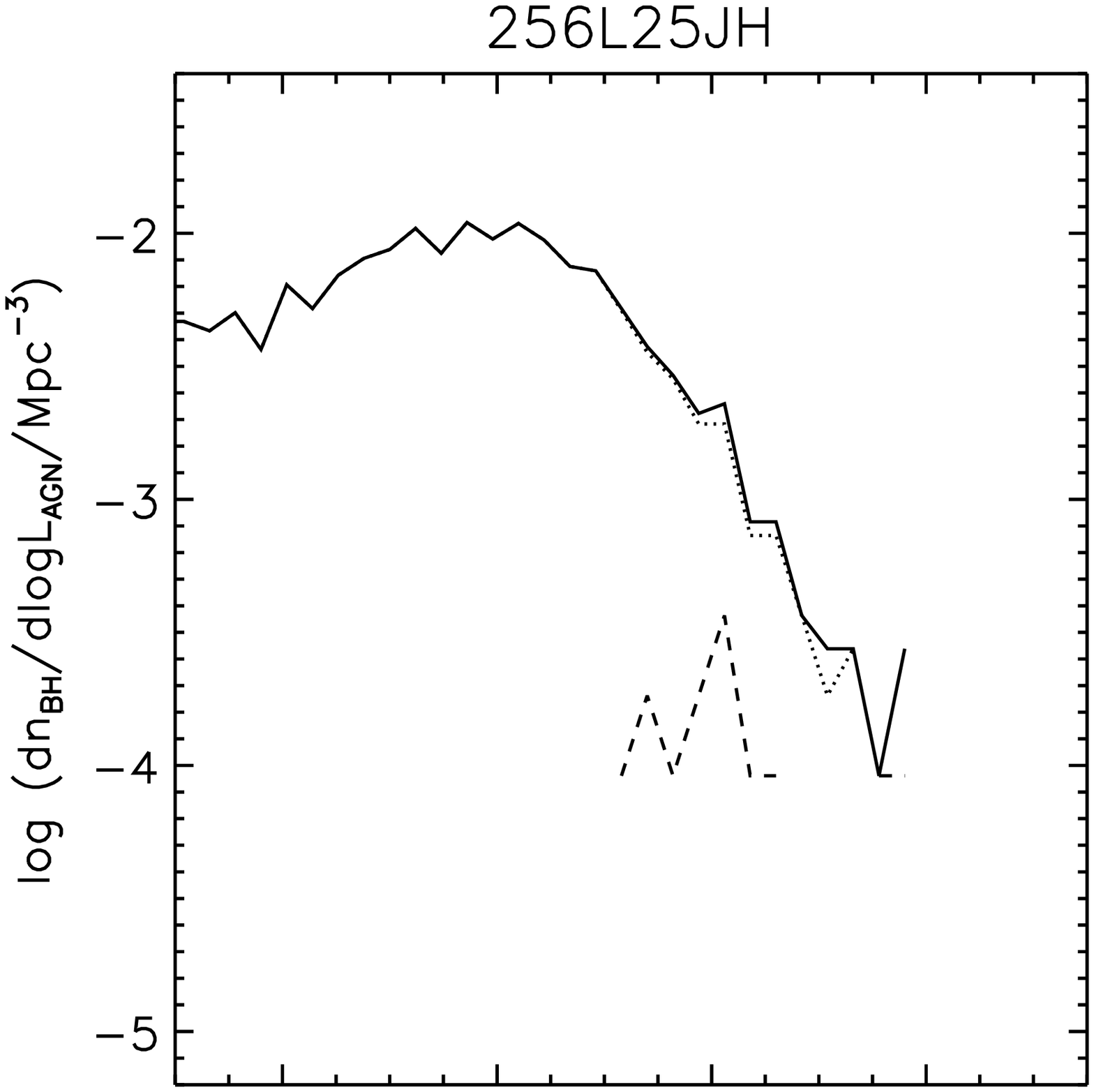}}\hspace{-1.4cm}}
  \centering{\resizebox*{!}{4.5cm}{\includegraphics{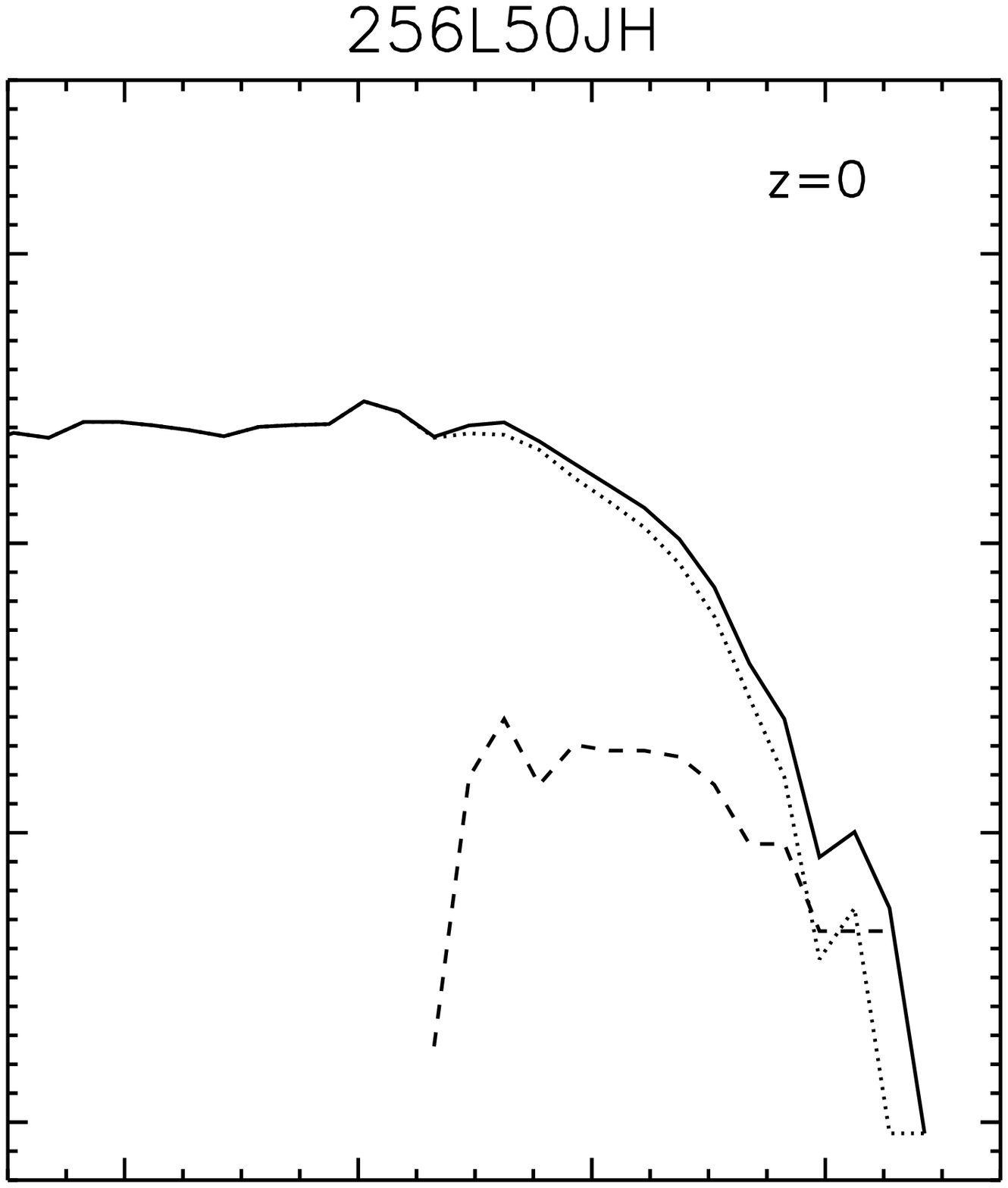}}\vspace{-0.97cm}}
  \centering{\resizebox*{!}{4.5cm}{\includegraphics{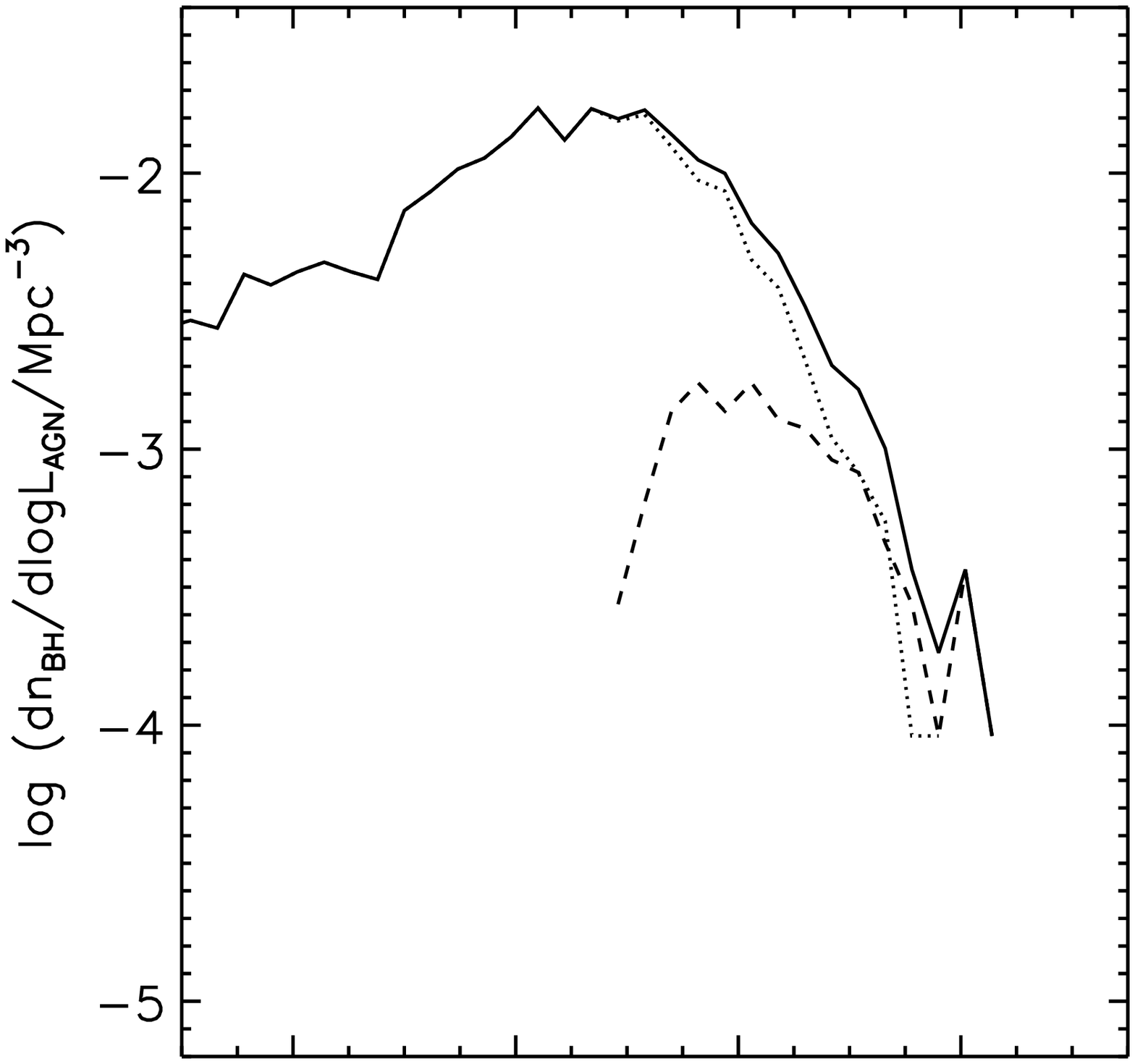}}\hspace{-1.4cm}}
  \centering{\resizebox*{!}{4.5cm}{\includegraphics{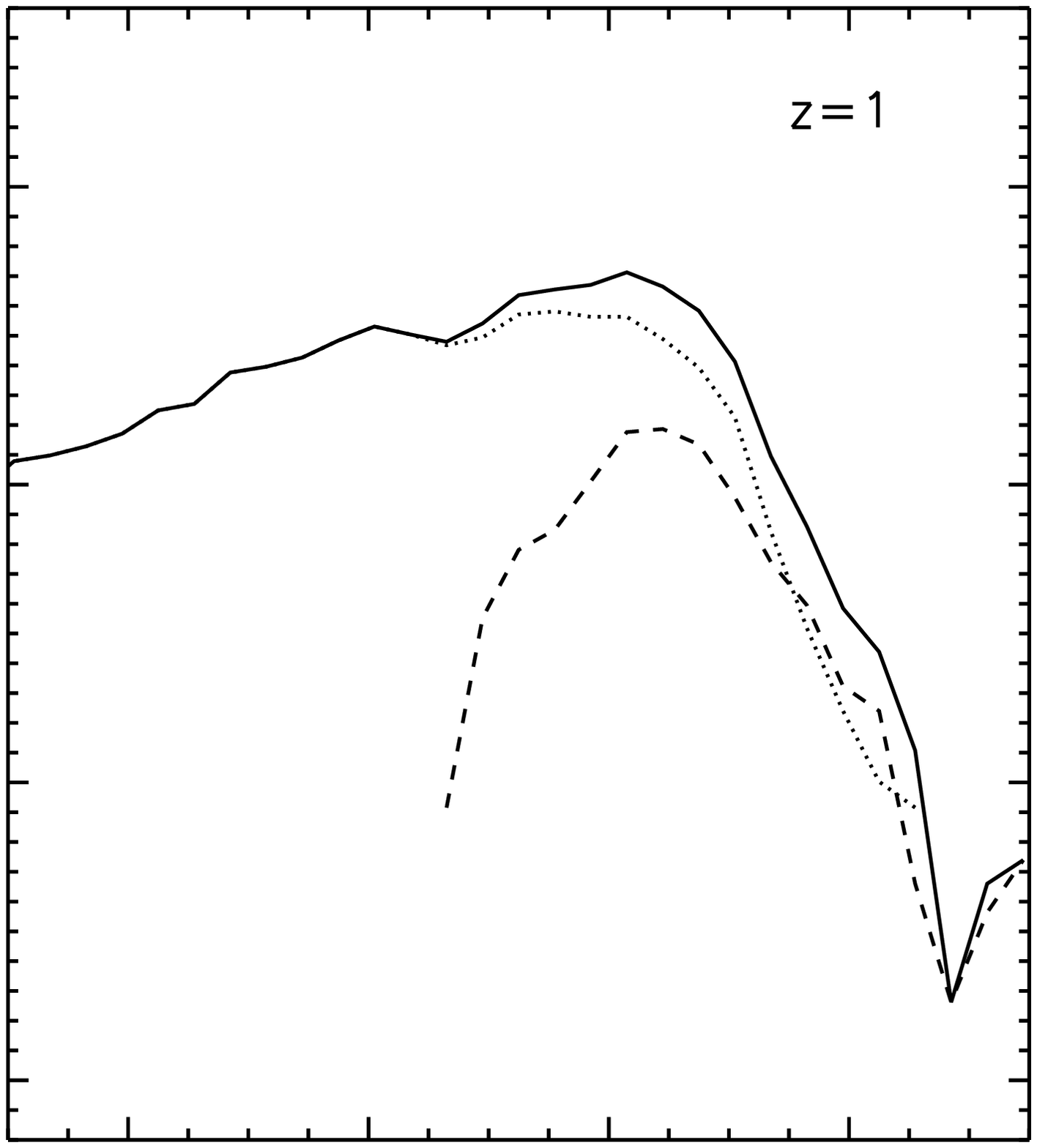}}\vspace{-0.97cm}}
  \centering{\resizebox*{!}{4.5cm}{\includegraphics{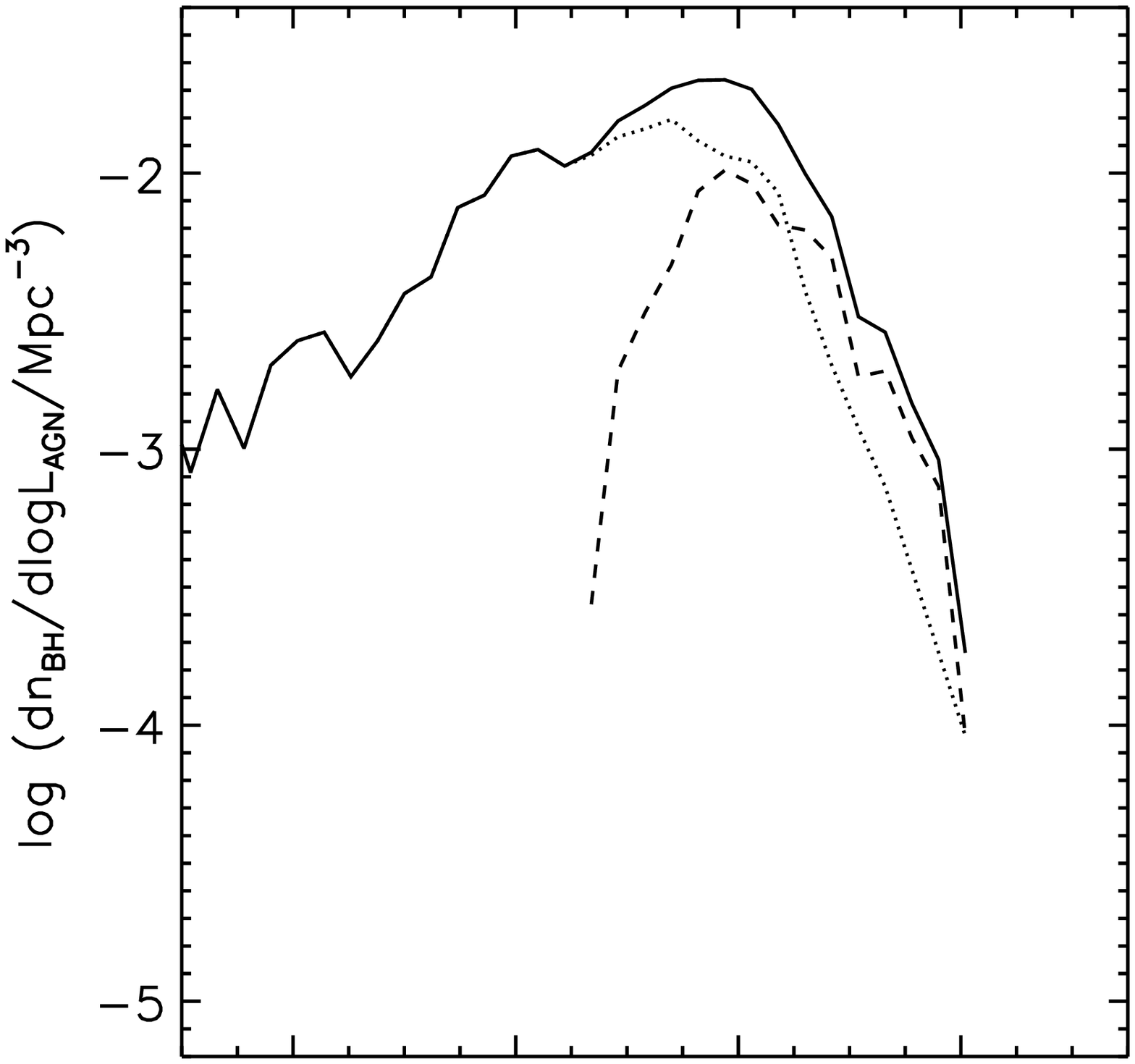}}\hspace{-1.4cm}}
  \centering{\resizebox*{!}{4.5cm}{\includegraphics{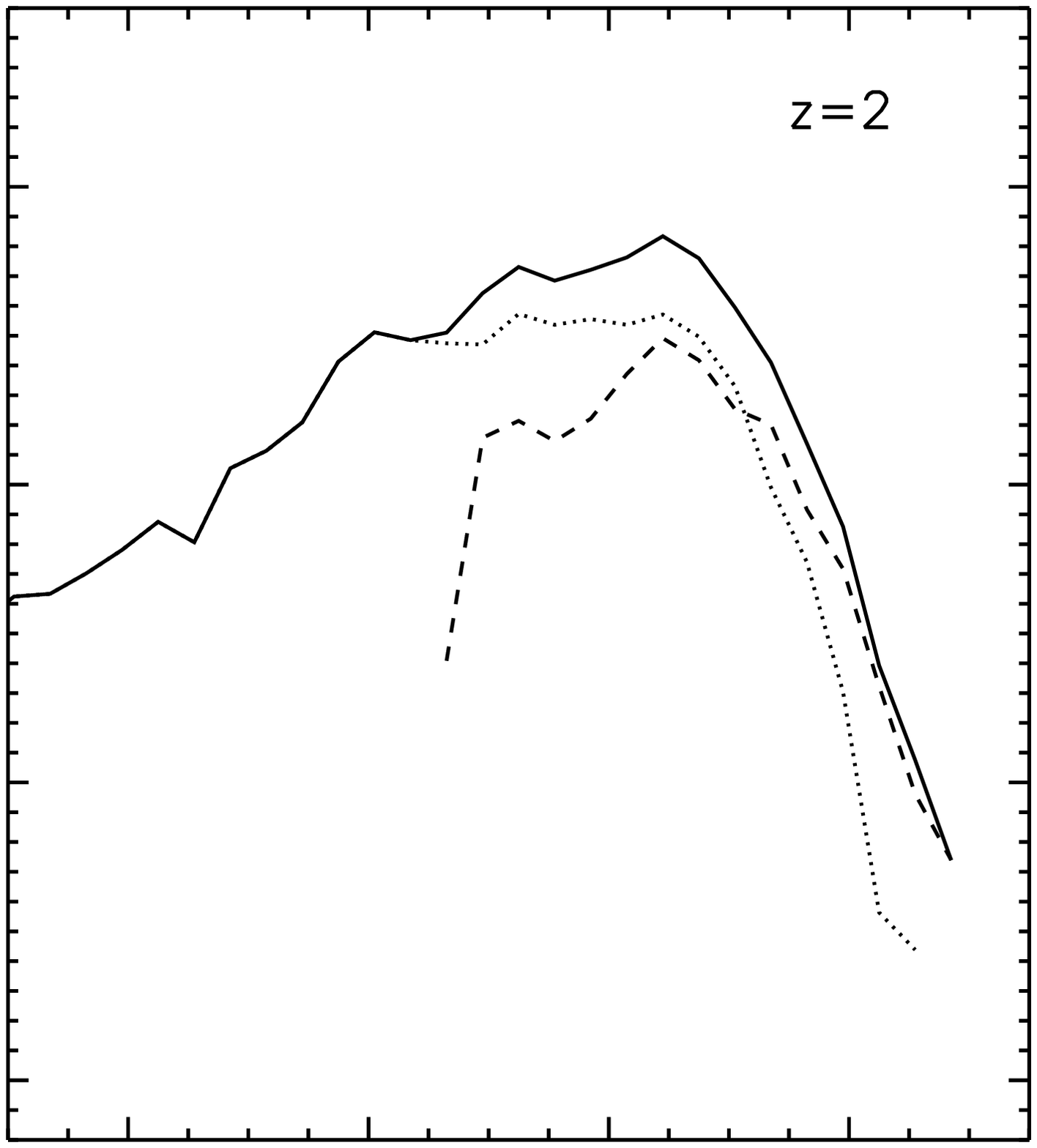}}\vspace{-0.97cm}}
  \centering{\resizebox*{!}{4.5cm}{\includegraphics{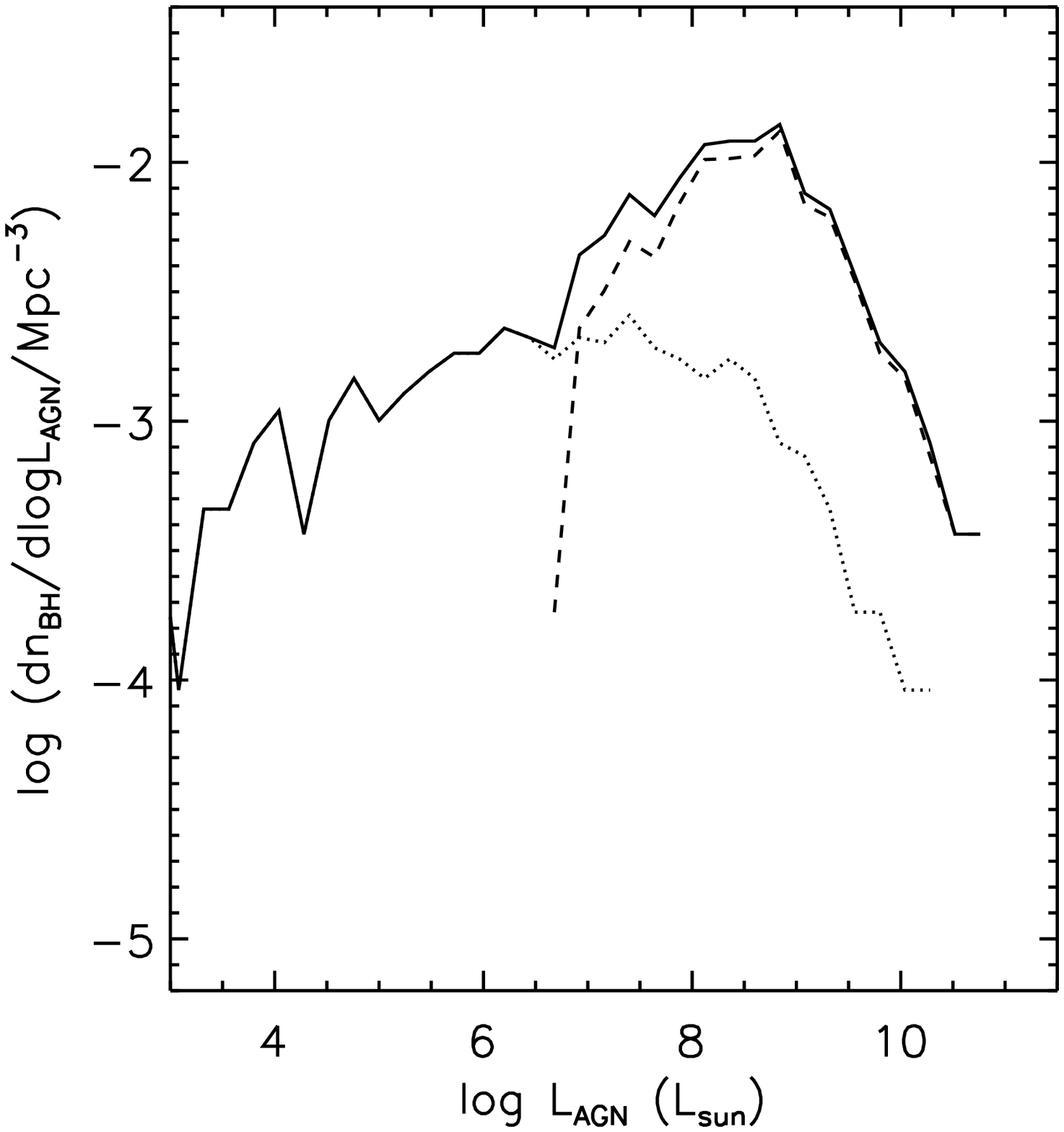}}\hspace{-1.4cm}}
  \centering{\resizebox*{!}{4.5cm}{\includegraphics{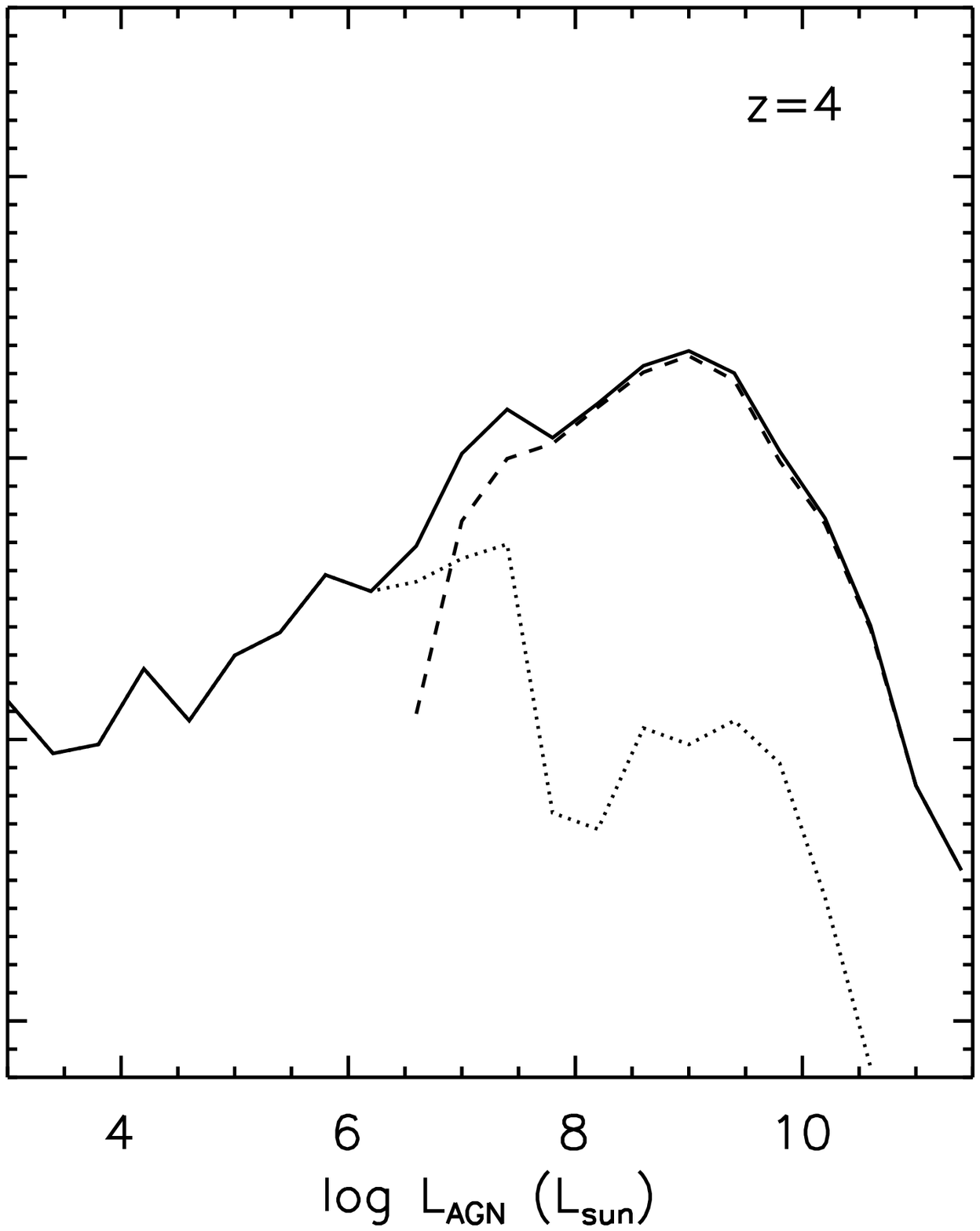}}}
  \caption{AGN luminosity functions at different redshifts for the 256L25JH simulation (left panels) and for the 256L50JH simulation (right panels). The total luminosity functions (solid lines) are decomposed into the contribution from the radio mode (dotted lines) and from the quasar mode (dashed lines).}
    \label{agn_lum_func}
\end{figure}

Fig.~\ref{agn_lum_func} shows the AGN luminosities $L_{\rm AGN}=\epsilon_f \epsilon_r \dot M_{\rm BH} c^2$ for different redshifts and two different box sizes 25 $\rm h^{-1}.Mpc$ and 50 $\rm h^{-1}.Mpc$ (256L25JH, 256L50JH). Recall that for the dual quasar/radio AGN mode,  $\epsilon_f$ depends on the accretion rate to Eddington ratio $\chi$.
The bright end of the high redshift ($z=4$) AGN luminosity function is dominated by the quasar mode because most of the BHs accrete gas at a high Eddington rate (see fig.~\ref{bh_lum_func_MBH_256L50JH}).
At intermediate redshifts, $z=2$ and $z=1$, the bright end of the AGN luminosity is  marginally dominated by the quasar mode, but the transition from the quasar mode to the radio mode appears at larger luminosities.
Emitting the largest amounts of energy, massive BHs start to strongly deplete the gas content of their host galaxies and enter a radio mode AGN regime.
At $z=0$ most of the BHs have reached a very quiescent phase for gas accretion and quasar mode feedback is almost  imperceptible.
The last remaining AGN quasars are intermediate mass BHs (see fig.~\ref{bh_lum_func_MBH_256L50JH}), with most of the very massive BHs in a radio mode.

It is worth noting that as a results of DM mass resolution, the AGN luminosity functions show extrema.
However the slope of the bright-end seems relatively independent of the resolution.

\section{Discussion and conclusions}
\label{conclusion}

We have designed a new self-consistent model of BH growth and AGN feedback for hydrodynamical cosmological simulations with a dual jet/heating mechanism that accounts for the radio mode and the quasar mode.
BHs are seeded at an early stage of the formation of galaxies.  They grow by successive mergers and by accretion of gas at a \cite{bondi52} rate.
Some of the rest-mass accreted energy is converted into energy for the AGN feedback mechanism.
At high accretion rates defined by the Eddington ratio $\chi>\chi_{\rm radio}$, BHs enter a quasar mode, with energy liberated as thermal energy.
At lower accretion rates $\chi \le \chi_{\rm radio}$, a radio mode is triggered with injection of mass, momentum and kinetic energy within a bipolar jet.

The parameters of the radio mode for the AGN feedback model are tested to reproduce the observations of the $z=0$ cosmic BH density and the \cite{magorrianetal98} relationships for $M_{\rm BH}$-$M_{\rm s}$ and $M_{\rm BH}$-$\sigma_{\rm s}$.
We find the following behavior of BH growth upon varying the parameters of the radio AGN feedback mode:
\begin{enumerate}
\item[-] AGN feedback efficiencies lower than $\epsilon_{\rm f}=1$ lead to larger and unrealistic BH masses that overshoot the observational predictions for the cosmic BH density and the Magorrian relationships.
BHs grow to larger masses to inject the same total amount of energy in order to self-regulate their growth.
\item[-] Introducing a time delay in the AGN feedback, as opposed to an instantaneous energy deposit, increases the effectiveness of AGN feedback effect on the gas and the growth of BHs, as a more important energy release occurs over a shorter period of time. 
However, this increased efficiency tends to exaggerate the effect of AGN feedback, i.e. it underestimates the BH density and overestimates the $M_{\rm BH}/M_{\rm s}$ ratios.
\item[-] The mass loading factor of the jet, i.e. the velocity of the jet, has a negligible impact on the results.
\item[-] The choice of the seed BH mass is relatively unimportant except at high redshift when BH masses are comparable to the total stellar mass of galaxies. But, the choice of the seed mass is quickly forgotten as BHs self-regulate their growth.
\item[-] The size of the jet for the radio mode, as well as the size of bubbles for the quasar mode, must be chosen carefully. Large jets and bubbles deposit energy far away from the galaxy and have 
a harder time impacting the dense gas, and, thus, self-regulating the growth of BHs. We find that the size of the energy injection region must be as close as possible to the minimum physical resolution of the code.
\item[-] We find that the radio mode alone requires a larger energy efficiency $\epsilon_{\rm f,r}=1$ than the quasar mode alone $\epsilon_{\rm f,q}=0.15$ to match the data from observations. The thermal mode couples more efficiently to the gas than the kinetic mode, and has more impact on the baryon content of galaxies.
\end{enumerate}

We also tested the convergence and the robustness of our model to the effect of finite numerical resolution using different box sizes, DM particle masses, and minimum cell sizes.
We obtain a satisfying convergence of the cosmic BH density, and $M_{\rm BH}$-$M_{\rm s}$ and $M_{\rm BH}$-$\sigma_{\rm s}$ relationships at $z=0$ as long as the DM mass resolution is $M_{\rm DM} \le 4.4 \, 10^8 \, \rm M_{\odot}$ and the minimum physical cell size is $\Delta x \le 1.52 \, \rm h^{-1}.kpc$.
However we have seen that the convergence of the cosmic SFR at high redshift is not reached because it is dominated by small halos.

These simulations have demonstrated the ability of AGN feedback to efficiently suppress the amount of stars and cold gas.
The removal of cold baryons begins at high redshift as soon as the first progenitors of massive galaxies collapse.
At low redshift, the effect of AGN feedback is amplified by the presence of more massive objects for which the impact of AGN feedback is stronger.
The presence of AGN feedback is a necessary ingredient of galaxy mass budgets as it leads to a better fit to observed SFRs that slowly converge at low redshift.

We have shown that quasars are increasingly important at high redshift.
Since cold gas is more abundant in galaxies at high redshift, BHs accrete gas at higher rates, and, as a consequence, the quasar mode of AGN feedback is more often triggered.
As the gas is consumed by star formation, and the accretion of new fresh gas is quenched by both the shock-heating of massive structures and the feedback from SN and AGN, the accretion onto BHs proceeds at lower rates.
Thus, as time goes by, the radio mode of AGN feedback becomes more and more dominant, and triggers jets in massive structures such as groups and clusters of galaxies.
Radio active galaxies comprise only a fraction of observed AGN \citep{bestetal05, smolcicetal09, smolcicetal11}. As the amount of radio emission varies a lot from object to object and depends on the gas accretion rate/mode onto the BH,  working out the reasons of quantitative (dis)agreement between our predictions and observations in detail, taking into account limitations on both sides, is a complex issue, 
beyond the scope of this paper. We therefore defer such an analysis to future work.

Our results are in good agreement with previous cosmological numerical simulations of the self-regulated growth of BHs.
The quasar mode employed here is extremely similar to that modeled with a different numerical technique by~\cite{booth&schaye09},  inspired by the model of~\cite{sijackietal07}. Our conclusions in terms of $M_{\rm BH}/M_{\rm s}$ relationships, and BH density evolution are, indeed, extremely comparable.
This is an important conclusion of this paper: cosmological codes that treat gravity and gas dynamics with different approaches produce extremely similar results for the co-evolution of BHs and galaxies with an identical set of sub-grid physics for galaxy formation.
We have implemented a radio mode for  AGN feedback and found a set of parameters that give results in good agreement with both observations and results produced by the well-tested quasar mode.
It proves that a pure kinetic mode is also capable of reducing the amount of baryons in galaxies and self-regulate the growth of BHs.
A simple combination of both modes, the dual radio/quasar mode, reproduces the observations as well as any one of the two single modes, and provides a more realistic approach to the treatment of AGN feedback.

\section*{Acknowledgments}

We thank J. Magorrian, L. Miller and A. Babul for useful discussions.
YD is supported by an STFC Postdoctoral Fellowship. The
simulations presented here were run on the TITANE cluster at the
Centre de Calcul Recherche et Technologie of CEA Saclay on allocated
resources from the GENCI grant c2009046197, and on the DiRAC facility jointly funded
by STFC, the Large Facilities Capital Fund of BIS and the University
of Oxford. This research is part of the Horizon-UK project. JD and AS' research is supported by Adrian Beecroft, the Oxford Martin
School and STFC.

\bibliographystyle{mn2e}
\bibliography{author}

\begin{appendix}

\section{Bulge and disc decomposition}
\label{AppendixA}

We decompose the bulge and the disc in our simulated galaxies with a similar procedure to what is used for decomposing observed stellar luminosity profiles.
The difference is that we do this decomposition directly on the stellar density on not on the luminosity.
Our major assumption  is that the relation between stellar surface-luminosity and stellar surface-density is linear.
The procedure is the following: 
\begin{enumerate}
\item[-] Galaxies with a minimum of 100 star particles are identified with a galaxy finder based on the Most massive Sub-node Method described in \cite{tweedetal09}, that allows for a clean separation of structures and sub-structures. This is particularly important for galaxy mergers, or for massive galaxies with extended stellar halos and galaxy satellites. 
\item[-] A stellar surface density profile is computed from the stars that belong to a given galaxy as defined by the galaxy finder, with the galaxy seen face-on. The face-on view of the galaxy is defined by the line-of-sight along the angular momentum axis defined by the rotation of the stars.
\item[-] A double decreasing exponential of the form $\propto\Sigma_i \exp{r/r_i}$ (the ``i'' subscript is for the bulge $\Sigma_{\rm b}$, $r_{\rm b}$, or for the disc $\Sigma_{\rm d}$, $r_{\rm d}$)  is fitted to the stellar surface-density profile using a least $\chi^2$ to separate the bulge from the disc component.
\item[-] The bulge mass is taken to be $M_{\rm b}=2\pi r_{\rm b}^2 \Sigma_{\rm b}$; the disc mass is taken to be   $M_{\rm d}=2\pi r_{\rm d}^2 \Sigma_{\rm d}$.
\item[-] If only one component fits the stellar density profile, we directly use the stellar mass given by the galaxy finder.
\end{enumerate}

We point out that, usually, cruder strategies are employed to separate the so-called ``bulge'' component from a galaxy's total stellar mass, by assuming that the half-mass equals the bulge mass~\citep{sijackietal07}, or using the total halo stellar mass as a proxy for the bulge stellar mass~\citep{dimatteoetal08, booth&schaye09}.
Even though our method is more consistent with the observational definition of a bulge, our tests suggest that at these kpc resolutions, other definitions lead to similar conclusions.

\begin{figure}
  \centering{\resizebox*{!}{6cm}{\includegraphics{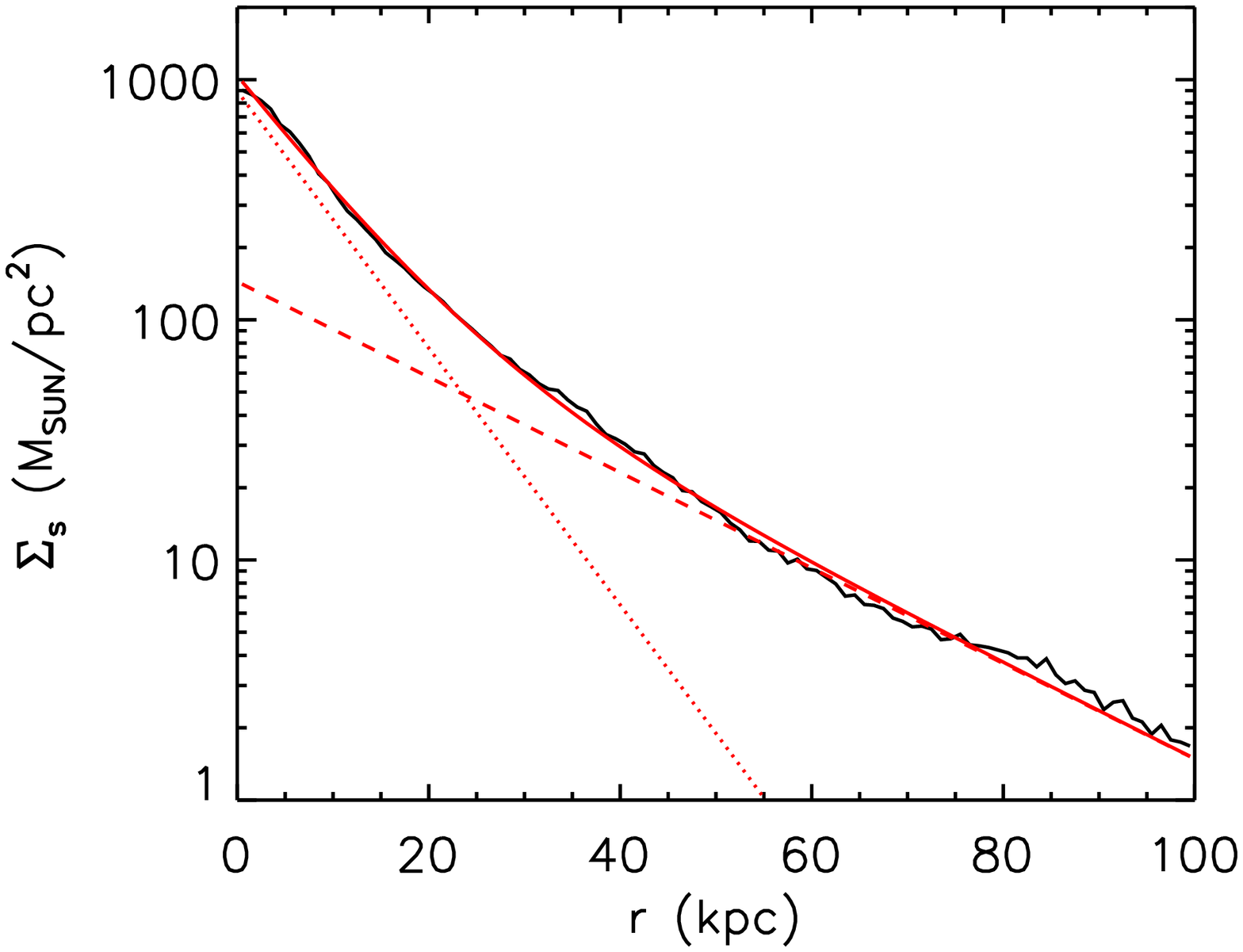}}}
  \caption{Stellar surface-density (black solid) for the most massive galaxy at $z=0$ in the 256L25JH simulation with the bulge (red dotted) and disc (red dashed) decomposition from the best fitting model (red solid).}
    \label{fit_bulge_disc}
\end{figure}

Fig.~\ref{fit_bulge_disc} shows the stellar surface-density profile obtained for the most massive galaxy in the 256L25JH simulation.
We can clearly observe an inner bulge component with size $r_{\rm b}=8.12$ kpc, and a disc component with larger characteristic radius $r_{\rm b}=21.85$ kpc.
This galaxy has a bulge mass $M_{\rm b}=3.71\, 10^{11}\, \rm M_{\odot}$ and a disc mass $M_{\rm d}=4.32\, 10^{11}\, \rm M_{\odot}$.
The sum of the two fits with the exponential forms is a good approximation of the stellar-density profile and gives a total stellar mass $M_{\rm s}=8.03\, 10^{11}\, \rm M_{\odot}$ similar to the total stellar mass obtained with the galaxy finder $M_{\rm gal}=8.34\, 10^{11}\, \rm M_{\odot}$.
We must stress that the disc component for such massive galaxies is not relevant, because this outer component of the galaxy corresponds to a large stellar halo rather than a rotating disc of stars.
However, in this paper, we do not discuss the properties of stellar discs or stellar halos of galaxies. Our main focus is to separate the bulge component from the total distribution of stars.

\end{appendix}

\label{lastpage}

\end{document}